\documentstyle[12pt]{article}
\makeatletter
\newdimen\normalarrayskip              
\newdimen\minarrayskip                 
\normalarrayskip\baselineskip
\minarrayskip\jot
\newif\ifold             \oldtrue            \def\new{\oldfalse}
\def\arraymode{\ifold\relax\else\displaystyle\fi} 
\def\eqnumphantom{\phantom{(\theequation)}}     
\def\@arrayskip{\ifold\baselineskip\z@\lineskip\z@
     \else
     \baselineskip\minarrayskip\lineskip2\minarrayskip\fi}
\def\@arrayclassz{\ifcase \@lastchclass \@acolampacol \or
\@ampacol \or \or \or \@addamp \or
   \@acolampacol \or \@firstampfalse \@acol \fi
\edef\@preamble{\@preamble
  \ifcase \@chnum
     \hfil$\relax\arraymode\@sharp$\hfil
     \or $\relax\arraymode\@sharp$\hfil
     \or \hfil$\relax\arraymode\@sharp$\fi}}
\def\@array[#1]#2{\setbox\@arstrutbox=\hbox{\vrule
     height\arraystretch \ht\strutbox
     depth\arraystretch \dp\strutbox
     width\z@}\@mkpream{#2}\edef\@preamble{\halign \noexpand\@halignto
\bgroup \tabskip\z@ \@arstrut \@preamble \tabskip\z@ \cr}%
\let\@startpbox\@@startpbox \let\@endpbox\@@endpbox
  \if #1t\vtop \else \if#1b\vbox \else \vcenter \fi\fi
  \bgroup \let\par\relax
\let\@sharp##\let\protect\relax
  \@arrayskip\@preamble}
\def\eqnarray{\stepcounter{equation}%
	      \let\@currentlabel=\theequation
	      \global\@eqnswtrue
	      \global\@eqcnt\z@
	      \tabskip\@centering
	      \let\\=\@eqncr
	      $$%
 \halign to \displaywidth\bgroup
    \eqnumphantom\@eqnsel\hskip\@centering
    $\displaystyle \tabskip\z@ {##}$%
    &\global\@eqcnt\@ne \hskip 1\arraycolsep
         $\arraymode{##}$\hfil
    &\global\@eqcnt\tw@ \hskip 1\arraycolsep
	 $\displaystyle\tabskip\z@{##}$\hfil
	 \tabskip\@centering
    &{##}\tabskip\z@\cr}
\makeatother

\def\theequation{\arabic{section}.\arabic{equation}}

\def\be{\begin{eqnarray}}
\def\ee{\end{eqnarray}}
\def\nn{\nonumber}
\def\n{\noindent}

\def\Bf#1{\mbox{\boldmath $#1$}}

\def\balpha{{\Bf\alpha}}
\def\bbeta{{\Bf\beta}}
\def\bnu{{\Bf\nu}}
\def\bmu{{\Bf\mu}}
\def\bphi{{\Bf\phi}}
\def\bN{{\Bf N}}

\def\bJ{{\Bf J}}

\def\bsN{{\Bf N}}

\def\bsalpha{{\Bf\alpha}}
\def\bsbeta{{\Bf\beta}}
\def\bsphi{{\Bf\phi}}

\begin{document}

\phantom. \hfill ITEP-M2/93\footnote{ITEP 117259 Moscow;
morozov@vxdesy.desy.de}    \\
\phantom. \hfill ITFA 93-10 \\
\phantom. \hfill {\it March 1993}           \\

\bigskip

\phantom. \hfill A.Morozov                    \\

\bigskip

\centerline{\Large\bf INTEGRABILITY AND MATRIX MODELS}

\bigskip

\bigskip

\bigskip

\centerline{ABSTRACT}

\bigskip

The theory of matrix models is reviewed from the point of view of its relation
to integrable hierarchies. Discrete 1-matrix, 2-matrix, "conformal"
(multicomponent)  and Kontsevich models are considered in some detail,
together with the Ward identites ("W-constraints"), determinantal formulas and
continuum limits, taking one kind of models into another. Subtle points and
directions of the future research are also discussed.

\bigskip
\bigskip
Lectures presented at the University of Amsterdam and \\
the Dutch Institute for High Energy Physics(NIKHEF)\\
in February-March 1993
\newpage

\bigskip

CONTENT
\bigskip

\n
1. Introduction  \\
2. Ward identities for the simplest matrix models

2.1. Ward identites versus equations of motion

2.2. Virasoro constraints for the discrete 1-matrix model

2.3. CFT formulation of matrix models

2.4. Gross-Newman equation

2.5. Ward identities for Generalized Kontsevich Model

2.6. Discrete Virasoro constraints for the Gaussian Kontsevich model

2.7. Continuous Virasoro constraints for the $V = \frac{X^3}{3}$ Kontsevich
       model

2.8. $\tilde W$-constraints for the asymmetric 2-matrix model

2.9. $\tilde W$-constraints for generic 2-matrix model

2.10. $\tilde W$-operators in Kontsevich model\\
3. Eigenvalue models

3.1. What are eigenvalue models

3.2. 1-matrix model

3.3. Itzykson-Zuber and Kontsevich integrals

3.4. Conventional Multimatrix models

3.5. Determinant formulas for eigenvalue models

3.6. Orthogonal polinomials

3.7. Two-component models in Miwa parametrization

3.8. Equivalence of the discrete 1-matrix and Gaussian Kontsevich
               models

3.9. Volume of unitary group\\
4. Integrable structure of eigenvalue  models

4.1. The concept of integrability

4.2. The notion of $\tau$-function

4.3. $\tau$-functions, associated with free fermions

4.4. Basic determinant formula for the free-fermion correlator

4.5. Toda-lattice $\tau$-function and linear
reductions of Toda-lattice hierarchy

4.6. Fermion correlator in Miwa coordinates

4.7. Matrix models versus $\tau$-functions.

4.8. String equations and generic concept of reduction

4.9. On the theory of GKM

4.10 1-Matrix model versus Toda-chain hierarchy\\
5. Continuum limits of discrete matrix models

5.1. What is continuum limit

5.2. From Toda-chain to KdV

5.3. Double-scaling limit of 1-matrix model

5.4. From Gaussian to $X^3$-Kontsevich model\\
6. Conclusion\\
7. Acknowledgements\\
8. References

\newpage
\setcounter{page}{1}

\section{Introduction}

\setcounter{equation}{0}

The purpose of these notes is to review one of the branches of modern string
theory: the theory of matrix models with the emphasize on their intrinsic
integrable structure. We begin with a brief description of the field and
its place in the closest environement within entire string theory.

The main content of the string theory
\footnote{See \cite{UFN} for a general review}
is the study of symmetries in the broadest possible sense of the word
by methods of the quantum field theory. The usual scheme is to start
from some symmetry and construct a field-theoretical model (usually
2-dimensional, for the reason that we do not discuss here),
which possesses this symmetry in some simple sense (e.g. as Noether
symmetry or as a chiral algebra). The main idea at this stage is to
find a model which is exactly solvable (if nothing but the symmetry
is given this is a nice principle to restrict dynamics). The next step
is to study the hidden symmetries of the model, which are somehow
responsible for its exact solvability and are usually much larger
than the original symmetry.

This "inverse" step: {\it model} $\longrightarrow$ {\it symmetry}
can be made at least with three different ideas in mind.

One can look for some hidden local (gauge) symmetry of the model,
which is fixed or spontaneously broken, i.e. identify it with some
other model which had more fields - $auxiliary$ from
the point of view of the smaller model and $gauge$ - from that of
the larger one. (Examples: gauged Wess-Zumi\-no-No\-vikov- Witten (WZNW)
model, topological theories in BRST formalism etc.)

One can take for a new (full) symmetry of the model just its operator
algebra (algebra of observables) (see \cite{KleP},
\cite{W(KleP)} and also \cite{Klepo} for the first results in this
direction). It deserves mentioning that $gauging$ of entire algebra
of observables gives rise to a "string field theory", associated
with original model (considered as a string model).

One can construct effective action of the theory by exact evaluation
of the functional integral.

As to direct step {\it symmetry} $\longrightarrow$ {\it model},
one can take as example the  best understood case, when
original symmetry is just a Lie algebra. Then the quantum mechanical
model can be constructed by geometrical quantization technique (see
\cite{AS} for the most important example of Kac-Moody algebra and
the WZNW model).

{}From the mathematical point of view the two elements of the above
scheme look like {\it algebra} (theory of symmetries) and
{\it Analysis} \& {\it Geometry} (field-theoretical models). The idea of
constructing models with a given symmetry (and nothing else relevant
for the dynamics) can be identified with the mathematical concept of
"universal objects".

\bigskip
\bigskip
\centerline{$|\ \ \ SYMMETRY \ \ \ | \ \ \ \ \ =
                         \ \ \ \ \ |\ \ \ {\bf ALGEBRA}\ \ \ |$}
\bigskip
\centerline{$\ \ \ \uparrow \ \ \ \downarrow\ \
                          \ \ \ \ \ \ \phantom{.................}$}
\bigskip
\centerline{$|\ \ \ \ \ \ \ \ \ MODEL \ \ \ \ \ \ \ \ \ | \ \ \ \ \ =
                   \ \ \ \ \ | \ \ \ {\bf ANALYSIS} \ \ \ \ \ \ |$}
\centerline{$|\ \ with\ this\ symmetry\ \ | \ \ \ \ \ \phantom.
                  \ \ \ \  \ \ \ \ | \ \ \ {\bf GEOMETRY} \ \ \ |$}
\bigskip
\centerline{\it Theory of everything}
\bigskip

The sequence of iterations of the two arrows in the picture leads to
a deaper understanding, enlargening and generalization of all the
notions involved: symmetry, exact solvability, field theory,
geometrical structures, quantization etc, thus stimulating considerable
progress both in physics and mathematics. If this iterative process
can somehow converge, the limit point will deserve the name of the
{\it theory of everything}, which will indeed unify all the
possible field theoretical models by embedding them into a huge, but
well structuired theory, which will be also exactly solvable in some,
yet unspecified sense of the word. We refer to \cite{UFN} for more details
about this semiphilosophical $programm$, known under the name of
(modern) $string$ $theory$, and now turn to a more narrow subject:
the theory of matrix models.

\bigskip

At the moment it is mainly associated with the theory of effective actions,
at least so far this is where the main results of the modern theory
of matrix models find their applications. This technique is especially
suited for the study of effective actions, obtained after integration
over 2-dimensional geometries (including the sum over genera) and
it produces non-perturbative (exact) partition functions of particular
string models. The main result of these studies points out that these
partition functions exhibit two remarkable (though expected \cite{GLM})
properties:

First, effective action for a given model is essentially the $same$ as for any
other model. In fact effective action is a function of coupling constants
("sources" in the old-fashioned terms), which are nothing but coordinates
in the $space\ of$ various $models$ (configuration space of entire string
theory): variation of couplings change one model for another.

Second, effective action possesses a huge additional symmetry, which is
somewhat similar to the general covariance in the space of all models
(the above mentioned configuration space) and in the simplest examples,
which have been studied so far,
can be expressed in terms of integrable hierarchies.
(This "general covariance" in the configuration space can after all turn
into the main dynamical principle of the string theory.)

Both these features seem to be very general, arising whenever the largest
possible Lagrangian with a given symmetry is considered (without restrictions
on the possible counterterms, imposed by requirements of renormalizability
or by locality-minimality "principles" - this is why this phenomenon is
not widely known to field theorists). An example of highly-nontrivial
calculations leading to similar conclusions can be found in ref.\cite{KoS}.

We hope that these remarks will become more clearer after some
specific examples will be considered below. Still they deserve being
fromulated in the full generality, not only to intrigue the reader, but also
because they can serve for better understanding of the ideas and outcomes of
generic string theory.

\bigskip

The "corner" of the string theory, associated with matrix models, can be
described by the following picture: see Fig.1.

\bigskip

The big blocks within the body of string theory, which are directly related
to matrix models are: theory of conformal models, that of the $N=2$
supersymmetry and the (loop-equation version of) the Yang-Mills theory (in any
dimension). Also the Einstein gravity should be related to the subject in a
way, similar to Yang-Mills theory, but these links are yet not clarified.

Both conformal theory and $N=2$ supersymmetry are sources of the concept of
"topological models" \cite{Wito}-\cite{BBRT}.
These arise after gauging of all continuous symmetries
of the WZNW models and/or as models with BRST-exact stress tensors, naturally
appearing in the context of $N=2$ supersymmetry. If formulated in a
self-consisted way in the "universal module space" (unification of module
spaces of all finite-genus Riemann surfaces and bundles over them)
these models turn into those of "topological gravity". Generating functionals
of topological gravity models in fact generate infinite sequences of
topological invariants of certain spaces (inverse definition is also
possible in some cases \cite{Wito}, though the universal (generic) algorithm
for the operation {\it topology of some space} $\longrightarrow$ {\it
topological gravity} is not yet formulated).

Alternative models of $2d$ quantum gravity arise straightforwardly from
conformal models through a procedure of "summation over geometries". There are
two essen\-tially different approaches to the problem. One ("Polyakov
approach") is to make use of the complex structure, intrinsic for conformal
theory \cite{BPZ} and sum over $Riemann$ $surfaces$, what involves integration
over module spaces and sum over genera. The main technique used in this
approach is the theory of free fields on Riemann surfaces \cite{KnUFN},
\cite{MP} and bosonization formalism for conformal field theories \cite{BosF},
\cite{GMMOS}.
This approach requires solution of Liouville theory, which still remains a
problem under intensive investigation (in turn related to conformal field
theory). Further progress in this direction should be related (or can be
expressed in terms of) the adequate theory of the {\it universal module
space}, {\it handle-gluing operators} etc. Similar objects arise in the
field-theoretical approach to the topological gravity (see \cite{LP} for a
recent review).

Alternative approach to summation over geometries does not refer at all to the
complex structure and instead involves a sum over random $equilateral$
triangulations \cite{Migrev}-\cite{mamo}.
\footnote{Its relation to the Polyakov approach is a separate
very interesting, important and badly understood problem, which allows a
non-trivial reformulation in terms of number theory (see \cite{LM}). The main
puzzle here is that equilateral triangulations are in fact $arithmetic$
Riemann surfaces - a dense discrete subset in the entire module space, with
interesting and deap algebraic properties. Equivalence of the two approaches
to $2d$ quantum gravity should imply the existence of some number-theoretical
background behind the scene, which would be very nice to discover in full
purity.}
This is the place where matrix models first appear in the context of string
theory. The random triangulation approach is by no means specific for
conformal models (since it ignores the complex structure) and can be applied
in many other situations - for example, to Yang-Mills (YM) theories in any
dimensions (where instead of summation over geometries one needs "simply" to
sum over ordinary Feynman diagramms).

Applications of the matrix-model method usually involve two steps: formulation
and the study of "discrete" model and then taking its "continuum limit",
giving rise to a new  - "continuous matrix model", which
sometimes can again be represented in a form of some  matrix integral.

One of the main dis\-coveries in the field of mat\-rix models is that
$continuous$
models arising finally from the random-equilateral-triangulation
description of the simplest (minimal with $c<1$) string models coincide with
the simplest ($CP^1$ Landau-Ginzburg)  models of topological gravity
\cite{WitTG},\cite{Ko}-\cite{MMM}: two
(classes of) theories are identical (this is not yet proved in full detail,
but is more than plausible).

So far $continuous$ models are actually found and somehow understood only for
string models, based on the $c<1$ minimal conformal theories (moreover, only
for $q=1$ in the $(p,q)$-seria). Conformal models with $c \geq 1$, which
are relevant for description of gauge theories in space-time dimension
$d \geq 2$ (which possess particles, rather than only topological degrees
of freedom), should give rise to the discrete matrix models with
"non-factorizable" integration over "angular variables", of which the
simplest (solvable) example is Kazakov-Migdal model \cite{KazMi}.
The issue of continuum limit for such models is yet not understood
(at least in terms of integrable structures, which should probably
generalize the familiar theory of Toda hierarchies).

The goal of the study of matrix models is three-fold. First of all, one can
look for the non-perturbative (exact) answers for the physical amplitudes in
the given model. This is the subject which attracts most attention in the
literature (for several obvious reasons). However, it is equally (and, perhaps,
even more) important to understand the mathematical structure behind the
matrix models (which involves topics like general theory of integrable
hierarchies, geometrical quanization, Duistermaat-Heckman theorem
("localization theory") etc). Also important for the purposes of string theory
is to use the results of the study of matrix models in order to unify $a\
priori$ different models (according to the above mentioned principle:
non-perturbative partition functions for different models differ by a change
of variables in the space of coupling constants). Matrix models already played
an important role in making this principle more clear and acceptable for many
string theorists.

\bigskip

Let us make the next step and look even closer at the field of matrix models,
especially, at its mostly studied domain, associated with the $d<2$
string models. Then the following structure will be seen: Fig.2.

\bigskip

The sample example of matrix model is that of 1-matrix integral
\be
Z_N\{t\} \equiv c_N\int_{N\times N}
dH e^{\sum_{k=0}^{\infty} t_k {\rm Tr} H^k},
\label{1mamo}
\ee
where the integral is over $N\times N$ Hermitian matrix $H$ and
$dH = \prod_{i,j} dH_{ij}$.
There are three directions in which one can proceed starting from
(\ref{1mamo}).

The first one \cite{GMMMO}
is to look for an invariant formulation of properties of the
functional $Z_N\{t\}$. It appears to satisfy the infinite set of differential
equations (in fact these are just Ward identities for the functional integral
(\ref{1mamo}) \cite{MM}):
\be
&L_n Z_N\{t\} = 0, \ \ \ n\geq -1, \nn \\
&L_n \equiv \sum_{k=0}^{\infty} kt_k\frac{\partial}{\partial t_{k+n}} +
    \sum_{k=0}^n \frac{\partial^2}{\partial t_k\partial t_{n-k}},
\label{vird} \\
&\frac{\partial}{\partial t_0}Z_N = NZ_N, \nn
\ee
which is known under the name of "discrete Virasoro constraints".
$Z_N\{t\}$ can be represented as a correlator of screening operators in
some auxiliary conformal model (of one free field on the "spectral surface"),
and Virasoro constraints (\ref{vird}) are of course related to the Virasoro
algebra in that conformal model.
Also $Z_N\{t\}$ is some $\tau$-function of integrable "Toda-chain" hierarchy
(in fact this statement should be a corollary of the Virasoro constraints,
but this relation is still not very well understood).

The most straightforward further developement \cite{GMMMO},\cite{MMMM}
is to take the continuum limit
of the Toda-\-chain hierarchy.
In the specially adjusted ("double-scaling") limit
\cite{mamo}) it gives rise to the KdV-hierarchy, and the corresponding
$\tau$-function appears subjected to the slightly different constraints
\cite{FKN},\cite{MMMM}
(which again form a Borel subalgebra of some other "continuous
Virasoro algebra"):
\be
&{\cal L}_{2n} {\cal Z}^{cont}\{T\} = 0, \ \ \ n\geq -1, \nn \\
&{\cal L}_{2n} \equiv \frac{1}{2}\sum_{{\rm odd}\ k=1}^{\infty}
       k(T_k + r_k)\frac{\partial}{\partial T_{k+2n}} +
 \frac{1}{4} \sum_{{\rm odd}\ k = 1}^{2n-1}
\frac{\partial^2}{\partial T_k\partial T_{2n-k}} + \nn \\
&\quad +   \frac{1}{16}\delta_{n,0} + \frac{1}{4} (T_1+r_1)^2\delta_{n,-1},
\label{virc}
\ee
where $r_k = -\frac{2}{3}\delta_{k,3}$.
In fact
\be
{\cal Z}^{cont}\{T\} \sim \left.\lim_{d.s.\{N \rightarrow\infty\}}
\sqrt{Z_N\{t\}} \right|_{t_{2k+1}=0},
\label{coverdi}
\ee
and $T$ are related to $t$ by linear transformation \cite{Kmamo},\cite{MMMM}:
\be
 T_k = \frac{1}{2}\sum_{m\geq \frac{k-1}{2}} \frac{g_m}{(m -\frac{k-1}{2})!}
      \frac{\Gamma(m+\frac{1}{2})}{\Gamma(\frac{k}{2}+1)},  \ \ \ k\ {\rm
odd}; \nn \\
   g_m = mt_{2m},\ m\geq 1;\ \ g_0 = 2N.
\label{Tt1mamo}
\ee
This ${\cal Z}^{cont}\{T\}$ can be again represented in the form of a matrix
integral (over $n\times n$ Hermitian matrix) \cite{Ko},\cite{GKM}-\cite{AvM}:
\be
{\cal Z}^{cont}\{T\} = {\cal Z}_V\{T\}
\label{T=Q}
\ee
with $V(X) = \frac{X^3}{3}$, where
\be
{\cal Z}_V\{T\} \sim {\cal F}_{V,n}\{ L\} \equiv
\int_{n\times n} dX e^{- {\rm tr} V(X) + {\rm tr} L X}
\label{GKM}
\ee
and
\be
T_k = \frac{1}{k} {\rm tr} L^{-k/2},\ \ \  k\ {\rm odd}.
\label{TtK}
\ee
The function ${\cal Z}_V\{T\}$ (but not ${\cal F}_{V,n}\{ L\}$)
is in fact independent of $n$: the only
thing that happens for finite values of $n$ is that the r.h.s. of (\ref{GKM})
can not describe ${\cal Z}_V\{T\}$ at $arbitrary$ points in the $T$-space,
in accordance with (\ref{TtK}). Continuous Virasoro constraints (\ref{virc})
are in fact equivalent to the trivial matrix-valued Ward identity
\be
\left( V'\left(\frac{\partial}{\partial L_{\rm tr}}
\right) -  L\right) {\cal F}_{V,n}\{ L\} = 0.
\label{WiK}
\ee

Another direction to proceed from the discrete 1-matrix model is to rewrite it
identically in the form of Kontsevich model: this time with
$V(X) = X^2$ and additional factor of $(det\ X)^N$ aunder the integral in
${\cal F}_{V,n}\{ L\}$
\cite{ChMa}. Then the double scaling limit can be
studied in internal terms of Kontsevich models \cite{Toda}.

The third direction is towards multimatrix models. In $continuous$
version they should provide $\tau$-functions of reduced KP-hierarchies
\cite{Doug} (KdV is the $p=2$ reduction), which are subjected to
"continuous $W$-constraints" \cite{FKN}. Matrix models of such
$\tau$-functions are Kontsevich models with $V(X) \sim X^{p+1}$
\cite{GKM}-\cite{AvM}. At $discrete$ level, however, things are not so
simple. The most popular discrete multimatrix models  \cite{mumamo}
are defined as the multiple matrix integrals of the form:
\be
\new
\begin{array}{c}
Z_N\{t^{(\alpha)}\} \equiv
 \\
c_N^{p-1}\int_{N\times N} dH^{(1)}...dH^{(p-1)}
\prod_{\alpha = 1}^{p-1} e^{\sum_{k=0}^{\infty}t_k^{(\alpha)}{\rm Tr}
H_{(\alpha)}^k} \prod_{\alpha = 1}^{p-2} e^{{\rm Tr}H^{(\alpha)}H^{(\alpha+1)}}
\label{mumamo}
\end{array}
\ee
(the form of the "interaction term" ${\rm Tr}H^{(\alpha)}H^{(\alpha+1)}$ is
restricted by the "solvability" principle, but not unambiguously).
In fact these models are particular examples of the "scalar-product
eigenvalue models"
and are not really distinguished except for the 1-matrix ($p=2$) and 2-matrix
($p=3$) cases. This is reflected in the absense of any reasonable Ward
identites and integrable structures for these models, which would somehow
involve their dependence on the variables $t^{(\alpha)}$ with $2\leq \alpha
\leq p-2$. Therefore the "multi-scaling continuum limit" of these models can
hardly be
investigated at any degree of rigourousness. (It is not so much important
for "physical" applications to have any discrete models associated with the
continuum ones, but this is an interesting problem for the "science for
science".)
For the 2-matrix ($p=3$) case the Ward identities can be
expressed in the form of "$\tilde W$-constraints" \cite{tildeW} and look like
\cite{GKM}
\be
\tilde W_{n-m}^{(m+1)}\{t\} Z_N\{t,\bar t\} = (-)^{m+n}
\tilde W_{m-n}^{(n+1)}\{\bar t\} Z_N\{t,\bar t\}
\label{Wtilde}
\ee
(here $t$ and $\bar t$ stand for $t^{(1)}$ and $t^{(2)}$, and $m,n$ are any
non-negative integers).

The really interesting set of discrete multimatrix models does exist, but it
is somewhat different from (\ref{mumamo}). These theories will be refered
to as "conformal matrix models", since they arise straightforwardly as
generalization of the "CFT-formulation" of the 1-matrix model \cite{comamo}: it
is enough to substitute discrete Virasoro constraints in the theory of one free
field by the $W_p$-constraints in the theory of $p-1$ free fields. Matrix
integral formulation then involves an "interaction term"
$Det \left(H^{(\alpha)}\otimes I - I \otimes H^{(\alpha+1)}\right)$ instead of
$e^{{\rm Tr}H^{(\alpha)}H^{(\alpha+1)}}$, which is not very easy to guess
$a\ priori$, but so defined models and their continuum limits can be examined
in a manner, quite parallel to the one-matrix case (though not all is already
done in this direction). Also this approach opens a possibility to formulate
discrete models for any set of constraints, e.g. assosiated with the more
exotic
$W$-algebras and with quantum groups (i.e.can help to solve the inverse
problem: $constraints\ \longrightarrow\ discrete\ matrix\ model$). This is an
option which also deserves further investigation. Another natural name for
this set of theoris is "multi-component eigenvalue models".

Kontsevich models should be also related to topological models of
Landau-\-Ginzburg gravity (LGG), though this relation is not yet clarified in
full detail (see, however \cite{Krich},\cite{LP}) .

Among the main unresolved puzzles in this whole field is the description of
generic $(p,q)$-models. Formally, Generalized Kontsevich model (\ref{GKM})
provides this description, but in fact the partition function
($\tau$-function) gets singular when the "phase transition" point where the
$q$ changes is approached, and Kontsevich model with $V(X) = polinomial\ of
\ degree\ p+1$ provides a nice description only of $(p,1)$-models.
Generically, Kontsevich integral describes a duality transformation between
$(p,q)$ and $(q,p)$ models: $(p,q) \longrightarrow (q,p)$ \cite{KhMa}, but not
any of these models separately. (The only exclusion are $(p,1)$-models because
they are related by Kontsevich transformation to the $(1,p)$ models which are
completely trivial.)

In fact continuous models have two $different$ sets of "time-variables".
Thus far we introduced $T$, which are  essentially expansion parameters
of the generating functional for correlation functions. More exact, these
parameters $\hat T$ depend on the particular model (vacuum), around which the
perturbation expansion is performed, and they differ slightly from the
model-independent $T$. Another set of "times",
$r_k = \frac{p}{k(p-k)}{\rm Res}\left(V'(\mu)\right)^{1-\frac{k}{p}}d\mu$,
parametrizes the shape of the polinomial "potential" $V_p(X)$ (of degree
$p+1$) and it describes the coordinates in the space of (matrix)
models. These two types of variables - parameters of the generating
functional and those labeling the shape of the Lagrangian -
are almost the same (in fact they would be
$just$ the same, if there were no loop (quantum) effects). This similarity
between $T$'s and $r$'s is reflected in the remarkable property of partition
function of the $(p,1)$ model  - it essentially
depends only on the {\it sum}  of  "times" $\hat T$ and $r$ \cite{Krich}:
\be
{\cal Z}_{V_p}\{T\} = f_p(r\mid \hat T_k + r_k)\tau_p\{\hat T_k + r_k\}
\label{Krich}
\ee
with some simple (and explicitly known) function $f_p$. (In eq.(\ref{TtK})
above for {\it monomial} cubic potential $V_3(x) = \frac{x^3}{3}$,
$\ \hat T_k = T_k = \frac{1}{k}{\rm tr}L^{-k/2}$, while $r_k =
-\frac{2}{3}\delta_{k,3}$.)

The last thing to be  mentioned in this general description of the field of
the matrix model theory is their relation to group theory. Generalized
Kontsevich model (\ref{GKM}) is intimately connected to the "integrable
nature" of group chracters and the coadjoint orbit integrals (characters of
all the irreducible representations of $U(N)$ are
usually KP $\tau$-functions \cite{chartau}). In fact some "discrete
(or quantum) version"
of Kontsevich integral is a sum over all unitary irreps of $U(n)$ ("integral"
over a $model$ of $U(n)$, or over the set of all coadjoint orbits):
\be
{\cal F}_V^{qu}\{G\} \equiv \sum_R d_R\chi_R(G) e^{-\sum_{k=0}^{\infty}
v_kC_k(R)},
\label{quGKM}
\ee
where $d_R$, $\chi_R$ and $C_k(R)$ stand for dimension, character and the
$k$-th Casimir of irreducible representation $R$ of $U(n)$.
Time variables $T_k \sim
\frac{1}{k}{\rm tr}G^k$, while potential $V(X) = \sum_{k=0}^{\infty}s_kX^k$.
This expression can be further generalized to
\be
{\cal F}_V^{qu}\{G\} \equiv \sum_R \chi_R(\bar G)\chi_R(G)
e^{-\sum_{k=0}^{\infty}v_kC_k(R)}.
\label{quGKM'}
\ee
Properties of these "quantum" Kontsevich models deserve further investigation
(objects like (\ref{quGKM}) are also known to arise in the localization theory,
in particular, in the study of the $d=2$ YM theory, see, for example,
\cite{Wit2YM} and \cite{NT}).

\bigskip

These notes are essentially a review of the views and results of the group,
working at Moscow (and Kiev). Since references will not be given every time,
I present here the list of people involved into these investigations:

L.Chekhov,
A.Gerasimov,
A.Losev,
S.Kharchev,
Yu.Makeenko,
A.Mar\-sha\-kov,
A.Mikhailov,
A.Mironov,
A.Orlov,
S.Pa\-kuliak,
I.Polyubin,
A.Zab\-rodin.

I also apologize for the somewhat sporadic references to the works of other
groups.

\






\bigskip

\section{Ward identities for the simplest matrix models}

\setcounter{equation}{0}

\subsection{Ward identites versus equations of motion}

We begin systematic consideration of matrix models from their simplest and at
the same time the most basic property: the Ward identites (WI) for partition
functions. Partition function is by definition a functional of the coupling
constants in the Lagrangian and WI will be understood here as (differentiual
or finite-difference) equations, imposed on this functional. If partition
function is represented in the form of a matrix integral,
\footnote{To avoid confusion we emphasize that such representation does $not$
need to exist, at least in any simple form. The more the theory of matrix
models develops, the less it has to do with $matrices$ and matrix integrals.
However (as in the case of entire $string\ theory$) original name has a
tendency to
survive. Anyhow, the main content of the theory of matrix models (at least of
its branch, analyzed in these notes) is the search for $invariant$
formulations of the properties of partition functions, while matrix integrals
(if at all existing) are considered as their particular realizations
(representations). Moreover, there can exist very different matrix integral
representations of the same partition function, the simplest example being
just the basic discrete 1-matrix model, which can be also represented in the
form of Kontsevich integral (see below).  }
the WI are usually implied by its invariance (or, better to say - covariance)
under the change of the integration variables (thus the name "WI").

In ordinary field theory we are usually dealing with models, where WI either
do not exist at all, or at most there is finite number of them - then they are
interpreted as reflecting the $symmetry$ of the theory. However, by no means
the finite set of these WI provides a $complete$ description of dynamics of
the theory: the number of (quantum) equations of motion (EqM) is usually
infinite and their solutions are never fixed by the WI. In fact this
difference between WI and EqM arises because the Lagrangians, considered in
the ordinary field theory are not of the most general form: they are usually
severely restricted by "principles" like renormalizability or minimality.
Because of this there is simply not many enough coupling constants in the
Lagrangian to describe the result of $any$ variation of integration variables
as that of the variation of coupling constants, and thus not every equation of
motion can be represented as (differential) equation for the partition
function. In other words, by restricting the shape of Lagrangian for
"non-symmetric" reasons one breaks the original huge "symmetry" (covariance)
of the model, which was enough to describe all the dynamics (all EqM) as
dictated by symmetry,- and a broader view is necessary in order to recognize
EqM as the WI, associated with that original high symmetry. This symmetry
(it is not of Noether type, of course) is peculiar property of all the
$quantum\  mechanical$ partition functions, since these usually arise from
the procedure of functional integration.

It happened so that matrix models appeared to be the first class of
quantum-mechanical systems (functional integrals) for which this identity:
\be
{\rm all\ EqM}\ \equiv\ {\rm all\ WI}
\nn
\ee
was not simply observed as a curious phenomenon, but became a subject of
intensive investigation and is identified as the source of exact solvability
(integrability) of the theory. Of course, significance of this observation
(and its implications) is quite universal, by no means restricted to the
field of matrix models themselves,
however, it is yet not enough appreciated by
the experts in other fields.
In any case, we are going to deal only with matrix
models in these notes.

We proceed to consideration of the WI according to the
following plan (not all the arrows will be actually discussed): see Fig.3.

\subsection{Virasoro constraints for the discrete 1-matrix model}

The basic example \cite{GMMMO},\cite{MM} which illustrates
the arguments from the previous subsection is provided by the 1-matrix model
\be
Z_N\{t\} \equiv c_N\int_{N\times N} dH e^{\sum_{k=0}^{\infty} t_k {\rm Tr}
H^k}.
\label{1mamo'}
\ee
This integral is invariant under any change of variables $H \rightarrow f(H)$.
It is convenient to choose the special basis in the space of such
transformations:
\be
\delta H = \epsilon_nH^{n+1}.
\ee
Here $\epsilon_n$ is some infinitesimal matrix and, of course, $n \geq -1$.
The value of integral can not change under the change of integration variable,
and we obtain the identity:
\be
\int_{N\times N} dH e^{\sum_{k=0}^{\infty} t_k {\rm Tr}H^k}
 = \int d(H + \epsilon_nH^{n+1})e^{\sum_{k=0}^{\infty} t_k {\rm Tr}
(H + \epsilon_nH^{n+1})^k},
\nn
\ee
i.e.
\be
\int dH e^{\sum_{k=0}^{\infty} t_k {\rm Tr}H^k} \left(
\sum_{k=0}^{\infty} kt_k {\rm Tr}H^{k+n} + {\rm Tr}
\frac{\delta H^{n+1}}{\delta H} \right) \equiv 0.
\label{vird1}
\ee
In order to evaluate the Jacobian
${\rm Tr}\frac{\delta H^{n+1}}{\delta H}$ let us restore the matrix indices:
\be
(\delta H^{n+1})_{ij} =
\sum_{k=0}^n (H^k \delta H H^{n-k})_{ij} =
\sum_{k=0}^n (H^k)_{il}( \delta H)_{lm}( H^{n-k})_{mj}.
\nn
\ee
In ${\rm Tr}\frac{\delta H^{n+1}}{\delta H}$ one should take $l=i$ and
$m=j$, so that
\be
{\rm Tr}\frac{\delta H^{n+1}}{\delta H} =
\sum_{k=0}^n {\rm Tr}H^k {\rm Tr} H^{n-k}.
\label{vird2}
\ee

Now we can note that since we started from Lagrangian of the most general form
(consistent with the symmetry  $H \rightarrow UHU^{\dagger}$), any correlation
function can be obtained as variation of the coupling constants (all possible
$sources$ are included as counterterms). In our particular example this is
just a trivial remark:
\be
<{\rm Tr} H^{a_1} ... {\rm Tr} H^{a_n}> &=
\int dH e^{\sum_{k=0}^{\infty} t_k {\rm Tr}H^k}
{\rm Tr} H^{a_1} ... {\rm Tr} H^{a_n}  = \nn \\
&= \frac{\partial^n}{\partial t_{a_1}...\partial t_{a_n}} Z_N\{t\}.
\label{vird3}
\ee
We can use this relation together with (\ref{vird2}) in order to rewrite
(\ref{vird1}) as:
\be
L_n Z_N\{t\} = 0, \ \ \ n\geq -1
\label{virdid}
\ee
with
\be
L_n \equiv \sum_{k=0}^{\infty} kt_k\frac{\partial}{\partial t_{k+n}} +
    \sum_{k=0}^n \frac{\partial^2}{\partial t_k\partial t_{n-k}}.
\label{virdop}
\ee
Note that according to the definition (\ref{1mamo'})
\be
\frac{\partial}{\partial t_0}Z_N = NZ_N. \nn
\ee

\bigskip

Several remarks are now in order.

First of all, expression in brackets in (\ref{vird1}) represents just $all$
the equations of motion for the model (\ref{1mamo'}), and (\ref{virdid}) is
nothing but another way to represent the same set of equations. This is an
example of the above-mentioned identification of EqM and WI.

Second, commutator of any two operators $L_n$ apearing in (\ref{virdid})
should also annihilate $Z_N\{t\}$. It is another indication (not a convincing
one, however) that we already
got a $complete$ set of constraints, that $L_n$'s form a closed
(Virasoro) algebra:
\be
\phantom. [ L_n, L_m] = (n-m) L_{n+m}, \ \ \ n,m\geq -1.
\label{virdal}
\ee
Third, (\ref{virdid}) can be considered as invariant formulation of what is
$Z_N$: it is a solution of this set of compatible differential equations.
{}From this point of view eq.(\ref{1mamo'}) is rather a particular
representation of $Z_N$  and it is sensible to look for
other representations as well (we shall later discuss two of them: one in
terms of CFT, another in terms of Kontsevich integrals).

Fourth, one can try to analyze the uniqueness of the solutions to
(\ref{virdid}). If there are not too many of them the set of constraints can
be considered complete. A natural approach to classification of solutions to
the algebra of constraints is in terms of the orbits of the corresponding
group \cite{GMMMMO}.
Let us consider an oversimplified example, which can still be usefull
to understand implications of the complete set of WI as well as clarify the
meaning of classes of universality and of integrability.

Imagine, that instead of (\ref{virdid}) with $L_n$'s defined in (\ref{virdop})
we would obtain somewhat simpler equations:
\footnote{One can call them "classical" approximation to (\ref{virdid}), since
they would arise if the variation of measure (i.e. a
"quantum effect") was not taken into account in the derivation of
(\ref{virdid}). Though  this concept is often
used in physics it does not have much sense in the present context, when we
are analyzing $exact$ properties of functional (matrix) integrals. }
\be
l_n Z = 0, \ \ n\geq 0 \ \ {\rm with} \ \
l_n = \sum_{k=1}^{\infty} kt_k\frac{\partial}{\partial t_{k+n}}. \nn
\ee
Then operator $l_1$ can be interpreted as generating the shifts
\be
t_2 \longrightarrow t_2 + \epsilon_1 t_1, \nn \\
t_3 \longrightarrow t_3 + 2\epsilon_1 t_2, \nn \\
.                    \nn
\ee
We can use it to shift $t_2$ to zero, and eq. $l_1Z = 0$
then implies that
\be
Z(t_1,t_2,t_3,...) = Z(t_1,0,\tilde t_3,...)
\nn
\ee
$(\tilde t_k = t_k - \frac{(k-1)t_2t_{k-1}}{t_1}, \ k\geq 3)$.

Next, operator $l_2$ generates the shifts
\be
t_3 \longrightarrow t_3 + \epsilon_2 t_1, \nn \\
t_4 \longrightarrow t_4 + 2\epsilon_2 t_2, \nn \\
. \nn
\ee
and does $not$ affect $t_2$. We can now use eq. $l_2Z = 0$ to
argue that
\be
Z(t_1,t_2,t_3,t_4,...) = Z(t_1,0,\tilde t_3,\tilde t_4,...) =
Z(t_1,0,0,\tilde{\tilde t}_4,...)  \nn
\ee
etc. Assuming that $Z$ is not very much dependent on $t_k$ with
$k \rightarrow \infty$,
\footnote{This, by the way,  is hardly correct in this particular example,
when the group has no compact orbits.}
we can conclude, that
\be
Z(t_1,t_2,t_3,...) = Z(t_1,0,0,...) =
Z(1,0,0,...) \nn
\ee
(at the last step we also used the equation $l_0Z = 0$ to rescale
$t_1$ to unity).

All this reasoning was correct provided $t_1 \neq 0$. Otherwise we would get
$Z(0,1,0,0,...)$, if $t_1 = 0,\ t_2\neq 0$, or
$Z(0,0,1,0,...)$, if $t_1 = t_2 = 0,\ t_3\neq 0$ etc.
In other words, we obtain classes of universality (such that the value of
partition function is just the same in the whole class), which in this
oversimplified example are labeled just by the first non-vanishing
time-variable. Analysis of the orbit structure for the actually important
realizations of groups, like that connected to eq.(\ref{virdop}) has never
been performed in the context of matrix model theory.
It may deserve emphasizing that the
constraints, as we saw, can actually allow one to eliminate (solve exactly)
all the dependence on the time-variables, in less trivial examples they
somehow imply the integrability structure, which is just a slightly more
complicated version of the same solvability phenomenon.

\subsection{CFT formulation of matrix models}

Given a complete set of the constraints on partition function of
infinitely many variables which form some closed algebra we can now ask an
inverse question: how these equations can be solved or what is the integral
representation of partition function. One approach to this problem
is analysis of orbits, briefly mentioned at the end of the previous section.
Now we turn to another technique \cite{comamo}, which makes use of
the knowledge from conformal field
theory. This constructions can have some meaning from the "physical"point of
view, which implies certain duality between the 2-dimensional world surfaces
and the spectral surfaces, associated to configuration space of the string
theory. However, our goal now is more formal: to use the means of CFT for
solution of the constraint equations.

This is very natural in the case when the algebra of constraints is Virasoro
algebra, as in the case of the 1-matrix model, or some other algebra if it is
known to arise naturally as chiral algebra in some simple conformal models.
In fact the approach which will be now discussed is rather general and can
be applied to construction of matrix models, associated with many different
algebraic structures.

We begin from the set of equations (\ref{virdid}) which we shall further refer
to as "discrete Virasoro constraints". The CFT formulation of interest should
provide the solution to these equations in the form of some correlation
function in some conformal field theory. Of course, it becomes natural, if we
somehow identify the operators $L_n$, which form Virasoro algebra with the
harmonics of the stress-tensor $T_n$, which satisfy the same algebra, and
manage to relate the constraint that $L_n$ annihilate the correlator to the
statement that $T_n$ annihilate the vacuum state. Thus the procedure is
naturally split into two steps: First we should find a $t$-dependent operator
("Hamiltonian") $H(t)$, such that
\be
L_n(t) \langle e^{H(t)} \ldots  = \langle e^{H(t)}T_n \ldots
\label{lcft1}
\ee
This will relate differential operators $L_n$ to $T_n$'s expressed through the
fields of conformal model. Second we need to enumerate the states, that are
annihilated by the operators $T_n$ with $n \geq -1$, i.e. solve equation
\be
T_n \mid G \rangle = 0
\ee
for the ket-states, what is an internal problem of
conformal field theory. If both ingredients $H(t)$ and $\mid G \rangle$ are
found, solution to the problem is given by
\be
\langle e^{H(t)}\mid G\rangle.
\ee

To be more explicit, for the case of the discrete Virasoro constraints we can
just look for solutions in terms of the simplest possible conformal model:
that of a one holomorphic scalar field
\be
\phi (z) =  \hat q + \hat p \log z  + \sum _{k\neq 0} {J_{-k}\over k}
z^{k}\nn\\
\  [J_n,J_m] = n\delta _{n+m,0},  \ \ \     [\hat q,\hat p] = 1.
\ee
Then the procedure is as follows:
 Define vacuum states
\be
J_k|0\rangle  &= 0, \ \ \  \langle N|J_{-k} = 0, \ \ \    k > 0\nn\\
\hat p|0\rangle  &= 0, \ \ \   \langle N|\hat p = N\langle N|,
\ee
the stress-tensor
\be
T(z) = {1\over 2}[\partial \phi (z)]^2 = \sum    T_nz^{-n-2},\quad
T_n = \sum _{k>0}J_{-k}J_{k+n} +
{1\over 2}\sum _{{a+b=n}\atop{a,b\geq 0}}J_aJ_b,
\ee
and the Hamiltonian
\be
H(t) &= {1\over \sqrt{2}} \sum _{k>0}t_kJ_k =
{1\over \sqrt{2}}\oint_{C_0}U(z)J(z)\nn\\
U(z) &= \sum _{k>0}t_kz^k, \ \  \   J(z) = \partial \phi (z).
\ee
It can be easily checked now that
\be
L_n\langle N|e^{H(t)}\ldots = \langle N|e^{H(t)}T_n\ldots
\ee
and
\be
 T_n|0\rangle  = 0,  \ \ \    n \geq  -1 .
\ee
As an immediate consequence, any correlator of the
form
\be
Z_N\{t\mid G\} = \langle N|e^{H(t)}G|0\rangle
\label{confsol}
\ee
gives a solution to (\ref{virdid}) provided
\be
[T_n,G] = 0, \ \ \  n \geq  -1.
\label{crGop}
\ee
In fact operators $G$ that commute with the stress tensor are well known:
these are just any functions of the "screening charges"
\footnote{For notational simplicity we omit the normal ordering signs, in
fact  the involved operators are  $:e^H:$ and $:e^{\pm \sqrt{2}\phi}:$}
\be
Q_\pm  = \oint J_\pm  = \oint
e^{\pm \sqrt{2}\phi }.
\ee
The correlator (\ref{confsol}) will be non-vanishing only if the matching
condition for zero-modes of $\phi$ is satisfied. If we demand the operator to
depend only on $Q_{+}$, this implies that only
one term of the expansion in powers of $Q_{+}$ will contribute to
(\ref{confsol}), so that the result is essentially independent on the choice
of the function $G(Q_+)$, we can for example take $G(Q_+) = e^{Q_+}$
and obtain:
\be
Z_N\{t\} \sim \frac{1}{N!}\langle N|e^{H(t)}(Q_+)^N|0\rangle .
\label{comamo1mm"}
\ee
This correlator is easy to evaluate using the Wick theorem and the propagator
$\phi(z)\phi(z')\sim \log(z-z')$ and finally we get
\be
Z_N\{t\} &= \frac{1}{N!} \langle N \mid
:e^{{1\over \sqrt{2}}\oint_{C_0}U(z)\partial\phi(z)}:
\prod_{i=1}^N \oint_{C_i} dz_i :e^{\sqrt{2}\phi(z_i) }: \mid 0  \rangle  =
\nn \\
&= \frac{1}{N!}\prod_{i=1}^N\oint_{C_i} dz_i e^{U(z_i)}
\prod_{i<j}^N (z_i-z_j)^2
\label{comamo1mm}
\ee
in the form of a multiple integral, which can in fact be directly related to
the matrix integral in (\ref{1mamo'}), see \cite{BIPZ} and
the next section.

Thus in the simplest case we resolved the inverse problem: reconstructed the
integral representation  from the set of discrete Virasoro constraints.
However, the answer we got seems a little more general than (\ref{1mamo'}):
the r.h.s. of eq.(\ref{comamo1mm}) still depends on the contours of
integration. Moreover, we can also recall that the operator $G$ above could
depend not only on $Q_+$, but also on $Q_-$. The most general formula is a
little more complicated than (\ref{comamo1mm}):
\be
\new
\begin{array}{c}
Z_{N}\{t\mid C_i, C_r\} \sim \frac{1}{(N+M)!M!}\langle N|e^{H(t)}(Q_+)^{N+M}
(Q_-)^M|0\rangle =     \\
= \frac{1}{(N+M)!M!}\prod_{i=1}^{N+M}\oint_{C_i} dz_i e^{U(z_i)}
\prod_{r=1}^M\oint_{C'_r} dz'_r e^{U(z'_r)}
\cdot \\
\cdot
\frac{\prod_{i<j}^{N+M} (z_i-z_j)^2  \prod_{r<s}^N (z'_r-z'_s)^2 }
{\prod_{i}^{N+M}\prod_r^M (z_i-z_r)^2}.
\end{array}
\label{comamo1mm'}
\ee
We refer to the papers \cite{comamo} for discussion of the issue of the
contour-dependence. In certain sense all these different integrals can be
considered as branches of the same analytical function $Z_N\{t\}$. Dependence
on $M$ is essentially eliminated by Cauchy integration around the poles in
denominator in (\ref{comamo1mm'}).

Above construction can be straightforwardly applied to other
algebras of constraints, provided:

(i) The free-field representation of the algebra is known in the
CFT-framework, such that the generators are $polinomials$ in the
fields $\phi$ (only in such case it is straightforward to construct a
Hamiltonian $H$, which relates CFT-realization of the algebra to that
in terms of differential operators w.r.to the $t$-variables; in fact
under this  condition $H$  is usually linear in $t$'s and $\phi$'s).
There are examples (like Frenkel-Kac representation of level $k=1$
simply-laced Kac-Moody algebras \cite{FK} or generic reductions of the WZNW
model \cite{GMMOS},\cite{BO}-\cite{GereF})
when generators are $exponents$ of free fields, then this construction
should be slightly modified.

(ii) It is easy to find vacuum, annihilated by the relevant generators (here,
for example, is the problem with application of this approach to the case of
"continuous" Virasoro and $W$-constraints). The resolution to this problem
involves consideration of correlates on Riemann surfaces with non-trivial
topologies, often - of infinite genus.

(iii) The free-field representation of the "screening charges", i.e. operators
that commute with the generators of the group within the conformal model, is
explicitly known.

These conditions are fulfilled in many case in CFT, including conventional
{\bf W}-algebras \cite{Zam} and ${\cal N} = 1$
\footnote{In the case of ${\cal N} = 2$ supersymmetry a problem arises because
of the lack of reasonable screening charges. At the most naive level the
relevant operator to be integrated over superspace (over $dzd^{\cal N}\theta$)
in order to produce screening charge has dimension $1-\frac{1}{2}{\cal N}$,
which $vanishes$ when ${\cal N} = 2$.
} supersymmetric models
\cite{AGS}.

For illustration purposes we present here several formulas from the last paper
of ref.\cite{comamo} for the case of the
${\bf W}_{r+1}$-constraints, associated with the simply-laced algebras ${\cal
A}$ of rank $r$.

Partition function in such "conformal multimatrix model" is a function of
"time-variables" $t_k^{(\lambda)},\ k = 0...\infty,\ \lambda = 1...r={\rm
rank}{\cal A}$, and also depends on the integer-valued $r$-vector
$\bN = \{N_1...n_r\}$.
The ${\bf W}_{r+1}$-constraints imposed on partition function are:
\be
W_n^{(a)}(t)Z_{\bN}^{\cal A}\{t\} = 0, \ \ n\geq 1-a, \ \ a= 2...r+1.
\ee
The form of the $W$-operators is somewhat complicated, for example, in the
case of $r+1=3$ (i.e. for ${\cal A} = A_2$ ($SL(3)$))
\be
\new
\begin{array}{c}
W^{(2)}_n =  \sum ^\infty _{k=0}(kt_k\frac{\partial}{\partial t_{k+n}} +
k\bar t_k\frac{\partial}{\partial \bar t_{k+n}}) + \\
+ \sum _{a+b=n}(\frac{\partial ^2}{\partial t_a\partial t_b} +
\frac{\partial ^2}{\partial \bar t_a\partial \bar t_b})
\end{array}
\ee
\be
\new
\begin{array}{c}
W^{(3)}_n = \sum _{k,l>0}(kt_klt_l\frac{\partial}{\partial t_{k+n+l}} -
k\bar t_kl\bar t_l\frac{\partial }{\partial t_{k+n+l}}
-2kt_kl\bar t_l\frac{\partial }{\partial \bar t_{k+n+l}})+ \\
+ 2\sum  _{k>0}\left[ \sum _{a+b=n+k}(kt_k\frac{\partial ^2}{\partial t_a
\partial t_b} - kt_k\frac{\partial ^2}{\partial \bar t_a\partial \bar t_b} -
2k\bar t_k\frac{\partial ^2}{\partial t_a\partial \bar t_b)}\right]  + \\
+ {4\over 3}\sum _{a+b+c=n}(\frac{\partial ^3}
{\partial t_a\partial t_b\partial t_c} -
\frac{\partial ^3}{\partial t_a\partial \bar t_b\partial \bar t_c}),
\end{array}
\label{wopex}
\ee
and two types of time-variables, denoted through $t_k$ and  $\bar t_k$.
are ossociated with two $orthogonal$ directions
in the  Cartan plane of $A_2$:
${\bf e} = {\balpha _1\over\sqrt{2}}$,
$\bar{\bf e} = {\sqrt{3}\bnu _2\over\sqrt{2}}$.
\footnote{Such orthogonal basis is especially convenient for discussion of
integrability properties of the model, these $t$ and $\bar t$ are linear
combinations of time-variables $t_k^{\lambda}$ appearing in eqs.
(\ref{hamAr}) and (\ref{comamoAr}).}

All other formulas, however, are very simple:
Conformal model is usually that of the $r$ free fields,
$S \sim \int\bar\partial\bphi\partial\bphi d^2z$,
which is used to describe representation
of the level one Kac-Moody algebra, associated with ${\cal A}$. Hamiltonian
\be
H(t^{(1)}\ldots t^{(r+1)}) =
\sum_{\lambda = 1}^{r+1}\sum _{k>0}t^{(\lambda )}_k\bmu _\lambda \bJ_k,
\label{hamAr}
\ee
where $\{\bmu_{\lambda}\}$ are associated with "fundamental weight" vectors
$\bnu_{\lambda}$ in Cartan hyperplane
and in the simplest case of ${\cal A} = A_r$ ($SL(r+1)$)  satisfy
$$
\bmu_{\lambda}\cdot \bmu_{\lambda'}=\delta_{\lambda\lambda'}-{1\over{r+1} },
  \ \ \ \sum_{\lambda=1}^{r+1} \bmu_{\lambda}=0,
$$
thus only $r$ of the time variables $t^{(1)}\ldots t^{(r+1)}$ are linearly
independent.
Relation between differential operators
$W_n^{(a)}(t)$ and operators ${\rm W}_n^{(a)}$ in the CFT is now defined by
\be
W^{(a)}_n\langle \bN|e^{H(t)}\ldots =
\langle \bN|e^{H(t)}{\rm W}^{(a)}_i\ldots\ , \nn \\   a=2,\ldots,p;  \ \ \
i\geq 1-a,
\ee
where
\be
{\rm W}^{(a)}_n = \oint z^{a+n-1}{\rm W}^{(a)}(z)\nn\\
{\rm W}^{(a)}(z) = \sum  _\lambda  [\bmu _\lambda \partial \bphi (z)]^a +
     \ldots
\ee
are spin-$a$ generators of the ${\bf W}^{\cal A}_{r+1}$ algebra.
The screening charges, that commute with all the ${\rm W}^{(a)}(z)$ are
given by
\be
Q^{(\alpha)}  = \oint J^{(\alpha)}  = \oint e^{\bsalpha \bsphi }
\ee
$\{\balpha \}$ being roots of finite-dimensional simply laced
Lie algebra ${\cal A}$.

Thus partition function arises in the form:
\be
Z^{\cal A}_{\bsN}\{t\} = \langle \bN|e^{H(t}G\{Q
^{(\alpha)} \}|0\rangle
\ee
where  $G$  is an exponential function of screening charges.
Evaluation of the free-feild correlator gives:
\be
\new
\begin{array}{c}
Z^{\cal A}_{\bsN}\{t\} \sim  \int   \prod  _\alpha
\left[ \prod ^{N_\alpha }_{i=1}dz^{(\alpha )}_i \exp \left(
\sum _{\lambda ;k>0}t^{(\lambda )}_k(\bmu _\lambda \balpha )(z^{(\alpha
)}_i)^k
\right) \right] \times \\
\times \prod _{(\alpha ,\beta )}\prod ^{N_\alpha }_{i=1}
\prod ^{N_\beta }_{j=1}(z^{(\alpha )}_i- z^{(\beta )}_j)^{\bsalpha \bsbeta }
\end{array}
\label{comamoAr}
\ee
In fact this expression can be rewritten in terms of an $r$-matrix integral -
a "conformal multimatrix model":
\be
Z^{\cal A}_{\bsN}\{t^{(\alpha)}\} =
c_N^{p-1}\int_{N\times N} dH^{(1)}...dH^{(p-1)}
\prod_{\alpha = 1}^{p-1} e^{\sum_{k=0}^{\infty}t_k^{(\alpha)}{\rm Tr}
H_{(\alpha)}^k}\cdot    \nn \\
\cdot\prod_{(\alpha ,\beta )}
{\rm Det} \left(H^{(\alpha)}\otimes I - I\otimes
H^{(\alpha+1)}\right)^{\bsalpha \bsbeta }
\label{comamo"}
\ee
In the simplest case of ${\bf W}_3$ algebra
eq.(\ref{comamoAr}) with insertion of only two (of the six) screenings
  $Q_{\alpha _1}$ and  $Q_{\alpha _2}$ turns into
\be
\new
\begin{array}{c}
Z^{A_2}_{N_1,N_2}(t,\bar t)  = {1\over N_1!N_2!}
\langle N_1,N_2|e^{H(t,\bar t)}(Q^{(\alpha _1)})^{N_1}
(Q^{(\alpha _2)})^{N_2}|0\rangle  =
 \\
= {1\over N_1!N_2!}  \prod_i \int   dx_i e^{U(x_i)}
\prod_j\int dy_j e^{\bar U(y_i)}  \Delta (x)\Delta (x,y) \Delta (y),
\end{array}
\label{comamo"3}
\ee
where $\Delta(x,y) \equiv \Delta(x)\Delta(y)\prod _{i,j}(x_i - y_j)$.
This model is associated with the algebra ${\cal A} = A_2\ (SL(3))$, while the
original 1-matrix model
(\ref{comamo1mm"})-(\ref{comamo1mm'})
- with ${\cal A} = A_1\ (SL(2))$.

The whole series of models (\ref{comamoAr}-\ref{comamo"})
for ${\cal A} = A_r\ (SL(r+1))$ is distinguished by its relation to the level
$k=1$ simply-laced Kac-Moody algebras. In this particular situation the
underlying conformal model has integer central charge $ c = r = {\rm rank}
 {\cal A}$ and can be "fermionized".\footnote{
This is possible only for very special Kac-Moody algebras, and such
formulation is important in order to deal with $conventional$ formulation of
integrability, which usually involves $commuting$ Hamiltonian flows (not just
a closed algebra of flows) and fermionic realization of the universal module
space (universal Grassmannian). In fact these restrictions are quite arbitrary
and can be removed (though this is not yet done in full details), see section
4 below for more detailed discussion.
}
The main feature of this formulation is that the Kac-Moody currents (which
after integration turn into "screening charges" in the above construction)
are quadratic in fermionic fields, while they are represented by exponents in
the free-boson formulation.

In fact fermionic (spinor) model naturally possesses $GL(r+1)$ rather than
$SL(r+1)$ symmetry (other simply-laced algebras can be embedded into larger
$GL$-algebras and this provides fermionic descriprion for them in the case of
$k=1$). The model contains $r+1$ spin-1/2 fields $\psi_i$ and their conjugate
$\tilde\psi_i$ ($b,c$-systems);
\be
S = \sum_{j=1}^{r+1} \int \tilde\psi_j\bar\partial\psi_j d^2z,
\nn
\ee
central charge $c=r+1$, and operator algebra is
\be
\tilde\psi_j(z)\psi_k(z') &= \frac{\delta_{jk}}{z-z'}\ +
:\tilde\psi_j(z)\psi_k(z'): \nn \\
\psi_j(z)\psi_k(z') &=  (z-z')\delta_{jk}:\psi_j(z)\psi_k(z'):
   + \ (1-\delta_{jk}):\psi_j(z)\psi_k(z'): \nn\\
\tilde\psi_j(z)\tilde\psi_k(z') &=
    (z-z')\delta_{jk}:\tilde\psi_j(z)\tilde\psi_k(z'):
   + \ (1-\delta_{jk}):\tilde\psi_j(z)\tilde\psi_k(z'): \nn
\ee
The Kac-Moody currents of level $k=1$ $GL(r+1)$ are just
$J_{jk} = :\tilde\psi_j\psi_k:\ \ j,k = 1\ldots r+1$, and screening charges
are
$Q^{(\alpha)} = iE_{jk}^{(\alpha)}\oint :\tilde\psi_j\psi_k:$, where
$E_{jk}^{(\alpha)}$ are representatives of the roots $\balpha$ in the matrix
representation of $GL(r+1)$. Cartan subalgebra is represented by $J_{jj}$,
while positive and negative Borel subalgebras - by $J_{jk}$ with $j<k$ and
$j>k$ respectively. In eq.(\ref{comamo1mm'})
$Q_+ = i\oint\tilde\psi_1\psi_2,\ \ Q_- = i\oint\tilde\psi_2\psi_1\ $
while in eq.(\ref{comamo"3})
$Q^{(\alpha_1)} = i\oint\tilde\psi_1\psi_2,\ \ Q^{(\alpha_2)} =
i\oint\tilde\psi_1\psi_3\ $
(and $Q^{(\alpha_3)} = i\oint\tilde\psi_2\psi_3,\ \ Q^{(\alpha_4)} =
i\oint\tilde\psi_2\psi_1,\ \ Q^{(\alpha_5)} = i\oint\tilde\psi_3\psi_1,\ \
Q^{(\alpha_6)} = i\oint\tilde\psi_3\psi_2$). $Q^{(\alpha_6)}$ can be
substituted instead of $Q^{(\alpha_2)}$ in (\ref{comamo"3}) without changing
the answer. For generic $r$ the similar choice of "adjacent" (not simple!)
roots (such that their scalar products are $+1$ or $0$)  leads to
selection of the following $r$ screening operators
$Q^{(1)} = i\oint\tilde\psi_1\psi_2\,\ \
Q^{(2)} = -i\oint\psi_2\tilde\psi_3,\ \
Q^{(3)} = i\oint\tilde\psi_3\psi_4,\ldots$, i.e.
$Q^{(j)} = i\oint\tilde\psi_j\psi_{j+1}$ for odd $j$ and
$Q^{(j)} = -i\oint\psi_j\tilde\psi_{j+1}$ for even $j$.


\subsection{Gross-Newman equation}

We turn now to consideration of the WI for another sort of matrix models. This
subject concerns at least two important classes: the conventional discrete
$two$-matrix models and Kontsevich models. As it was explained in the
Introduction the theories of the second type arise in consderation of the
$(p,1)$ continuous matrix models, as well as in the study of topological
Landau-Ginzburg theories, while the two-matrix model is believed to exhibit a
rich pattern of continuous limits and is capable to provide representatives of
all the $(p,q)$ universality classes (this line of reasoning, however, has
never been really developed and we shall not discuss it in these notes).

The starting point and the basic example is provided by the integral
\be
{\cal F}_{V,n}\{ L\} \equiv
\int_{n\times n} dX e^{- {\rm tr} V(X) + {\rm tr} L X}
\label{KI}
\ee
over $n\times n$ Hermitean matrix, which we shall further refer to as
"Kontsevich integral", keeping in mind its most important application (though
this obvious quantity has been of course considered by many other people).
It may seem that
the action in this integral is $not$ of the most general type and we can no
longer perform $arbitrary$ change of variables $X \longrightarrow f(X)$
without changing the functional form of the integral.
In fact this is incorrect, because "external field" $ L$ is matrix valued
and coupled to $X$ linearly, and therefore $any$ correlator of $X$-fields can
be represented through $ L$-derivatives. Consider again the
shift $X \rightarrow X + \epsilon_nX^{n+1},\ n \geq -1$.
Invariance of the integral implies:
\be
\int dX e^{- {\rm tr} V(X) + {\rm tr} L X}
{\rm tr}\ \epsilon_n \left(-X^{n+1} V'(X) +  L X^{n+1} +
\sum_{k=0}^n X^k {\rm tr}X^{n-k} \right)  = 0,
\nn
\ee
which can be rewritten as
\footnote{The obvious relation is used here:
$X_{\gamma\delta} e^{{\rm tr} L X} =
\frac{\partial}{ L_{\delta\gamma}} e^{{\rm tr} L X}$.
Note that the order of matrix indices $\gamma\delta$
is reversed at the r.h.s. as
compared to the l.h.s., i.e. derivatives are in fact w.r.to $transponed$
matrix $ L$: $\ f(X)e^{{\rm tr} L X} = f(\frac{\partial}{\partial
 L_{tr}}) e^{{\rm tr} L X}$ (at least for any function $f(x)$,
which can be represented as a formal seria in integer powers of $X$).
}
\be
&{\rm tr}\ \epsilon_n \left(
\left(-\frac{\partial}{\partial L_{\rm tr}}\right)^{n+1}
V'\left(\frac{\partial}{\partial L_{\rm tr}}\right) +  L
\left(-\frac{\partial}{\partial L_{\rm tr}}\right)^{n+1} +
\right. \nn \\ &\left.
\sum_{k=0}^n \left(-\frac{\partial}{\partial L_{\rm tr}}\right)^k
{\rm tr}\left(-\frac{\partial}{\partial L_{\rm tr}}\right)^{n-k}
\right) {\cal F}_V\{ L\} =  \nn \\
&=  {\rm tr}\ \epsilon_n \left(-\frac{\partial}{\partial L_{\rm
tr}}\right)^{n+1}
\left(V'\left(\frac{\partial}{\partial L_{\rm tr}}\right) -
 L\right){\cal F}_V\{ L\} = 0
\label{preGNe}
\ee
This system is in fact equivalent to a single matrix-valued equation
\be
\left(V'\left(\frac{\partial}{\partial L_{\rm tr}}\right) -
 L\right)
{\cal F}_V\{ L\} = 0.
\label{GNe}
\ee
As well as I know this equation was first written down in \cite{GN}, therefore
it will be refered to as the Gross-Newman (GN) equation. It was rediscovered
and implications for the theory of matrix models were investigated in
\cite{MMM},\cite{GKM},\cite{tildeW}.

There are essentially two types of corollaries, which will be discussed in the
next two subsections. First, GN equation can be used to characterize the
function ${\cal F}_V\{ L\}$ itself. This will lead us to consideration of
Kontsevich models. Second, it can be used to derive equations for the 2-matrix
model, which arises after ${\cal F}_V\{ L\}$ is further integrated with
some weight over $ L$.

\subsection{Ward identities for Generalized Kontsevich Model}

Being just the complete set of equation of motion the GN equation (\ref{GNe})
provides complete information about the function ${\cal F}_V\{ L\}$.
However, this statement needs to be formulated more carefully. A need for this
comes, for example, from the observation that operators
\be
{\rm tr}  L^m
\left(V'\left(\frac{\partial}{\partial L_{\rm tr}}\right) -
 L\right)
\ee
do not form a closed algebra: their commutators have some different functional
form. One of the reasons for these complications is that eq.(\ref{GNe}) does
not account explicitly for a very important property of ${\cal
F}_V\{ L\}$: this function in fact depends only on the eigenvalues of
$ L$. This information should be still added somehow to the GN equation.
We shall analyze this issue of eigenvalue-dependence in more details in next
sections. For our current purposes this argument implies  that one should
try to express equation (\ref{GNe}) in terms of eigenvalues. Here, however,
one should be carefull again. Clearly, not only ${\cal F}_V\{ L\}$
depends on eigenvalues, it depends on their "symmetric" (Weyl-group invariant)
combinations, i.e. it rather depends on quantities like
${\rm tr} L^a$ then on particular eigenvalues.
Moreover, powers $a$ here should be negative and fractional.

Indeed, integrals like
(\ref{KI}) are usually understood as analytical continuation from some values
of parameters in the potential $V$, when integral is convergent. They can be
also related to the formal (perturbation) seria arising when integrand is
expanded around a stationary point.
To begin with it is reasonable to take $n=1$
i.e. consider just an ordinary integral. For the sake of simplicity also take
particular $V(x) = -\frac{x^{p+1}}{p+1}$. Then the stationary point is at
$x = \lambda^{\frac{1}{p}}$ and
\be
\int dx e^{-\frac{x^{p+1}}{p+1}+ l x} \sim
l^{-\frac{p-1}{2}} e^{\frac{p}{p+1}l^{\frac{p+1}{p}}}
\sum_{k \ge 0} c_kl^{-\frac{k}{p}}.
\label{KIper}
\ee

It is now easy to understand what should be done in the general
situation with matrices and arbitrary potentials. First of all, one needs to
solve equation for the stationary point, $V'(X) =  L$. For this purpose
it is most convenient to introduce a new matrix variable $\Lambda$ instead of
$ L$, which by definition satisfies $V'(\Lambda) =  L$. Then stationary
point is just $X=\Lambda$. Second, one should separate the analogue of the
complicated prefactor (quasiclassical contribution):
\be
{\cal C}_V\{\Lambda\}  = (2\pi)^{n^2/2}
\frac{e^{{\rm tr}\left(\Lambda V'(\Lambda) - V(\Lambda)\right)}}
{\sqrt{{\rm det} V''(\Lambda)}},
\label{cvfactor}
\ee
Then the function that describes pure "quantum" contribution
\footnote{The "classical action" in (\ref{cvfactor}) can be also represented
as ${\rm tr}\left(\Lambda V'(\Lambda) - V(\Lambda)\right) =
{\rm tr}\int \Lambda dV'(\Lambda)$.
Determinant of quadratic fluctuations is defined as
$$(2\pi)^{n^2/2}
\left({\rm det} V''(\Lambda)\right)^{-1/2}
\sim \int dY e^{-{\rm tr}V_2(\Lambda,Y)}, \nn $$
where
$V_2(\Lambda,Y) \equiv {\rm lim}_{\epsilon \rightarrow 0}\frac{1}{\epsilon^2}
\left(V(\Lambda+\epsilon Y) - V(\Lambda) - \epsilon V'(\Lambda) Y\right)$.
For $V(\Lambda) = \frac{\Lambda^{p+1}}{p+1}\ $ we have $V''(\Lambda) =
( \sum_{k=0}^{p-1} \Lambda^k \otimes \Lambda^{p-k-1} )$.
One could easily choose an "opposite"
parametrization in eq.(\ref{Miwatimes}):
$T_k = -\frac{1}{k}{\rm tr}\Lambda^{-k}$. Though not quite obvious, this never
influences any results (see
section 2.10 for an example). Our choice of signs is motivated by
simplification of formulas for the GKM, including the relation between
$ L$ and $\Lambda$. Instead, some sign factors appear in formulas, related to
Toda-like representations of partition functions and those involving
$\tilde W$-operators.
}
\be
{\cal Z}_V\{T\} \equiv {\cal C}_V\{\Lambda\}^{-1} {\cal F}_V\{V'(\Lambda)\}
\label{GKMdef}
\ee
to be refered as partition function of the Generalized Kontsevich Model (GKM)
\cite{GKM}, can be represented as a formal (perturbation) series expansion
in variables
\be
T_k = \frac{1}{k}{\rm tr}\Lambda^{-k}.
\label{Miwatimes}
\ee

GN (\ref{GNe}) equation can be now rewritten as a set of differential
equations for ${\cal Z}_V\{T\}$. Indeed, we already have:
\be
{\cal C}_V^{-1}
          \left(V'\left(\frac{\partial}{\partial L_{\rm tr}}\right) -
 L\right)
{\cal C}_V {\cal Z}_V\{T\} = 0,
\label{riv1}
\ee
but it is still necessary to express the operator at the l.h.s. in terms of
$T$. This is in fact possible to do, using the relation:
\be
\frac{\partial}{\partial L_{tr}}{\cal Z}_V\{T\} =
\sum_k \frac{\partial T_k}{\partial L_{tr}}
\frac{\partial Z}{\partial T_k}
\ee
and substituting the $traces$ of $\Lambda$-matrices, which can arise in the
process of calculation, by $T$'s. It is important only that $\Lambda$'s
usually appear in negative powers: this is already achieved by the choice of a
proper normalization factor ${\cal C}_V\{\Lambda\}$. For monomial potential
$V_p(X) = \frac{X^{p+1}}{p+1}$ this is especialy simple: $ L = \Lambda^p$
and $\frac{\partial T_k}{\partial L_{tr}} = -\frac{1}{p}\Lambda^{-p-k}$.

This reasoning allows one to rewrite eq.(\ref{riv1}) identically in the form
\be
\sum_{l} \Lambda^{-l} {\cal O}_l(T) {\cal Z}_V\{T\} = 0,
\ee
where ${\cal O}_l$ are some differential operators, depending on the shape of
$V$, but independent on the size $n$ of the matrix (as all the above reasoning
never refered to particular values of $n$, except for a sample $example$ at
the very beginning). It remains to use the fact that matrix $ L$ can be
arbitrary large and have arbitrarily many independent entries, in order to
conclude that we derived a set of constraints on ${\cal Z}_V$ in the form
\be
{\cal O}_l(T) {\cal Z}_V\{T\} = 0.
\ee

For potential $V$ of degree $p+1$ these appear to be exactly the "continuous
Virasoro constraints".
See refs.\cite{MMM} and \cite{GKM} for detailed analysis of the Virasoro case
$p=2$ (associated with the pure topological gravity and with the
double-scaling limit of the 1-matrix model), and \cite{Mikh} for the
exhaustive presentation of the case of $p=3$.


\subsection{Discrete Virasoro constraints for the Gaussian Kontsevich model}

As a simplest illustration of the technique, described in the
previous subsection, we derive now the constraints for the Gaussian Kontsevich
model \cite{ChMa} with potential $V(X) = \frac{1}{2}X^2$:
\be
{\cal Z}_{\frac{X^2}{2}}\{N,T\} =
\frac{e^{-{\rm tr}\frac{ L^2}{2}}}{({\rm det} L)^N}
\int dX ({\rm det}X)^N e^{-{\rm tr}\frac{X^2}{2} +  L X}.
\label{gako}
\ee
In this case $ L = V'(\Lambda) = \Lambda$, and the time-variables are just
\be
T_k = \frac{1}{k}{\rm tr} \Lambda^{-k} = \frac{1}{k}{\rm tr} L^{-k}.
\label{gakoT}
\ee
To make the model non-trivial an extra "zero-time" variable $N$ \cite{Toda}
is introduced, which was not included into the previous definition
(\ref{GKMdef}). Now note that the $N$-dependence of Kontsevich integral
(\ref{KI}) can be described simply as an extra term in the potential:
$V(X) \rightarrow \hat V(X) = V(X) - N\log X$ (though this can $not$ be done
neither in the quasiclassical factor ${\cal C}_V$ nor in the definition of
time-variables $T$). Since the GN equation depends only on Kontsevich
equation, we can use it with $V$ substituted by $\hat V$. Then we have
instead of (\ref{riv1}):
\be
\new
\begin{array}{c}
\frac{e^{-{\rm tr}\frac{ L^2}{2}}}{({\rm det} L)^N}
\left(\frac{\partial}{\partial L_{tr}}\right)^{n+1}\cdot
\left( \frac{\partial}{\partial L_{tr}} -
N\left(\frac{\partial}{\partial L_{tr}}\right)^{-1} -  L
\right) \cdot \\
\cdot
({\rm det} L)^Ne^{+{\rm tr}\frac{ L^2}{2}}
{\cal Z}_{\frac{X^2}{2}}\{N,T\} = 0.
\end{array}
\label{gako2}
\ee
In order to get rid of the integral operator
$(\frac{\partial}{\partial L})^{-1}$
one should take here $n \geq 0$ rather than $n \geq -1$. In fact all the
equations with $n > 0$ follow from the one with $n=0$, and we restrict
our consideration to the last one. For $n=0$ we obtain from
(\ref{gako2}):
\be
\left(\left( \frac{\partial}{\partial L_{tr}} + \frac{N}{ L} +
{ L}\right)^2 - 2N -  L
\left( \frac{\partial}{\partial L_{tr}} + \frac{N}{ L} +
{ L}\right) \right) {\cal Z} = 0 \nn
\ee
or
\be
\left(\left( \frac{\partial}{\partial L_{tr}}\right)^2 +
     \left( L +
\frac{2N}{ L}\right)\frac{\partial}{\partial L_{tr}}
+ \frac{N^2}{ L^2} - \frac{N}{ L}{\rm tr}\frac{1}{ L}
\right) {\cal Z} = 0,
\label{gako3}
\ee
and it remains to substitute:
\be
\frac{\partial{\cal Z}}{\partial L_{tr}} &=
-\sum_{k=0}^{\infty} \frac{1}{ L^{k+1}}
\frac{\partial{\cal Z}}{\partial T_k};
\nn \\
\frac{\partial^2{\cal Z}}{\partial L_{tr}^2}  &=
\sum_{k=1}^{\infty}
\left( \sum_{a=1}^{k+1}\frac{1}{ L^{k+2-a}}
{\rm tr}\frac{1}{ L^a} \right)
\frac{\partial{\cal Z}}{\partial T_k}
+ \sum_{k,l=1}^{\infty} \frac{1}{ L^{k+l+2}}
\frac{\partial^2{\cal Z}}{\partial T_k\partial T_l} = \nn \\
&= \sum_{m=-1}^{\infty} \frac{1}{ L^{m+2}}
\left( \sum_{k>{\rm max}(m,0)}
\left({\rm tr}\frac{1}{ L^{k-m}}\right)
\frac{\partial{\cal Z}}{\partial T_k} +
   \sum_{k=1}^{m-1} \frac{\partial^2{\cal Z}}{\partial T_k\partial T_{m-k}}
\right)
\nn
\ee
and finally obtain:
\be
\sum_{m=-1}^{\infty} &\frac{1}{ L^{m+2}}
\left( \sum_{k=1+\delta_{m,-1}}^{\infty}
\left({\rm tr}\frac{1}{ L^{k}}\right)
\frac{\partial}{\partial T_{k+m}} +
  \sum_{k=1}^{m-1} \frac{\partial^2}{\partial T_k\partial T_{m+k}} -
\right. \nn \\ &\left.
-  \frac{\partial}{\partial T_{m+2}} - 2N\frac{\partial}{\partial T_{m}} +
N^2\delta_{m,0} - N \left({\rm tr}\frac{1}{ L}\right) \delta_{m,-1}
\right) {\cal Z} = \nn
\\
&= \sum_{m=-1}^{\infty} \frac{1}{ L^{m+2}}
e^{NT_0}  L_m(T+r) e^{-NT_0} {\cal Z}
= 0.
\label{gako4}
\ee
Here $L_m(t)$ are just the generators (\ref{virdop}) of the discrete Virasoro
algebra (\ref{virdid}):
\be
e^{Nt_0}  L_m(t) e^{-Nt_0} =
e^{Nt_0}  \left( \sum_{k=1}^{\infty} kt_k\frac{\partial}{\partial t_{k+m}} +
\sum_{k=0}^m \frac{\partial^2}{\partial t_k\partial t_{m-k}}
\right) e^{-Nt_0}.
\ee
and at the r.h.s. of (\ref{gako4}) $r_k =
-\frac{1}{2}\delta_{k,2}$.\footnote{
This small correction is a manifestation of a very general
phenomenon: from the point of view of symmetries (Ward identities)
it is more natural to consider $Z_V$ not as a function of $T$-variables, but
of some more complicated combination $\hat T_k + r_k$, depending on the shape
of potential $V$. If $V$ is a polinomial of degree $p+1$, $\hat T_k =
\frac{1}{k}{\rm tr} (V'(\lambda))^{-k/p},$ while $r_k = \frac{p}{k(p-k)}{\rm
Res}\left(V'(\mu)\right)^{1-\frac{k}{p}}d\mu$. For monomial potentials these
expressions become very simple: $\hat T_k = T_k$ and $r_k =
-\frac{p}{p+1}\delta_{k,p+1}$. See \cite{comamo} and section 4.9 below for
more details. In most places in these notes we prefer to use invariant
potential-independent times $T_k$, instead of $\hat T_k$, but then Ward
identites acquire some extra terms with $r_k$ (which in fact will be
very simple in our examples, which are all given for monomial potentials).
}

Thus we found that the WI of the Gaussian Kontsevich model (\ref{gako})
coincide with those of the ordinary 1-matrix model, moreover the size of the
matrix $N$ in the latter model is associated with the "zero-time" in the
former one. This result \cite{ChMa} of course implies, that the two models
are identical:
\be
e^{-NT_0}{\cal Z}_{\frac{X^2}{2}}\{N,T_1,T_2,\ldots\} \sim
Z_N\{T_0,T_1,T_2,\ldots\}.
\ee
We shall discuss direct connection between these two matrix integrals
(\ref{1mamo'}) and (\ref{gako}) in the next section, after some more details
will be presented about the structure of "eigenvalue" matrix models.

\subsection{Continuous Virasoro constraints for the $V = \frac{X^3}{3}$
Kontsevich model}

This example is a little more complicated than that in the previous
subsection, and we do not present calculations in full details (see
\cite{MMM} and \cite{GKM}). Our goal is to demonstrate that the
constraints which arise in this model,
though still form  (Borel subalgebra of) some Virasoro
algebra, are $different$  from (\ref{virdid}). From the point of view of the
CFT-formulation the relevant model is that of the $twisted$ (in this
particular case - antiperiodic) free fields. These so called "continuous
Virasoro constraints" give the simplest illustration of the difference between
discrete and continuous matrix models: this is essentially the difference
between "homogeneous" (Kac-Frenkel) and "principal" (soliton vertex operator)
representation of the level $k=1$ Kac-Moody algebra. From the point of view of
integrable hierarchies this is the difference between Toda-chain-like and
KP-like hierarchies. We shall come back to a more detailed discussion of this
difference later, when the "multi-scaling continuum limit" will be considered.

Another (historical) aspect of the same relation also deserves mentioning,
since it also illustrates the interrelation between different models.
The discrete 1-matrix model arises naturally in description of quantum $2d$
gravity as sum over 2-geometries in the formalism of random equilateral
triangulations. The model, however, decribes only lattice approximation to
$2d$ gravity and (double-scaling) coninuum limit should be taken in order to
obtain the real (continuous) theory of $2d$ gravity. This limit was originally
formulated in terms of the contraint algebra (equations of motion or "loop"
or "Schwinger-Dyson" equations - terminology is taste-dependent), leaving open
the problem of what is the form of partition function ${\cal Z}^{cont}\{T\}$
of continuous theory. Since the relevant algebra appeared to be just the WI
for  Kontsevich model (with $V(X) = \frac{X^3}{3}$), this proves that the
latter one is exactly the continuous theory of pure $2d$ gravity. At the same
time, Kontsevich model itself can be naturally introduced as a theory of
$topological$ gravity (in fact this is how the model was originally discovered
in \cite{Ko}). From this point of view the constraint algebra, to be discussed
below in this subsection, plays central role in the proof of equivalence
between pure $2d$ quantum gravity and pure topological gravity (in both cases
"pure" means that "matter" fields are not introduced).

After these introductory remarks we proceed to  calculations. Actually they
just repeat those for the Gaussian model, performed in the previous
subsection, though formulas get somewhat more complicated. This time we do not
include zero-time $N$ and just use eq.(\ref{GNe}) with $V(X) = \frac{X^3}{3}$.
Now it is also much more tricky (though possible)
to work in  matrix notations (because
fractional powers of $ L$ will be involved) and we rewrite everything in
terms of the eigenvalues of $ L$.

We substitute
\be
\new
\begin{array}{c}
{\cal C}_{\frac{X^3}{3}} = \frac{\prod_\delta
e^{\frac{2}{3}\lambda_\delta^{3/2}}}
{\sqrt{\prod_{\gamma,\delta}
              (\sqrt{\lambda_\delta} + \sqrt{\lambda_\gamma})}},  \\
\left( \frac{\partial^2}{\partial L_{tr}^2}\right)_{\gamma\gamma} =
 \frac{\partial^2}{\partial\lambda_\gamma^2} +
     \sum_{\delta\neq\gamma}\frac{1}{\lambda_\gamma-\lambda_\delta}
     \left(\frac{\partial}{\partial\lambda_\gamma}
                 - \frac{\partial}{\partial\lambda_\delta}\right)
\end{array}
\nn
\ee
and introduce a special notation for
\be
\frac{{\cal D}}{{\cal D}\lambda_\gamma} \equiv
{\cal C}_{\frac{X^3}{3}}^{-1} \frac{\partial}{\partial \lambda_\gamma}
         {\cal C}_{\frac{X^3}{3}} =
 \frac{\partial}{\partial \lambda_\gamma} + \sqrt{\lambda_\gamma} -
\frac{1}{4\lambda_\gamma}
- \frac{1}{2} \sum_{\delta\neq\gamma}\frac{1}{\sqrt{\lambda_\gamma}
(\sqrt{\lambda_\delta} +
           \sqrt{\lambda_\gamma})}. \nn
\ee
Then (\ref{GNe}) turns into
\be
\left( \left(\frac{{\cal D}}{{\cal D}\lambda_\gamma}\right)^2 +
\sum_{\delta\neq\gamma}\frac{1}{\lambda_\gamma-\lambda_\delta}
\left(\frac{{\cal D}}{{\cal D}\lambda_\gamma} - \frac{{\cal D}}{{\cal
D}\lambda_\delta}\right)
\right) {\cal Z}_{\frac{X^3}{3}}\{T\} = 0.
\label{GNX3}
\ee
Now we need explicit expression for $T$:
\be
T_k = \frac{1}{k} L^{-k},
\label{KMX3T}
\ee
and as we already know from the previous subsection we  also need
\be
r_k = - \frac{2}{3}\delta_{k,3}.
\ee
It will not be explained untill we turn to consideration of integrable
structure of Kontsevich model in the following sections,
but ${\cal Z}_{\frac{X^3}{3}}\{T\}$ is
in fact independent of all time-variables with $even$ numbers
(subscripts). Therefore we can take only $k=2a+1$ in (\ref{KMX3T}),
\be
T_{2a+1} &= \frac{1}{2a+1} \sum_\delta \lambda_\delta^{-a-\frac{1}{2}}, \nn \\
r_{2a+1} &= - \frac{2}{3}\delta_{a,1}
\ee
and
\be
\frac{\partial}{\partial \lambda_\gamma} {\cal Z}_{\frac{X^3}{3}}\{T\} &=
\sum_{a=0}^{\infty} \frac{\partial T_{2a+1}}{\partial \lambda_\gamma}
   \frac{\partial{\cal Z}}{\partial T_{2a+1}} =
  -\frac{1}{2} \sum_{a=0}^{\infty} \lambda_\gamma^{-a-\frac{3}{2}}
                 \frac{\partial{\cal Z}}{\partial T_{2a+1}}; \nn \\
\frac{\partial^2}{\partial \lambda_\gamma^2} {\cal Z}_{\frac{X^3}{3}}\{T\} &=
  \frac{1}{4}\sum_{a,b=0}^{\infty} \lambda_\gamma^{-a-b-3}
                 \frac{\partial{\cal Z}}{\partial T_{2a+1}\partial T_{2b+1}} +
   \frac{1}{2} \sum_{a=0}^{\infty} (a + \frac{3}{2})
\lambda_\gamma^{-a-\frac{5}{2}}
                 \frac{\partial{\cal Z}}{\partial T_{2a+1}}. \nn
\ee
These expressions should be now substituted into (\ref{GNX3}) and we obtain:
\be
\new
\begin{array}{c}
\frac{1}{4}\sum_{a,b=0}^{\infty} \lambda_\gamma^{-a-b-3}
                 \frac{\partial{\cal Z}}{\partial T_{2a+1}\partial T_{2b+1}} +
 \\
+ \sum_{a=0}^{\infty} \left[
\frac{1}{2} \sum_{a=0}^{\infty}
(a + \frac{3}{2})\lambda_\gamma^{-a-\frac{5}{2}} -
\frac{1}{2} \sum_{\delta\neq\gamma}\frac{1}{\lambda_\gamma-\lambda_\delta}
\left(\lambda_\gamma^{-a-\frac{3}{2}} - \lambda_\delta^{-a-\frac{3}{2}}
\right) -
\right.
 \\
\left.
- \left( \sqrt{\lambda_\gamma} - \frac{1}{4\lambda_\gamma}
- \frac{1}{2} \sum_{\delta\neq\gamma}
\frac{1}{\sqrt{\lambda_\gamma}(\sqrt{\lambda_\delta} +
           \sqrt{\lambda_\gamma})}
\right)\lambda_\gamma^{-a-\frac{3}{2}}
\right]\frac{\partial{\cal Z}}{\partial T_{2a+1}} +  \\
+ \left[\ldots\right] {\cal Z}  \ \ \
= \ \ \ \sum_{n=-1}^{\infty} \frac{1}{\lambda_\gamma^{n+2}}{\cal L}_n{\cal Z}
\end{array}
\label{vircder}
\ee
with
\be
\new
\begin{array}{c}
{\cal L}_{2n} =
\sum_{a=0}^{\infty}
    \left(a+\frac{1}{2}\right)\left(T_{2a+1}+r_{2a+1}\right) \times \\ \times
\frac{\partial}{\partial T_{2a+2n+1}} +
\frac{1}{4} \sum_{\stackrel{a+b=n-1}{a,b\geq 0}}
\frac{\partial^2}{\partial T_{2a+1}\partial T_{2b+1}}
+  \frac{1}{16}\delta_{n,0} + \frac{1}{4} T_1^2\delta_{n,-1} = \\
= \frac{1}{2}\sum_{{\rm odd}\ k=1}^{\infty}
       k(T_k+r_k)\frac{\partial}{\partial T_{k+2n}} +
 \frac{1}{4} \sum_{{\rm odd}\ k = 1}^{2n-1}
\frac{\partial^2}{\partial T_k\partial T_{2n-k}} +
  \frac{1}{16}\delta_{n,0} + \frac{1}{4} T_1^2\delta_{n,-1}.
\end{array}
\label{vircidder}
\ee
Factor $\frac{1}{2}$ in front of the first term at the r.h.s. in
(\ref{vircidder}) is important for ${\cal L}_{2n}$ to satisfy the properly
normalized Virasoro algebra:\footnote{
Therefore it could be reasonable to use a different notation: ${\cal L}_n$
instead of ${\cal L}_{2n}$. We prefer ${\cal L}_{2n}$, because it emphasises
the property of the model to be 2-reduction of KP hierarchy (to KdV), see
section 4 below.
}
\be
\phantom. [{\cal L}_{2n}, {\cal L}_{2m}] = (n-m){\cal L}_{2n+2m}.
\nn
\ee
Coefficient $\frac{1}{4}$ in front of the second term can be eliminated by
rescaling of time-variables: $T \rightarrow \frac{1}{2}T$, then the last term
turns into $\frac{1}{16}T_1^2\delta_{n,-1}$.

We shall not actually discuss evaluation of the coefficient in front of
${\cal Z}$ (with no derivatives), which is denoted by $[\ldots]$ in
(\ref{vircder}) (see \cite{MMM} and \cite{GKM}). In fact almost all the terms
in original complicated expression cancel, giving finally
\be
\left[ \ldots \right] =
\frac{1}{16\lambda_\gamma^2} + \frac{T_1^2}{4\lambda_\gamma},
\nn
\ee
and this is represented by the terms with $\delta_{n,0}$ and $\delta_{n,-1}$
in expressions (\ref{vircidder}) for the Virasoro generators ${\cal L}_{2n}$.

The term with the double $T$-derivative in (\ref{vircder}) is already of the
necessary form. Of intermidiate complexity is evaluation of the coefficient in
front of $\frac{\partial{\cal Z}}{\partial T_{2a+1}}$ in (\ref{vircder}),
which we shall briefly describe now.
First of all, rewrite this coefficient, reordering the items:
\be
\frac{1}{2}\left[ (a + \frac{3}{2})\lambda_\gamma^{-a-\frac{5}{2}} -
 \sum_{\delta\neq\gamma}\frac{1}{\lambda_\gamma-\lambda_\delta}
\left(\lambda_\gamma^{-a-\frac{3}{2}} -
\lambda_\delta^{-a-\frac{3}{2}} \right)\right] +
\nn \\
+ \left[ \frac{1}{4}\lambda_\gamma^{-a-\frac{5}{2}} +
       \frac{1}{2} \sum_{\delta\neq\gamma}
\frac{\lambda_\gamma^{-a-2}}{\sqrt{\lambda_\delta} +
           \sqrt{\lambda_\gamma}} \right]
- \lambda_\gamma^{-a-1}.
\label{vircder2}
\ee
The first two terms together are equal to the sum over $all$ $j$ (including
$j=i$):
\be
- \frac{1}{2} \sum_\delta \frac{1}{\lambda_\gamma-\lambda_\delta}
\left(\lambda_\gamma^{-a-\frac{3}{2}} - \lambda_\delta^{-a-\frac{3}{2}} \right)
 =  \frac{1}{2} \sum_\delta \frac{\lambda_\gamma^{a+\frac{3}{2}} -
             \lambda_\delta^{a+\frac{3}{2}}}{\lambda_\gamma-\lambda_\delta}
  \cdot\frac{1}{\lambda_\gamma^{a+\frac{3}{2}}\lambda_\delta^{a+\frac{3}{2}}} =
\nn \\
= \frac{1}{2\lambda_\gamma^{a+2}} \sum_\delta
\frac{\lambda_\gamma^{a+2} - \lambda_\gamma^{\frac{1}{2}}
\lambda_\delta^{a+\frac{3}{2}}}
{\lambda_\gamma-\lambda_\delta}\cdot\frac{1}{\lambda_\delta^{a+\frac{3}{2}}}.
\nn
\ee
Similarly, the next two terms can be rewritten as
\be
\frac{1}{2} \sum_\delta \frac{\lambda_\gamma^{-a-2}}{\sqrt{\lambda_\gamma} +
\sqrt{\lambda_\delta}} =
\frac{1}{2\lambda_\gamma^{a+2}}
\sum_\delta \frac{\sqrt{\lambda_\gamma} - \sqrt{\lambda_\delta}}
{\lambda_\gamma-\lambda_\delta} = \nn \\
= \frac{1}{2\lambda_\gamma^{a+2}} \sum_\delta
\frac{\lambda_\gamma^{\frac{1}{2}} \lambda_\delta^{a+\frac{3}{2}} -
\lambda_\delta^{a+2}}
{\lambda_\gamma-\lambda_\delta}\cdot\frac{1}{\lambda_\delta^{a+\frac{3}{2}}}.
\nn
\ee
The sum of these two expressions is equal to
\be
\frac{1}{2\lambda_\gamma^{a+2}} \sum_\delta
\frac{\lambda_\gamma^{a+2} - \lambda_\delta^{a+2}}
{\lambda_\gamma-\lambda_\delta}
\cdot\frac{1}{\lambda_\delta^{a+\frac{3}{2}}}.
\nn
\ee
Note that powers $a+2$ are already integer and the remaining ratio can be
represented as a sum of $a+2$ terms. Adding also the last term from the l.h.s.
of (\ref{vircder2}), we finally obtain:
\be
-\frac{1}{\lambda_\gamma^{a+1}} + \frac{1}{2} \sum_{n=-1}^a
\frac{1}{\lambda_\gamma^{n+2}}\sum_\delta
\frac{1}{\lambda_\delta^{a-n+\frac{1}{2}}} =   \nn \\ =
\frac{1}{2}\sum_{n=-1}^a \frac{1}{\lambda_\gamma^{n+2}}
(2a-2n+1)(T+r)_{2a-2n+1}
\nn
\ee
in accordance with (\ref{vircder}) and (\ref{vircidder}).

\subsection{$\tilde W$-constraints for the asymmetric 2-matrix model}

We turn now to a very different application \cite{tildeW} of the GN equation
(\ref{GNe}). Namely, we shall now consider ${\cal F}_{V,n}\{ L\}$
as a building block in construction of conventional discrete two-matrix model
\be
Z_N\{t,\bar t\} \equiv c_N^2 \int dHd\bar H
e^{\sum_k (t_k {\rm Tr}H^k + \bar t_k {\rm Tr} \bar H^k) + {\rm Tr}H\bar H} =
\nn \\
= \int d L e^{\sum_k t_k {\rm Tr} L^k}
{\cal F}_{\bar U,N}\{ L\}.
\label{2mamo}
\ee
Now $ L$ plays the role of $H$ and $\bar U(\bar H) =
\sum_k\bar t_k  \bar H^k$.

We can now use GN equation to derive a relation for $Z_N\{t,\bar t\}$.
Take (\ref{GNe}),
\be
\left(\bar U (\frac{\partial}{\partial L_{tr}}) +  L\right)
{\cal F}_{\bar U,N}\{ L\} = 0,
\label{GNe'}
\ee
multiply it by  $e^{{\rm Tr}U( L)} = e^{\sum_k t_k {\rm Tr} L^k}$
and integrate over $ L$.
In order to express this relation in terms of $t$-derivatives of $z$ it is
necessary to have some "scalar" rather than $matrix$ equations, therefore
we"ll actually need to take trace of (\ref{GNe'}). However, in order not to
loose any information, we first multiply (\ref{GNe'}) by $ L^n$ and
$then$ take the trace. In this way we obtain:
\be
\int d L e^{\sum_k t_k {\rm Tr} L^k}
{\rm Tr}  L^n
\left(\bar U (\frac{\partial}{\partial L_{tr}}) +  L\right)
{\cal F}_{\bar U}\{ L\}  = 0.
\nn
\ee
Integration by parts gives:
\be
\int d L {\cal F}_{\bar U}\{ L\}
{\rm Tr}\left(\bar U (-\frac{\partial}{\partial L_{tr}}) +  L\right)
 L^n e^{\sum_k t_k {\rm Tr} L^k}.
\label{gana1}
\ee

Now we need to introduce a new class of operators \cite{tildeW}.
Consider the action of
${\rm Tr} \frac{\partial^m}{\partial  L_{tr}^m}  L^n$
on  $e^{{\rm Tr}U( L)} = e^{\sum_k t_k {\rm Tr} L^k}$. It
gives some linear combination of terms like
\be
{\rm tr} L^{a_1} ...{\rm tr} L^{a_l} e^{{\rm tr} U( L)} =
\frac{\partial^l}{\partial t_{a_1}...\partial t_{a_l}} e^{-{\rm tr} U( L)}
\nn
\ee
i.e. we obtain a combination of differential operators with $t$-derivatives,
to be denoted $\tilde W(t)$:
\be
\tilde W_{n-m}^{(m+1)}(t)
e^{{\rm tr} U( L)} \equiv
{\rm Tr} \frac{\partial^m}{\partial  L_{tr}^m}  L^n
e^{{\rm tr} U( L)}, \ \ \ m,n \geq 0.
\label{twop}
\ee
For example,
\be
\new
\begin{array}{c}
\tilde W_n^{(1)} = \frac{\partial}{\partial t_n}, \ \ n\geq 0;  \\
\tilde W_n^{(2)} = \sum_{k=0}^{\infty} kt_k\frac{\partial}{\partial t_{k+n}} +
    \sum_{k=0}^n \frac{\partial^2}{\partial t_k\partial t_{n-k}},
\ \ n\geq -1;   \\
\tilde W_n^{(3)} = \sum_{k,l = 1}^{\infty} kt_klt_l
\frac{\partial}{\partial t_{k+l+n}}
+ \sum_{k=1}^{\infty} kt_k
  \sum_{a+b=k+n}\frac{\partial^2}{\partial t_a\partial t_b}  + \\
+ \sum_{k=1}^{\infty} kt_k
  \sum_{a+b=n+1}\frac{\partial^2}{\partial t_a\partial t_{b+k-1}}
+ \sum_{a+b+c=n} \frac{\partial^3}{\partial t_a\partial t_b\partial t_c}
+ \frac{(n+1)(n+2)}{2}\frac{\partial}{\partial t_n};
 \\
\ldots
\end{array}
\label {twopex}
\ee
Note, that while $\tilde W_n^{(1)}$ and $\tilde W_n^{(2)}$
are just the ordinary $(U(1)$-Kac Moody and Virasoro
operators respectively, the higher $\tilde
W^{(m)}$-operators do $not$ coincide with the generators of the
{\bf W}-algebras: already
\be
\tilde W_n^{(3)} \neq W^{(3)} =
\sum_{k,l = 1}^{\infty} kt_klt_l
\frac{\partial}{\partial t_{k+l+n}}
+ 2\sum_{k=1}^{\infty} kt_k
  \sum_{a+b=k+n}\frac{\partial^2}{\partial t_a\partial t_b} \nn \\
+ \frac{4}{3}
  \sum_{a+b+c=n} \frac{\partial^3}{\partial t_a\partial t_b\partial t_c}.
\nn
\ee
$\tilde W$-operators (in variance with ordinary $W$-operators) satisfy
recurrent relation:
\be
\tilde W_n^{(m+1)} = \sum_{k=1}^{\infty} kt_k\tilde W_{n+k}^{(m)} +
          \sum_{k=0}^{m+n-1} \frac{\partial}{\partial t_k}
                      \cdot \tilde W_{n-k}^{(m)}, \ \ \ n\geq -m.
\ee
Actually not too much is already known about the ${\tilde W}$ operators and
the structure of $\tilde{\bf W}$-algebras (in particular it remains unclear
whether the negative harmonics $\tilde W_n^{(m+1)}$ with $n < -m$ can be
introduced in any reasonable way), see \cite{tildeW} for some preliminary
results.

Equation (\ref{gana1}) can now be represented in terms of the $\tilde
W$-operators:
\be
\int d L {\cal F}_{\bar U}\{ L\}
\left(\sum_{k\geq 1} k\bar t_k
\left(-\frac{\partial}{\partial L_{tr}}\right)^{k-1}
+  L\right) L^n e^{{\rm Tr}U(t)} = \nn \\
= \left(\sum_{k\geq 1} (-)^{k-1}k\bar t_k \tilde W_{n+1-k}^{(k)} +
     \tilde W_{n+1}^{(1)} \right) Z_N\{t,\bar t\} = 0.
\label{gana2}
\ee
This relation is highly asymmetric in $t$ and $\bar t$, and in fact it
provides a suitable description of the WI only in the somewhat peculiar case
when potential $\bar U(\bar H)$ is a polinomial of $finite$ degree. See
refs.\cite{GaNa} and \cite{tildeW} for discussion of such asymmetric models.

\subsection{$\tilde W$-constraints for generic 2-matrix model}

When both potentials $U$ and $\bar U$ in (\ref{2mamo}) are generic formal
seria, eqs.(\ref{gana2}) represent only a one-parametric subset of the
2-parametric family of WI. Before we describe the whole set, let us emphasize
that the two-matrix model (\ref{2mamo}) is the one, where the action is $not$
of the most general form, consistent with some symmetry. Therefore it is not
covariant under arbitrary change of variables $H,\bar H \longrightarrow
f(H,\bar H), \bar f(H,\bar H)$, and our usual method of derivation of Ward
identities does not work. The reason why generic 2-matrix model with action
containing all the possible combinations ${\rm Tr} (H^{a_1}\bar H^{b_1}
H^{a_2}\bar H^{b_2} ...)$  is never considered seriously is essentially our
poor understanding of the unitary-matrix interals for "non-eigenvalue"
theories, to which class such generic model belongs.
For reasons to be explained in the next section such problems do not arise for
the models of the form (\ref{2mamo})
or (\ref{comamo"}), and this is why they attracted most attention so far.
Hopefully the problems with the unitary-matrix integrals are temporal and
this restricted class of multimatrix models will be unlarged, this should be
especially easy to do in the part of the theory dealing with constraint
algebras, but this subject is beyond the scope of the present notes.

In order to derive the complete set of WI for the model (\ref{2mamo}), we
apply the following semi-artificial trick. Note that exponential
$e^{{\rm Tr}H\bar H}$ satisfies:
\be
\left( {\rm Tr} H^n\frac{\partial^m}{\partial H^m_{tr}} -
{\rm Tr} \bar H^m \frac{\partial^n}{\partial \bar H^n_{tr}}\right)
e^{{\rm Tr} H\bar H} = 0.
\label{wt2m1}
\ee
Let us integrate this identity over $H$ and $\bar H$ with the weight
$e^{{\rm Tr}U(H) + {\rm Tr}\bar U(\bar H)}$ and then integrate by parts.
We obtain an identity:
\be
\new
\begin{array}{c}
\int dHd\bar H e^{{\rm Tr} H\bar H} \cdot \\ \cdot
\left( {\rm Tr} \left(-\frac{\partial}{\partial H_{tr}}\right)^mH^n  -
{\rm Tr} \left(-\frac{\partial}{\partial \bar H_{tr}}\right)^n\bar H^m\right)
e^{{\rm Tr}U(H) + {\rm Tr}\bar U(\bar H)} = 0,
\end{array}
\label{wt2m2}
\ee
which can be represented in terms of $\tilde W$ operators \cite{GKM}:
\footnote{Relations (\ref{wt2m1}) and thus (\ref{twid}) are in the obvious
sense associated with $\ {\rm Tr} H^n\bar H^m$. Of course there are similar
relations, in the same sense associated with any object like
${\rm Tr} (H^{a_1}\bar H^{b_1}H^{a_2}\bar H^{b_2} ...)$
and with products of such traces: it is enough to substitute all
$\bar H \rightarrow \frac{\partial}{\partial H_{tr}}$ to obtain the l.h.s. of
the equation and substitute all $H \rightarrow \frac{\partial}{\partial
\bar H_{tr}}$ to obtain its r.h.s. (one should only remember that such
substitution is possible, say in the l.h.s. if all the $\bar H$ are put
to the right of all $H$, in order to restore the matrix-product form of the
relation, one should carefully take into account all the commutators, arising
when $\frac{\partial}{\partial H_{tr}}$ is carried back to original position
of the corresponding $\bar H$).
All such relations can appear to be just implications of eq.(\ref{twid}).
}
\be
\tilde W^{(m+1)}_{n-m}(t) Z\{t,\bar t\} =
(-)^{m-n}\tilde W^{(n+1)}_{m-n}(\bar t) Z\{t,\bar t\}, \ \ {\rm for\ all}\
m,n\geq 0.
\label{twid}
\ee

This is the full(?) set of WI for the 2-matrix model.
When one of potentials (say, $U(t)$ is polinomial of finite degree, the
most of this symmetry is "spontaneously broken", the surviving part being
described by eqs.(\ref{gana2}).

Among other things eq.(\ref{twid})
reveals an amusing automorphsim of the $\tilde{\bf W}_{\infty}$
algebra:
\be
\tilde W^{(m+1)}_{n-m} \longleftrightarrow  \tilde W^{(n+1)}_{m-n}, \ \ \
m,n\geq 0,
\ee
for example, Virasoro's Borel subalgebra is formed not only by operators
$\tilde W^{(2)}_n$, but also by $\tilde W^{(n+2)}_{-n}, \ n\geq -1$
(while $U(1)$ Borel subalgebra -  not only by $\tilde W^{(1)}_n =
\frac{\partial}{\partial t_n}$, but also by $\tilde W^{(n+1)}_n, \ n\geq 0$).

One can attempt to apply the same procedure and derive $\tilde W$-identites
for the conventional $(p-1)$-matrix models with $p-1 > 2$. In principle, this
is possible, but unfortunately the arising equations neither have a nice form
nor is there many enough of them. However, for illustrational purposes we shall
scetch some relevant fromulas in the rest of this subsection.

Consider the multimatrix integral
\be
\new
\begin{array}{c}
Z  = \int dH_1...dH_{p-1} \cdot
\\
\cdot e^{{\rm Tr}U_1(H_1) + \ldots + {\rm
Tr}U_{p-1}(H_{p-1})}\ldots e^{{\rm Tr}(H_1H_2 + H_2H_3 + \ldots
+H_{p-2}H_{p-1})}
\end{array}
\label{mumamosc}
\ee
Acting on $Z$, operator
$\ \displaystyle{\tilde W^{(m+1)}_{n-m}(t^{(1)})}\ $ produces an
insertion of
$\ \displaystyle{{\rm Tr} H_1^n \stackrel{\longleftarrow}
{\left(\frac{\partial}{\partial H_{1,tr}}\right)^m}}$
at the position of $\ldots$ in (\ref{mumamosc}).
Integration by parts gives:
\be
{\rm Tr} H_1^n \stackrel{\longrightarrow}
{\left(-\frac{\partial}{\partial H_{1,tr}}\right)^m}
\longrightarrow (-)^m{\rm Tr} H_1^nH_2^m = (-)^m{\rm Tr} H_2^mH_1^n
\nn
\ee

In the case of $p-1=2$, that we discussed above, this can be rewriten as
$\displaystyle{(-)^m{\rm Tr} H_1^m \stackrel{\longrightarrow}
{\left(\frac{\partial}{\partial H_{2,tr}}\right)^n}}$  and integration
by parts gives
$\displaystyle{(-)^m{\rm Tr} H_2^m \stackrel{\longleftarrow}
{\left(-\frac{\partial}{\partial H_{2,tr}}\right)^n}}$,
what is equivalent to the action of $(-)^{m+n}\tilde W^{(n+1)}_{m-n}(t^{(2)})$
on $Z$: we reproduce equation (\ref{twid}).

However, for $p-1>2$ things are more complicated. Insertion of ${\rm Tr}
H_2^mH_1^n$ is equivalent to that of
$\displaystyle{{\rm Tr} H_2^m\left(\stackrel{\longrightarrow}
{\frac{\partial}{\partial H_{2,tr}}} - H_3\right)^n}$, which after
integration by parts and acts on $e^{U_2(H_2)}$ and gives:
\be
\new
\begin{array}{c}
{\rm Tr} H_2^m\left(-\stackrel{\longleftarrow}
{\frac{\partial}{\partial H_{2,tr}}} - H_3\right)^n \sim
\\ \sim
{\rm Tr} H_2^m\left(\sum_k kt_k^{(2)}H_2^{k-1} - H_3\right)
\left(-\stackrel{\longleftarrow}
{\frac{\partial}{\partial H_{2,tr}}} - H_3\right)^{n-1} \sim \ldots
\end{array}
\ee
Derivatives remaining at the r.h.s. should be carried through the first
bracket and than act on $e^{U_2(H_2)}$ etc. After all we get
some linear combination of terms like
${\rm Tr} H_2^{b_1}H_3^{c_1}H_2^{b_2}H_3^{c_2}\ldots $
with $t^{(2)}$-dependent coefficients.

Now, if we are dealing with the $p-1=3$-matrix model, every $H_2$
standing to the right
of all $H_3$'s can be substituted by $\frac{\partial}{\partial H_{3,tr}}$,
otherwise one should also include terms with commutators when this
$\frac{\partial}{\partial H_{3,tr}}$ is carried back to the place where
$H_2$ was standing. In this way we obtain a combination of insertions of
the form
\be
\new
\begin{array}{c}
{\rm Tr} \stackrel{\longrightarrow}
{\left(\frac{\partial}{\partial H_{3,tr}}\right)^{b_1}}H_3^{c_1}
   \stackrel{\longrightarrow}
{\left(\frac{\partial}{\partial H_{3,tr}}\right)^{b_2}}H_3^{c_2}\ldots  \sim
\\
\sim {\rm Tr} \stackrel{\longleftarrow}
{\left(-\frac{\partial}{\partial H_{3,tr}}\right)^{b_1}}H_3^{c_1}
 \stackrel{\longleftarrow}
{\left(-\frac{\partial}{\partial H_{3,tr}}\right)^{b_2}}H_3^{c_2}\ldots
\end{array}
\ee
The resulting operator can be expressed through $\tilde W(t^{(3)})$ and we
obtain an identity, saying that some algebraic combination of
$\tilde W(t^{(1)})$ and $\tilde W(t^{(3)})$ with $t^{(2)})$-dependent
coefficients annihilates the partition function.

For $p-1>3$ insertion of $H_2$ is equivalent to that of
$\frac{\partial}{\partial H_{3,tr}} - H_4$
rather than  $\frac{\partial}{\partial H_{3,tr}}$, and the
procedure should be repeated again and again.
Finaly one arrives at constraints where the
operators are algebraic combinations of
$\tilde W(t^{(1)})$ and $\tilde W(t^{(p-1)})$ with the coefficients, which
depend on $t^{(2)},\ldots,t^{(p-2)}$ (moreover these are $infinite$
seria in $\tilde W$ operators, unless all the intermediate potentials
$U_2,\ldots,U_{p-2}$ are polinomials of $finite$ degree.

This is of course not a too illuminating procedure and in fact it was never
worked through to get concrete identities in any nice form.
Instead it can serve to
illustrate the problems, peculiar for the class of conventional multimatrix
models (at least for $p-1>2$). It can also emphasize the beauty of $conformal$
multimatrix models, which have clear advantages already at the level of
Ward identites.

\subsection{$\tilde W$-operators in Kontsevich model}

After $\tilde W$-operators are introduced, we can also rewrite the GN equation
(\ref{riv1})
for Kontsevich models in terms of $\tilde W$'s. Namely, we shall prove the
following identity
\cite{tildeW}:
\be
\left(\frac{\partial}{\partial
\Lambda_{tr}}\right)^{m+1} {\cal Z}\{ T_k\} =
(\pm)^{m+1}
\sum_{l\geq 0} \Lambda^{-l-1} \tilde W_{l-m}^{(m+1)}( T)
{\cal Z}\{ T_k\},
\label{tiwko}
\ee
valid for $any$ function ${\cal Z}$ which depends on
$ T_k = \mp\frac{1}{k}{\rm tr}\Lambda^{-k},\ \ k\geq 1\ $ and
$ T_0 = \pm{\rm tr}\log \Lambda$
with $n\times n$ matrix $\Lambda$.
Application of the identity (\ref{tiwko}) is most straightforward in the
Gaussian model (\ref{gako}), e.g. for transformation of eq.
(\ref{gako3}) into eq.(\ref{gako4})
(we remind that $ L = \Lambda$ in this case).
In other cases calculations with the use of identity
(\ref{tiwko}), accounting for the quasiclassical factor
${\cal C}_V\{ L\}$ and the difference between $ L = V'(\Lambda)$ and $\Lambda$
become somewhat more involved, though still seem enough straightforward.
Also for particular potentials
$V(X)$ partition function ${\cal Z}_V\{T\}$  is actually independent of
certain (combinations of) time-variables (for example, if
$V(X) = \frac{X^{p+1}}{p+1}$ it is independent of all $T_{pk},\ k\in Z_+$),
and this is
important for appearence of the constraints in the standard from like
eqs.{\ref{vircder}), (\ref{vircidder}),
i.e. for certain $reduction$ of $\tilde W$-constraints to the ordinary
$W$-constraints. This relation between $\tilde W$- and $W$-operators deserves
further investigation.

The proof of eq.(\ref{tiwko}) is provided by the following trick. Let us make
a sort of Fourier transformation:
\be
{\cal Z}\{ T\} =
\int dH\ {\cal G}\{H\} e^{\sum_{k=0}^{\infty} T_k{\rm Tr}H^k},
\label{mafoutr}
\ee
where integral is over $N\times N$ Hermitean matrix $H$.\footnote{
Here it is for the first time that we encounter an important idea: matrix
models
- the ordinary 1-matrix model (\ref{1mamo'})
in this case - can be considered as defining integral transformations. This
view on matrix models can to large extent define their role in the future
developement of string theory.
}
Then it is clear that once the identity (\ref{tiwko}) is established for
${\cal Z}\{ T\}$ substituted by $ e^{{\rm Tr} U(H)},\ \
 U(H) = \sum_{k=0}^{\infty} T_k{\rm Tr}H^k$, with any matrix $H$,
it is valid for $any$ function ${\cal Z}\{ T\}$. The advantage of such
substitution is that we can now make use of the definition (\ref{twop})
of the $\tilde W$ operators in order to rewrite (\ref{tiwko}) in a very
explicit form:
\be
\left(\frac{\partial}{\partial \Lambda_{tr}}\right)^{m+1}
e^{{\rm Tr} U(H)} =
(\pm)^{m+1}\sum_{l \geq 0}^{\infty} \Lambda^{-l-1}
 \tilde W_{l-m}^{(m+1)}( T) e^{{\rm Tr} U(H)} = \nn \\
= (\pm)^{m+1}
\sum_{l \geq 0}^{\infty} \Lambda^{-l-1} {\rm Tr}
\left(\frac{\partial}{\partial H_{tr}}\right)^m H^l
e^{{\rm Tr} U(H)} = \nn \\
= (\pm)^{m+1}
{\rm Tr}\left(\frac{\partial}{\partial H_{tr}}\right)^m
\frac{1}{\Lambda\otimes I - I\otimes H} e^{{\rm Tr} U(H)}.
\label{tiwkoder1}
\ee
Now expression for $ T$'s in terms of $\Lambda$ should be used. Then
\be
e^{{\rm Tr} U(H)} =  Det^{\pm 1} (\Lambda\otimes I - I\otimes H)
\nn
\ee
and substituting this into (\ref{tiwkoder1}) we see that (\ref{tiwko}) is
equivalent to
\be
\left( \left(\frac{\partial}{\partial \Lambda_{tr}}\right)^{m+1} -
(\pm)^{m+1} I\cdot{\rm Tr} \left(\frac{\partial}{\partial H_{tr}}\right)^m
\cdot \frac{1}{\Lambda\otimes I - I\otimes H}
\right)\cdot \nn \\
\cdot  Det^{\pm 1} (\Lambda\otimes I - I\otimes H) = 0
\label{tiwkoder2}
\nn
\ee
Here "${\rm Tr}$" stands for the trace in the $H$-space only, while
$Det = {\rm Det}\otimes {\rm det}$ - for determinant in both $H$ and
$\Lambda$ spaces.
After one $\Lambda$-derivative is taken explicitly, we get:
\be
\left(I\otimes{\rm Tr}\right) \left( \left(\frac{\partial}{\partial
\Lambda_{tr}}\right)^m\otimes I -  I\otimes \left(\pm\frac{\partial}{\partial
H_{tr}}\right)^m
\right)
\cdot \nn \\ \cdot
\frac{Det^{\pm 1} (\Lambda\otimes I -  I\otimes H)}
{\Lambda\otimes I - I\otimes H} = 0.
\label{tiwkoder3}
\ee
This is already a matrix identity,
valid for any $\Lambda$ and $H$ of the sizes
$n\times n$ and $N\times N$ respectively. For example, if $m=0$ ($\tilde
W^{(1)}$-case), it is obviously satisfied. If both $n=N=1$, it is also
trivially true, though for different reasons for different choice of signs:
for the upper signs, the ratio at the l.h.s. is just unity and all derivatives
vansih; for the lower signs we have:
\be
\left(\frac{\partial}{\partial \lambda }\right)^m  -
\left(-\frac{\partial}{\partial h}\right)^m =
\left(\sum_{\stackrel{a+b=m-1}{a,b\geq 0}}
\left(\frac{\partial}{\partial \lambda }\right)^a
\left(-\frac{\partial}{\partial h}\right)^b \right)
\left(\frac{\partial}{\partial \lambda }  + \frac{\partial}{\partial h}\right),
\nn
\ee
and this obviously vanishes since $(\frac{\partial}{\partial \lambda } +
\frac{\partial}{\partial h})f(\lambda -h) \equiv 0$ for any $f(x)$.

If $m>0$ and $\Lambda,\ H$ are indeed {\it matrices},
direct evaluation becomes much more
sophisticated. We present the first two nontrivial examples: $m=1$ and $m=2$.
The following relartions will be usefull. Let
$Q \equiv  \frac{1}{\Lambda\otimes I - I\otimes H}$. Then
\be
Det^{\pm 1}Q \frac{\partial}{\partial \Lambda_{tr}}  Det^{\mp 1} Q =
\pm\left[(I\otimes {\rm Tr})Q\right]; \nn \\
Det^{\pm 1}Q \frac{\partial}{\partial H_{tr}}  Det^{\mp 1} Q =
\mp\left[({\rm tr}\otimes I)Q\right]; \nn \\
\left(\frac{\partial}{\partial \Lambda_{tr}}\otimes I\right) Q =
-\left[({\rm tr}\otimes I)Q\right]Q;
\nn \\
\left(I\otimes\frac{\partial}{\partial H_{tr}}\right) Q =
\left[(I\otimes {\rm Tr})Q\right]Q.
\label{tiwkoder5}
\ee

This is already enough for the proof in the case of $m=1$. Indeed:
\be
Det^{\pm 1}Q \left(\frac{\partial}{\partial \Lambda_{tr}}\otimes I \mp
    I\otimes \frac{\partial}{\partial H_{tr}}\right)Q Det^{\mp 1} Q = \nn \\
= \{-\left[({\rm tr}\otimes I)Q\right]Q
  \pm \left[(I\otimes {\rm Tr})Q\right]Q \}\mp \nn \\
\mp\{ \left[(I\otimes {\rm Tr})Q\right]Q \mp
 \left[({\rm tr}\otimes I)Q\right]Q \} = 0. \nn
\ee
The first two terms at the r.h.s. come arise from $\Lambda$-, while the
last two - from $H$-derivatives.

In the case of $m=2$ one should take derivatives once again. This is a little
more tricky, and the same compact notation are not sufficient. In addition to
(\ref{tiwkoder5}) we now need:
\be
\left(\frac{\partial}{\partial \Lambda_{tr}}\otimes I\right)
\left[({\rm tr}\otimes I)Q\right]Q =
- \left[({\rm tr}\otimes I)Q\right]^2 Q - {\cal B}.
\label{tiwkoder7}
\ee
Here
\be
\left[({\rm tr}\otimes I)Q\right]^2 =
\left[({\rm tr}\otimes I)\left[({\rm tr}\otimes I)Q\right]Q\right],
\label{tiwkoder8}
\ee
while in order to write ${\cal B}$ explicitly we need to restore matrix
indices (Greek for the $\Lambda$-sector and Latin - for the $H$ one). The
$(\alpha i,\gamma k)$-component of (\ref{tiwkoder7}) looks like:
\be
\left(\frac{\partial}{\partial \Lambda_{\beta\alpha}}\delta^{im}\right)
Q_{\delta\delta}^{mj}Q_{\beta\gamma}^{jk} =
- Q_{\delta\delta}^{ij}Q_{\beta\beta}^{jl}Q_{\alpha\gamma}^{lk} -
  Q_{\delta\beta}^{il}Q_{\alpha\delta}^{lj}Q_{\beta\gamma}^{jk}
\ee
and appearence of the second term at the r.h.s. implies, that
${\cal B}_{\alpha\gamma}^{ik} =
Q_{\delta\beta}^{il}Q_{\alpha\delta}^{lj}Q_{\beta\gamma}^{jk}$.
Further,
\be
&\left(\frac{\partial}{\partial \Lambda_{tr}}\otimes I\right)
\left[(I\otimes {\rm Tr})Q\right]Q =  \nn \\
&-\left[(I\otimes {\rm Tr})\left[({\rm tr}\otimes I)Q\right]Q\right]Q
-\left[(I\otimes {\rm Tr})\left[(I\otimes {\rm Tr})Q\right]Q\right]Q; \nn \\
&\left( I\otimes \frac{\partial}{\partial H_{tr}}\right)
\left[({\rm tr}\otimes I)Q\right]Q = \nn \\
&+ \left[({\rm tr}\otimes I)\left[(I\otimes {\rm Tr})Q\right]Q\right]Q +
  \left[(I\otimes {\rm Tr})\left[({\rm tr}\otimes I)Q\right]Q\right]Q; \nn \\
&\left( I\otimes \frac{\partial}{\partial H_{tr}}\right)
\left[(I\otimes {\rm Tr})Q\right]Q =
+ \left[(I\otimes {\rm Tr})\left[(I\otimes {\rm Tr})Q\right]Q\right]Q
+{\cal B}.
\ee
It is important that ${\cal B}$ that appears in the last relation
in the form of
${\cal B}_{\alpha\gamma}^{ik} =
Q_{\alpha\delta}^{lj}Q_{\delta\beta}^{il}Q_{\beta\gamma}^{jk}$
is exactly the same ${\cal B}$ as in eq.(\ref{tiwkoder7}).

Now we can prove (\ref{tiwkoder3}) for $m=2$:
\be
&Det^{\pm 1}Q \left(\left(\frac{\partial}{\partial
\Lambda_{tr}}\right)^2\otimes I -
I\otimes \left(\frac{\partial}{\partial H_{tr}}\right)^2\right)
Q Det^{\mp 1} Q =  \nn \\
&= \left\{
\pm\left[(I\otimes{\rm Tr})Q\right]\left(
- \left[({\rm tr}\otimes I)Q\right]Q \pm \left[(I\otimes{\rm Tr})Q\right]Q
\right) - \right. \nn \\ &\left. - \left(
- \left[({\rm tr}\otimes I)\left[({\rm tr}\otimes I)Q\right]Q\right]Q - {\cal
B}
\right) \pm \right. \nn \\ &\left. \pm \left(
- \left[(I\otimes{\rm Tr})\left[({\rm tr}\otimes I)Q\right]Q\right]Q -
  \left[({\rm tr}\otimes I)\left[(I\otimes{\rm Tr})Q\right]Q\right]Q
\right)\right\} - \nn \\
&- \left\{
\mp\left[({\rm tr}\otimes I)Q\right]\left(
\left[(I\otimes{\rm Tr})Q\right]Q \mp \left[({\rm tr}\otimes I)Q\right]Q
\right) + \right. \nn \\ &\left. + \left(
\left[(I\otimes{\rm Tr})\left[(I\otimes{\rm Tr})Q\right]Q\right]Q + {\cal B}
\right) \mp \right. \nn \\ &\left. \mp \left(
\left[({\rm tr}\otimes I)\left[(I\otimes{\rm Tr})Q\right]Q\right]Q +
\left[(I\otimes{\rm Tr})\left[({\rm tr}\otimes I)Q\right]Q\right]Q
\right)\right\}
\ee
where the terms 1,2,3,4,5,6 in the first braces cancel the terms 1,3,2,4,6,5
in the second braces and identity (\ref{tiwkoder8}) and its counterpart with
$({\rm tr}\otimes I) \rightarrow  (I\otimes {\rm Tr})$ is used.

Explicit proof of eq.(\ref{tiwkoder3}) for generic $m$ is unknown.

\






\bigskip

\section{Eigenvalue models}

\setcounter{equation}{0}

\subsection{What are eigenvalue models}

\bigskip

Given the present state of knowledge we need to consider in most cases
only the narrow class of the "eigenvalue" models.
These models have the property of being associated with conventional
integrable hierarchies (of (multicomponent) KP and Toda type), where
integrable flows just commute (instead of forming less trivial closed
algebras), and thus with the level-1 Kac-Moody algebras (by artificial tricks,
familiar from the bosonization formalism in conformal field theory
\cite{Turb} these can be
sometimes generalized to particular other levels like $k=2$). This means that
the models are essentially associated with abelian Cartan subalgebras rather
than with full matrix algebras.
\footnote{Groups, arising in the theory of matrix models and integrable
hierarchies are not just those of matrices,
appearing in the integral representations: the latter ones are in the best
case related to the zero-modes of the former. Moreover, even this relation is
not usually simple to reveal. This remark is important to avoid confusion in
the next paragraphs.}
In CFT-formulation (see below) this means that the eigenvalue models can be
represented in terms of the free fields, which bosonize the Cartan subalgebra
of the whole group in the WZNW model (the remaining $(\beta,\gamma)$-fields
\cite{GMMOS} being (almost) neglected - their remnants are observed in the
form of "cocycle" factors in the Frenkel-Kac formulas \cite{FK}, see
\cite{Turb}).
In the matrix-integral representations the integrals for the eigenvalue models
are in fact reduced to those over diagonal matrices (consisting of eigenvalues
of original matrices, thus the name "eigenvalue models").

Most important, from the physical point of view eigenvalue models describe
only $topological$ (discrete) degrees of freedom, but not any propagating
$particles$.
\footnote{ Particles are always related to the "angular (unitary-) matrix"
integrals (as everybody knows from example of the Wilson lattice QCD) which
are highly less trivial to deal with, though these are also integrable in some
broader sense of the word - within the (yet non-existing) generalization of
integrable hierarchies from the fields in the Cartan subalgebra to the entire
WZNW model.}
This can be understood if one notes that matrix models usually possess gauge
symmetry, associated with the unitary rotation of matrices,
$M_{\alpha} \longrightarrow U_{\alpha}^{\dagger}M_{\alpha}U_{\alpha}$,
i.e. matrix models are usually $gauge$ theories. In the case of eigenvalue
models this symmetry is realized without "gauge fields" $V_{\alpha\beta}$,
which would depend on pairs of indices $\alpha$, $\beta$ and transform like
$V_{\alpha\beta} \longrightarrow
U_{\alpha}^{\dagger}V_{\alpha\beta}U_{\beta}$.
In other words, eigenvalue models are gauge theories without gauge fields,
i.e. are pure topological. Thus it is not a surprise that they usually live in
the space time of dimension $d<2$,
\footnote{Let us remind that in the Polyakov formulation which is the least
counterintuitive for interpretation of  what happens in the space-time
(target space), string models usually involve Liouville field, identified as a
time-variable in the target-space formalism. (Note that for this reason there
is usually (at least one) $time$ in the string theory, while space can be of
any dimension (at least between 0 and 25), not obligatory integer.) Because of
this extra Liouville field the space-time dimension $d$ usually differs by $1$
from the central charge of the CFT model, which is coupled to $2d$-gravity to
from a string model: $d = c+1$ and $d<2$ is the same as $c<1$. }
since for $d>2$ there $should\ be$ particles, associated with the gauge
fields. At the "boundary" lies the model of "$d=2\ (c=1)$ string", which has
one particle-like degree of freedom (dilaton, which becomes tachyon in the
$d>2$ models). This very interesting model
is much worse understood than the $d<2$
models, at least its properties are already somewhat different from other
eigenvalue models (especially in the most interesting "compactified" case),
and it will not be discussed in these notes.
Later we"ll return to the subject of non-eigenvalue ($d>2$) theories, though
not
too much is yet known about them, but now we are going to concentrate on the
eigenvalue models.

\subsection{1-matrix model}

Hermitean matrix integrals are usually transformed to the eigenvalue form by
separation of angular and eigenvalue variables. As usually, the simplest is
the case of the 1-matrix model
\be
Z_N\{t\} \equiv c_N\int_{N\times N} dH e^{\sum_{k=0}^{\infty} t_k {\rm Tr}
H^k},
\label{1mamo"}
\ee
where this separations does not involve any information about unitary-matrix
integrals. Take
\be
H = U^{\dagger}DU,
\label{diag}
\ee
where $U$ is a unitary matrix and
diagonal matrix $D = {\rm diag}(h_1...h_N)$ has eigenvalues of $H$ as its
entries. Then integration measure
\be
dH = \prod_{i,j=1}^N dH_{ij} = \frac{[dU]}{[dU_{Cartan}]} \prod_{i-1}^N dh_i
\Delta^2(h),
\label{Dyson}
\ee
where "Van-der-Monde determinant" $\Delta (h) \equiv det_{(ij)} h_i^{j-1} =
\prod_{i>j}^N (h_i - h_j)$
and $[dU]$ is Haar measure of integration over unitary matrices.

The way to derive eq.(\ref{Dyson}) is to consider the norm of infinitesimal
variation
\be
\mid\mid \delta H\mid\mid^2 &\equiv \sum_{i,j=1}^N \mid \delta H_{ij}\mid^2 =
 \sum_{i,j=1}^N \delta H_{ij}\delta H_{ji} = {\rm Tr} (\delta H)^2 = \nn \\
&= {\rm Tr}\left(-U^{\dagger}\delta UU^{\dagger}DU + U^{\dagger}D\delta U
+ U^{\dagger}\delta DU \right)^2 = \nn \\
&= {\rm Tr} (\delta D)^2 + 2i{\rm Tr}{\delta u}[\delta D,D] +
2{\rm Tr}\left(-{\delta u} D {\delta u} D + (\delta u)^2D^2 \right), \nn
\ee
where $\delta u \equiv \frac{1}{i}{\delta U}U^{\dagger} = \delta u^{\dagger}$
and $\delta D = {\rm diag}(\delta h_1\ldots \delta h_N)$.
The second term at the r.h.s. vanishes because both $D$ and $\delta D$ are
diagonal and commute. Therefore
\be
\mid\mid \delta H\mid\mid^2 = \sum_{i=1}^N (\delta h_i)^2 +
\sum_{i,j=1}^N (\delta u)_{ij}(\delta u)_{ji}(h_i-h_j)^2. \nn
\ee
Now it remains to recall the basic relation between the infinitesimal norm and
the measure:
${\rm if}\ \ \mid\mid \delta l \mid\mid^2 = G_{ab}\delta l^a\delta l^b \ \
{\rm then}\ \ [dl] = \sqrt{{\rm det}_{ab}G_{ab}} \prod_a dl^a, $
to obtain eq.(\ref{Dyson}) with Haar measure
$[dU] = \prod_{ij}^N du_{ij}$ being associated with the infinitesimal norm
\be
\mid\mid \delta u\mid\mid^2 = {\rm Tr}(\delta u)^2 = \sum_{i,j=1}^N \delta
u_{ij}\delta u_{ji} = \sum_{i,j=1}^N \mid \delta u_{ij}\mid^2 \nn
\ee
and $[dU_{Cartan}] \equiv \prod_{i=1}^N du_{ii}$.

Coming back to the 1-matrix model it remains  to note that the "action"
${\rm Tr} U(H) \equiv \sum_{k=0}^{\infty} t_k {\rm Tr}H^k$ with $H$ substituted
in the form (\ref{diag}) is independent of $U$:
\be
{\rm Tr} U(H) = \sum_{i=1}^N U(h_i). \nn
\ee
Thus
\be
Z_N\{t\} = \frac{1}{N!} \prod_{i=1}^N \int dh_i e^{U(h_i)} \prod_{i>j}^N
(h_i-h_j)^2 = \nn \\
= \frac{1}{N!} \prod_{i=1}^N \int dh_i e^{U(h_i)} \Delta^2(h),
\label{1mamoev}
\ee
provided $c_N$ is chosen to be
\be
c_N^{-1} = N!\frac{{\rm Vol}_{U(N)}}{({\rm Vol}_{U(1)})^N},
\label{cN}
\ee
where the volume of unitary group in Haar measure is equal to
\be
{\rm Vol}_{U(N)} = \frac{(2\pi)^{N(N+1)/2}}{\prod_{k=1}^N k!}.
\label{volun}
\ee
A simple way to derive eq.(\ref{volun}) will be described at the end of this
section, as an example of application of orthogonal polinomials technique.

\subsection{Itzykson-Zuber and Kontsevich integrals}

Let us proceed now to Kontsevich integral,
\be
{\cal F}_{V,n}\{ L\} =
\int_{n\times n} dX e^{- {\rm tr} V(X) + {\rm tr} L X}.
\label{KI'}
\ee
We shall see shortly that it in fact depends only on the eigenvalues of the
matrix $ L$ (this fact was already used in the previous section),
however, this time somewhat more sophisticated unitary matrix integrals will
be involved.

Substitute $X = U_X^{\dagger}D_XU_X;  \ \  L = U_{ L}^{\dagger}
D_{ L}U_{ L}$ in (\ref{KI'}) and denote
$U \equiv U_XU_{ L}^{\dagger}$. Then
\be
&{\cal F}_{V,n}\{ L\} = \nn \\
&= \prod_{i=1}^n \int dx_i e^{-V(x_i)}
\Delta^2(x)
\int_{n\times n} \frac{[dU]}{[dU_{Cartan}]}
\exp\left({\sum_{\gamma,\delta =1}^n x_\gamma l_\delta\mid
U_{\gamma\delta}\mid^2} \right).
\label{KI'2}
\ee
In order to proceed further we need to evaluate the integral over unitary
matrices, which appeared at the r.h.s.

This integral can actually be represented in two different ways:
\be
I_n\{X, L\} \equiv  \int_{n\times n} \frac{[dU]}{[dU_{Cartan}]}
e^{{\rm tr} XU L U^{\dagger}} =
\label{IZa} \\
=  \int_{n\times n} \frac{[dU]}{[dU_{Cartan}]}
e^{\sum_{\gamma,\delta =1}^n x_\gamma l_\delta\mid U_{\gamma\delta}\mid^2}
\label{IZb}
\ee
(the U's in the two integrals are related by the transformation
$U \longrightarrow U_XUU_{ L}^{\dagger}$ and Haar measure is both left
and right invariant).
Formula (\ref{IZa}) implies that $I_n\{X, L\}$ satisfies a set of simple
equations \cite{Migeq}:
\be
\left( {\rm tr} \left(\frac{\partial}{\partial X_{tr}}\right)^k -
       {\rm tr}  L^k \right) I_n\{X, L\} = 0, \ \ k\geq 0, \nn\\
\left( {\rm tr} \left(\frac{\partial}{\partial  L_{tr}}\right)^k -
       {\rm tr} X^k \right) I_n\{X, L\} = 0, \ \ k\geq 0,
\label{Migeq}
\ee
which by themselves are not very restrictive. However, another formula,
(\ref{IZb}), implies that $I_n\{X, L\}$ in fact depends only on the
eigenvalues of $X$ and $ L$, and for $such$ $I_n\{X, L\} =
\hat I\{x_\gamma, l_\delta\}$ eqs.(\ref{Migeq}) become very restrictive
\footnote{When acting on $\hat I$, which depends only on eigenvalues,  matrix
derivatives turn into:
\be
{\rm tr} \frac{\partial}{\partial X_{tr}} \hat I &=
   \sum_\gamma \frac{\partial}{\partial x_\gamma} \hat I; \nn \\
{\rm tr} \frac{\partial^2}{\partial X_{tr}^2} \hat I &=
   \sum_\gamma \frac{\partial^2}{\partial x_\gamma^2} \hat I  +
\sum_{\gamma\neq\delta} \frac{1}{x_\gamma-x_\delta}
\left(\frac{\partial}{\partial x_\gamma} -
\frac{\partial}{\partial x_\delta}\right) \hat I;
\nn
\ee
etc.
} and allow to
determine $\hat I\{x_\gamma, l_\delta\}$
unambigously (at least if $\hat I\{x_\gamma, l_\delta\}$
is expandable in a formal power seria in $x_\gamma$ and $ l_\delta$).
The final solution is
\be
I_n\{X, L\} = \frac{(2\pi)^{\frac{n(n-1)}{2}}}{n!}
\frac{{\rm det}_{\gamma\delta}
   e^{x_\gamma l_\delta}}{\Delta(x)\Delta( l)}.
\label{IZ}
\ee
Normalization constant can be defined by taking $ L = 0$, when
\be
I_n\{X,L=0\} =
\frac{{\rm Vol}_{U(n)}}{({\rm Vol}_{U(1)})^n} =
\frac{(2\pi)^{\frac{n(n-1)}{2}}}{\prod_{k=1}^n k!},
\nn
\ee
and using the fact that
\be
\left.\frac{{\rm det}_{\gamma\delta} f_\gamma(l_\delta)}{\Delta(l)}\right|
_{\{l_\delta = 0\}} =
\left(\prod_{k=0}^{n-1}\frac{1}{k!} \right) {\rm det}_{\gamma\delta}
\partial^{\delta-1}f_\gamma(0).
\nn
\ee

Eq.(\ref{IZ}) is usually refered to as the Itzykson-Zuber formula
\cite{IZ}. In mathematical literature it was earlier derived by
Kharish-Chandra \cite{KhCh}, and in fact the integral (\ref{IZa} is the basic
example of the coadjoint orbit integrals \cite{STS}-\cite{AS}, which can be
exactly evaluated with the help of the Duistermaat-Heckmann theorem \cite{DH},
\cite{NiDH},\cite{Wit2YM},\cite{NT}.
This calculation is the simplest example of the very
important technique of $exact$ evaluation of $non-Gaussian$ unitary-matrix
integrals, which is now doing its first steps (see \cite{KMSW1}-\cite{Shata})
and will be discussed at the end of these notes.

Now we turn back to the eigenvalue formulation of the GKM. Substitution of
(\ref{IZ}) into (\ref{KI'2}) gives:
\be
{\cal F}_{V,n}\{ L\} = \frac{(2\pi)^{\frac{n(n-1)}{2}}}{\Delta( l)}
\prod_{\delta=1}^n \int dx_\delta e^{-V(x_\delta)} \Delta(x)
\frac{1}{n!} {\rm det}_{\gamma\delta} e^{x_\gamma l_\delta} = \nn \\
= \frac{(2\pi)^{\frac{n(n-1)}{2}}}{\Delta( l)}
\prod_{\delta =1}^n \int dx_\delta e^{-V(x_\delta) + x_\delta l_\delta}
\Delta(x),
\label{KIev}
\ee
where we used antisymmetry of $\Delta(x)$ under permutations of $x_\gamma$'s in
order to change $\frac{1}{n!} {\rm det}_{\gamma\delta}
e^{x_\gamma l_\delta}$ for
$e^{\sum_\delta x_\delta l_\delta}$
under the sign of the $x_\delta$ integration.

We can now use the fact that $\Delta(x) = {\rm
det}_{\gamma\delta}x_\delta^{\gamma -1}$ in order to
rewrite the r.h.s. of (\ref{KIev}):
\be
{\cal F}_{V,n}\{ L\} = (2\pi)^{\frac{n(n-1)}{2}}\frac{{\rm det}_{\gamma\delta}
\hat\varphi_\gamma( l_\delta)}{\Delta( l)},
\label{KIev'}
\ee
where
\be
\hat\varphi_\gamma( l) \equiv \int dx x^{\gamma -1}e^{-V(x)+ l x},
\ \ \ \gamma\geq 1.
\label{hatvarphi}
\ee
These functions $\hat\varphi( l)$ satisfy a simple recurrent relation:
\be
\hat\varphi_\gamma = \frac{\partial\hat\varphi_{\gamma -1}}{\partial  l} =
\left(\frac{\partial}{\partial  l}\right)^{\gamma -1}\hat\Phi
\label{preKasch}
\ee
with
\be
\hat\Phi( l)\equiv \hat\varphi_1( l) =
\int dx e^{-V(x)+ l x}.
\label{preKasch'}
\ee

Note also that if the "zero-time" $N$ is introduced (see subsection 2.6 above
and \cite{Toda}), then
\be
{\cal F}_{V,n}\{N\mid L\} \equiv {\cal F}_{V(X)-N\log X,n}\{ L\} =
(2\pi)^{\frac{n(n-1)}{2}}\frac{{\rm det}_{\gamma\delta}
\hat\varphi_{\gamma +N}( l_\delta)}{\Delta( l)}
\label{KIev'N}
\ee
with just the $same$ $\hat\varphi_\gamma( l)$ and
$\gamma,\delta = 1\ldots n$.
If we divide by the  quasiclassical factor
${\cal C}_V\{\Lambda \}({\rm det} \Lambda )^N,\
 L = V'(\Lambda )$, in order to transform Kontsevich integral into Kontsevich
model (see section 2.5), we get:
\be
{\cal Z}_V\{N,T\} = \frac{1}{({\rm det} \Lambda )^N}\cdot\frac{{\rm
det}_{\gamma\delta} \varphi_{\gamma +N}(\lambda _\delta)}{\Delta(\lambda )}.
\label{KIev"N}
\ee
The role of ${\cal C}_V\{\Lambda \}$ is to convert $\hat\varphi( l)$ into
the properly normalized expansions in the negative iteger powers of $\lambda $:
\be
\varphi_\gamma(\lambda )  =
\frac{e^{-\lambda V'(\lambda )+V(\lambda )}\sqrt{V''(\lambda )}}{\sqrt{2\pi}}
\hat\varphi_\gamma(V'(\lambda )) =
\lambda ^{\gamma -1}(1 + {\cal O}(\lambda ^{-1})),
\label{varphiko}
\ee
and to change $\Delta( l) = \Delta(V'(\lambda ))$ in the denominator of
(\ref{KIev'N}) for  $\Delta(\lambda )$ in (\ref{KIev"N}).
Instead of the simple recurrent relations (\ref{preKasch}) for $\hat\varphi$
the normalized functions $\varphi$ satisfy:
\be
\varphi_\gamma(\lambda ) = {\cal A}\varphi_{\gamma -1}(\lambda ) =
{\cal A}^{\gamma -1}\Phi(\lambda ),
\label{Kasch}
\ee
where $\Phi(\lambda ) = \varphi_1(\lambda )$ and operator
\be
{\cal A} = \frac{1}{V''(\lambda )}\cdot\frac{\partial}{\partial \lambda }
-\frac{1}{2}\frac{V'''(\lambda )}{(V''(\lambda ))^2} + \lambda
\label{Kaschop}
\ee
now depends on the potential $V(x)$.

\subsection{Conventional Multimatrix models}

The multimatrix integrals of the form
\be
Z_N\{t^{(\alpha)}\} \equiv \nn \\
\equiv c_N^{p-1}\int_{N\times N} dH^{(1)}...dH^{(p-1)}
\prod_{\alpha = 1}^{p-1} e^{\sum_{k=0}^{\infty}t_k^{(\alpha)}{\rm Tr}
H_{(\alpha)}^k} \prod_{\alpha = 1}^{p-2} e^{{\rm
Tr}H^{(\alpha)}H^{(\alpha+1)}}
\label{mumamo'}
\ee
can be rewritten in the eigenvalue form using the same Itzykson-Zuber formula
(\ref{IZ}). Indeed, substituting
$H^{(\alpha)} = \left.U^{(\alpha)}\right.^{\dagger}D^{(\alpha)} U^{\alpha)}$
and then defyining $U^{(\alpha)}\left.U^{(\alpha+1)}\right.^{\dagger}
\equiv \tilde U^{(\alpha)}$, we
obtain:
\be
\new
\begin{array}{c}
Z_N\{t^{(\alpha)}\} =\\
= \frac{1}{N!} \prod_{\alpha = 1}^{p-1}
\prod_{i=1}^N \int dh_i^{(\alpha)} e^{-V(h_i^{(\alpha)})}
\Delta^2(h^{(\alpha)})
\prod_{\alpha = 1}^{p-2} I_N\{H^{(\alpha)}, H^{(\alpha+1)}\} =  \\
= \frac{1}{N!} \prod_{\alpha = 1}^{p-1}
\prod_{i=1}^N \int dh_i^{(\alpha)} e^{-V(h_i^{(\alpha)})}
\prod_{\alpha = 1}^{p-2} e^{h_i^{(\alpha)}h_i^{(\alpha+1)}}
\Delta(h^{(1)})\Delta(h^{(2)}),
\end{array}
\label{mumamoev}
\ee
where the same trick is done with the substitution of
$\frac{1}{N!} {\rm det}_{ij} e^{h_i^{(\alpha)}h_j^{(\alpha+1)}}$ for
$e^{\sum_{i=1}^N h_i^{(\alpha)}h_i^{(\alpha+1)}}$ under the sign of the
$h_i^{(\alpha)}$-integration  (step by step: first - for $\alpha = 1$,
then for $\alpha = 2$ and so on). Note that all the Van-der-Monde determinants
disappeared from the final formula at the r.h.s. of eq.(\ref{mumamoev}), except
for those at the ends of the matrix chain (at $\alpha = 1$ and $\alpha =
p-1$).

If the chain was closed rather than open, i.e. there was an additional factor
of $e^{{\rm Tr}H^{(p-1)}H^{(1)}}$ under the integral in (\ref{mumamo'}),
then the trick with separation of all angular-variable (unitary-matrix)
integrations would not work so simply: in addition to the Itzykson-Zuber
integral the much more involved quantities would be required, like
\be
\new
\begin{array}{c}
I_n\{X_1,X_2; L\} \equiv  \\
\equiv c_n \int_{n\times n}
\frac{[dU_1]}{[dU_{1,Cartan}]}  \frac{[dU_2]}{[dU_{2,Cartan}]}
\\ \cdot
\exp\left({{\rm tr} X_1U_1 L U_1^{\dagger} + {\rm tr} X_2U_2 L U_2^{\dagger}
+ {\rm tr}X_1(U_1U_2^{\dagger})X_2(U_2U_1^{\dagger})}\right)
\end{array}
\ee
This (so far unresolved) closed chain model (lattice Potts model) is an
example of non-eigenvalue models, in the $p=\infty$ case it turns into
"compactified" $c=1$ model. This theory is $more$ complicated then the
so far simplest class of non-eigenvalue models of "induces Yang-Mills theory",
known as Kazakov-Migdal models.

\subsection{Determinant formulas for eigenvalue models}

We are now prepared to make the crucial step towards understanding of
mathematical structure behind eigenvalue models, which distinguishes their
partition functions in the entire variety of arbitrary $N$-fold integrals.
This structure expresses itself in the form of determinantal formulas, which
we are now going to discuss. In the next section 4 these formulas will be
identified as examples of $\tau$-functions of KP and Toda hierarchies.

Looking at the relevant
integrals (\ref{1mamoev}), (\ref{mumamoev}) one can notice that
integrals over different eigenvalues with non-trivial measures which depend
on the shape of potentials $U$ or $V$, are almost separated, the only
"interaction" between different eigenvalues being defined by $universal$
(potential-independent) quantities, made from the Van-der-Monde determinants.
This feature is intimately related both to its origin (decoupling of angular
variables in original matrix integral) and to its most important implication
(integrability). The main property of the Van-der-Monde determinant is that it
is at the same time a $Pfaffian$ (and it is in this quality that it arises
from matrix integrals) and a $determinant$ (and this is the feature that
implies integrability):
\be
\prod_{i>j}(h_i-h_j) = \Delta(h) = {\rm det}_{ij}h_i^{j-1}.
\label{vdmd}
\ee
We already used this property above, when going from eq.(\ref{KIev}) to
eq.(\ref{KIev'}), which as we shall see later is the crucial step in the proof
of integrability of Kontsevich model. In that case determinantal formula
(\ref{KIev'}) for partition function was trivial to derive, because the
integrand was $linear$ in Van-der-Monde determinants. Now we turn to slightly
more complicated situations, involving products of Van-der-Monde determinants.

Consider an eigenvalue model of the form:
\be
Z_N = \frac{1}{N!} \prod_{k=1}^N \int d\mu_{h_k,\bar h_k} \Delta(h)\Delta(\bar
h),
\label{meas1}
\ee
to be refered to as "scalar-product" model.
All conventional multimatrix models (\ref{mumamo'}) belong to this class.
In the case of the 1-matrix model (\ref{1mamoev})
\be
d\mu_{h,\bar h} = dhd\bar h e^{U(h)}\delta(h-\bar h),
\ee
while for conventional multimatrix models
(\ref{mumamoev})
\be
d\mu_{h^{(1)}, h^{(p-1)}} = dh^{(1)}dh^{(p-1)}\prod_{\alpha=2}^{p-2}
\int dh^{(\alpha)} \prod_{\alpha =1}^{p-1} e^{U_{\alpha}(h^{(\alpha)})}
\prod_{\alpha = 1}^{p-2} e^{h^{(\alpha)}h^{(\alpha+1)}}
\label{measmumamo}
\ee
If $d\mu_{h,\bar h} =\delta(h-\bar h)d\bar h d\mu_h$ we call this measure
local. The  main feature of local measure is that operator of multiplication
by $H$ (or any function of $h$) is Hermitean. Thus measure is local in the
1-matrix model, but is non-local for all $p-1>1$.
In the latter case the measure is defined to depend only on $h = h^{(1)}$ and
$\bar h = h^{(p-1)}$, all other
$h^{(\alpha)},\ \alpha = 2\ldots p-2$ being integrated out, what makes
the "interaction" between $h$ and $\bar h$ more complicated than just
$\delta(h-\bar h)$ in the one-matrix ($p=2$) and $e^{h\bar h}$ in the
two-matrix ($p=3$) cases. In no sense the set of particular formulas
(\ref{measmumamo}) for $p>3$ is distinguished among other scalar-product
models,
and from now on we shall not consider conventional multimatrix models with
$p-1 > 2$ as a separate class of theories.

Eqs.(\ref{meas1}) and (\ref{vdmd}) imply together that
\be
Z_N = \frac{1}{N!} \prod_{k=1}^N \int d\mu_{h_k,\bar h_k}
 {\rm Det}_{ik} h_k^{i-1} {\rm Det}_{jk} \bar h_k^{j-1} =   \nn \\
= {\rm Det}_{ij} \int d\mu_{h,\bar h} h^{i-1}\bar h^{j-1} =
      {\rm Det}_{ij} \langle h^{i-1} \mid  \bar h^{j-1} \rangle,
\label{dimamodet}
\ee
where an obvious notation is introduced for the scalar product
\be
\langle f(h)\mid g(\bar h)\rangle \equiv \int d\mu_{h,\bar h} f(h)g(\bar h).
\nn
\ee

We can now be a little more specific and introduce time-variables $t_k$ and
$\bar t_k$ so, that
\be
d\mu_{h,\bar h} = e^{U(h)+\bar U(\bar h)}{d\hat\mu}_{h,\bar h}, \nn \\
U(h) = \sum_{k=-\infty}^{\infty} t_kh^k, \ \
\bar U(\bar h) = \sum_{k=-\infty}^{\infty}
\bar t_k\bar h^k,
\label{tbartmeas}
\ee
and $\hat{d\mu}_{h,\bar h}$ is already independent of $h$ and $\bar h$. If we
now denote ${\cal H}^f(t,\bar t) \equiv \langle 1\mid 1\rangle$, then
\be
{\cal H}^f_{ij} \equiv \langle h^{i}\mid \bar h^{j} \rangle
= \frac{\partial^2}{\partial t_{i}\partial \bar t_{j}}{\cal H}^f(t,\bar t) =
\nn \\ \stackrel{{\rm if}\ i,j\geq 0}{=}
\left(\frac{\partial}{\partial t_1}\right)^{i}
\left(\frac{\partial}{\partial\bar t_1}\right)^{j}{\cal H}^f(t,\bar t),
\label{todamatr}
\ee
and
\be
Z_N = {\rm Det}_N {\cal H}^f_{ij},
\label{todarepr}
\ee
where $\ {\rm Det}_N$ stands for determinant of the $N\times N$ matrix ${\cal
H}_{i-1,j-1}$ (which is defined itself for {\it any} integers $i,j$)
with $i,j = 0,\ldots, N-1$.
Characteristic property of ${\cal H}_{ij}^f$ is its
peculiar time-dependence:
\be
\frac{{\cal H}^f_{ij}}{\partial t_k} = {\cal H}^f_{i+k,j}; \ \ \
\frac{{\cal H}^f_{ij}}{\partial \bar t_k} = {\cal H}^f_{i,j+k}.
\ee

Eq.(\ref{todarepr}) provides the determinantal formula for all scalar-product
models. The case of the local measure - for  1-matrix
model - is a little special. In this case $U(h)$ contains the whole
information about the measure: $d\mu_{h,\bar h} = \delta(h-\bar h)d\mu_h,\
d\mu_h = e^{U(h)}dh$, and there is no $\bar U(\bar h)$ (or $\bar t$ simply
coincide with $t$). Then (\ref{todarepr}) is still valid, but
\be
{\cal H}^f_{ij} = \left.\langle h^{i}\mid \bar h^{j} \rangle\right|
_{d\mu_{h,\bar h}}
= \left.\langle h^{i+j} \rangle\right|_{d\mu_h} =
\frac{\partial}{\partial t_{i+j}}{\cal H}^f(t) = \nn \\
\stackrel{{\rm if}\ i,j\geq 0}{=}
\left(\frac{\partial}{\partial t_1}\right)^{i+j}{\cal H}^f(t).
\label{todachainmatr}
\ee
The same formula (\ref{todachainmatr}) can be also derived as a limit of
eq.(\ref{KIev'}) for Kontsevich integral. Indeed,
\be
Z_N\{t\} = c_N\int_{N\times N}dH e^{{\rm Tr}U(H)} =
\lim_{ L \rightarrow 0} {\cal F}_{U,N}\{ L\} =
\nn \\
= \lim_{\{ l_j\}\rightarrow 0}
 \frac{{\rm Det}_{ij}\hat\varphi_i^{\{U\}}( l_j)}{\Delta( l)} =
{\rm Det}_{ij}\frac{\partial^{j-1}\hat\varphi_i^{\{U\}}( l_j)}
{\partial l^{j-1}}(0) = {\rm Det}_{ij} {\cal H}^f_{i-1,j-1},
\ee
where this time
\be
{\cal H}^f_{i-1,j-1} \stackrel{i,j>0}{=}
\frac{\partial^{j-1}\hat\varphi_i^{\{U\}}( l_j)}
{\partial l^{j-1}}( l =0) \stackrel{(\ref{KIev'})}{=}
\left(\frac{\partial}{\partial  l}\right)^{i+j-2} \hat\Phi^{\{U\}}
\mid_{ l = 0}.
\ee
Now we note, that the action of $\frac{\partial}{\partial  l}$ on
$\hat\Phi^{\{U\}}( l) = \int dx e^{U(x) +  l x}$ is equivalent to
that of $\left( \frac{\partial}{\partial t_1}\right)$, since this is no
longer a matrix integral, and thus
\be
{\cal H}^f_{ij} =  \left( \frac{\partial}{\partial t_1}\right)^{i+j}
\hat\Phi^{\{U\}}(0),
\label{todachainmatr1}
\ee
i.e. ${\cal H}^f(t) = \hat\Phi^{\{U\}}(0)$.

$Conformal$ multimatrix models were introduced in section 2.3 above just
as eigenvalue models. For the $A_{p-1}$ series partition functions are defined
to be
\be
\new
\begin{array}{c}
Z_{N_1\ldots N_{p-1}}^{A_{p-1}}\{t^{(1)} \ldots t^{(p-1)}\} =   \\
= \prod_{\alpha =1}^{p-1} c_{N_\alpha}\int_{N_\alpha\times N_\alpha}
dH^{(\alpha)} e^{{\rm Tr}U_\alpha(H^{(\alpha)})} \cdot \\ \cdot
\prod_{\alpha =1}^{p-2} {\rm Det}
\left( H^{(\alpha)}\otimes I - I\otimes H^{(\alpha +1)}\right) =  \\
= \prod_{\alpha = 1}^{p-1} \frac{1}{N_\alpha !}\prod_{i=1}^{N_\alpha}
\int dh_i^{(\alpha)} e^{U_\alpha (h_i^{(\alpha)})} \Delta^2(h^{(\alpha)})
\prod_{\alpha = 1}^{p-2} \prod_{i,k}
(h_i^{(\alpha)} - h_k^{(\alpha +1)}).
\label{comamoap}
\end{array}
\ee
This expression does not have a form of eq.(\ref{meas1}), thus conformal
matrix models for $p-1>1$ are not of the "scalar-product" type. We
shall sometimes call them $(p-1)$-component models, because they are related
to the multi-component integrable hierarchies.
The simplest way to proceed with their investigation is
to use the same trick with Kontsevich integral, which was just applied in the
1-matrix case.

Let us start from a very general $(p-1)$-component model:
\be
Z = \prod_{\alpha = 1}^{p-1} \int_{N_\alpha\times N_\alpha}
dH^{(\alpha)} e^{{\rm Tr} U_\alpha (H^{(\alpha)})}
K(H^{(1)}\ldots H^{(p-1)}).
\label{mucomammo}
\ee
It can be also represented in terms of Kontsevich integrals:
\be
Z = \left.K\left(\frac{\partial}{\partial L^{(1)}_{tr}},\ldots,
\frac{\partial}{\partial L^{(p-1)}_{tr}}\right)
\prod_{\alpha = 1}^{p-1} {\cal F}_{U_\alpha,
N_\alpha}\{ L^{(\alpha)}\}\right|_{ L^{(\alpha)} = 0}.
\label{mucomamo1}
\ee
This representation is not very usefull, since the limit $ L \rightarrow
0$ is not easy to take, unless $K$ is a polinomial in the eigenvalues of all
its arguments. However, this is exactly the case for our confromal models
(\ref{comamoap}). Indeed,
\be
K^{A_{p-1}} = \prod_{\alpha =1}^{p-2} {\rm Det}
\left(\frac{\partial}{\partial  L^{(\alpha)}_{tr}}\otimes I -
      I \otimes \frac{\partial}{\partial  L^{(\alpha+1)}_{tr}}\right).
\ee
Still this is not very convenient, because representation (\ref{KIev'}) for
${\cal F}$ contain $\Delta( L)$ in denominator, which are not very
pleasant to differentiate. Simplification can be achieved if instead we
rewrite the original expression at the r.h.s. of (\ref{comamoap}) as follows:
\be
&Z_{N_1\ldots N_{p-1}}^{A_{p-1}}\{t^{(1)} \ldots t^{(p-1)}\} =
\nn \\ &=
\Delta\left(\frac{\partial}{\partial l^{(1)}} \right)
\prod_{\alpha =1}^{p-2}
\Delta\left(\frac{\partial}{\partial l^{(\alpha)}},
\frac{\partial}{\partial l^{(\alpha+1)}} \right)
\Delta\left(\frac{\partial}{\partial l^{(p-1)}} \right)
\times \nn \\ &\times
\left.
\prod_{\alpha =1}^{p-1}\left( \frac{1}{N_\alpha !}\prod_{i=1}^{N_\alpha}\int
dh_i^{(\alpha)}
e^{U_\alpha (h_i^{(\alpha)})+ l_i^{(\alpha)}h_i^{(\alpha)}}
\right)\right|_{ l^{(\alpha)} = 0}
\label{mucomamo2}
\ee
where
$\Delta(h,h') \equiv \prod_{i>j}^N(h_i-h_j)\prod_{k>l}^{N'}(h'_k-h'_l)
\prod_{i=1}^N\prod_{k=1}^{N'}(h'_k-h_i)$.
This formula already takes the specific form of $K$ into account.
The product of integrals in brackets at the r.h.s. of (\ref{mucomamo2})
is equal (for every fixed $\alpha$) to
\be
\frac{1}{N_\alpha !}
\prod_{j=1}^{N_\alpha}\hat\Phi^{\{U_\alpha\}}( l_j^{(\alpha)})
\ee
(compare with eq.(\ref{todachainmatr1})).

In order to simplify the notation we shall further denote
$\hat\Phi^{\{U_\alpha\}}( l) \equiv \int dx e^{U_\alpha(x)+ l x}$
through $\hat\Phi_\alpha( l)$, and
$\left(\frac{\partial}{\partial t_1^{(\alpha)}}\right)^k
\hat\Phi^{\{U_\alpha\}}( l^{(\alpha)}) =
\left(\frac{\partial}{\partial  l^{(\alpha)}}\right)^k
\hat\Phi^{\{U_\alpha\}}( l^{(\alpha)})$ - through
$\partial^k \hat\Phi_\alpha( l)$. Thus
\be
&Z_{N_1\ldots N_{p-1}}^{A_{p-1}}\{t^{(1)} \ldots t^{(p-1)}\} =
\nn \\
&= \Delta\left(\frac{\partial}{\partial l^{(1)}} \right)
\prod_{\alpha =1}^{p-2}
\Delta\left(\frac{\partial}{\partial l^{(\alpha)}},
\frac{\partial}{\partial l^{(\alpha+1)}} \right)
\Delta\left(\frac{\partial}{\partial l^{(p-1)}} \right)
\times \nn \\ &\times \left.
\prod_{\alpha =1}^{p-1}\left(   \frac{1}{N_\alpha !} \prod_{j=1}^{N_\alpha}
\hat\Phi_\alpha( l_j^{(\alpha)})\right) \right|_{ l^{(\alpha)} = 0}.
\label{mucomamo4}
\ee
If $p-1=1$, differential operator is just a square of determinant
$\Delta(\partial/\partial l)$ and we can use the relation
\be
\Delta^2(h) &= \sum_P {\rm Det}_{ij} h_{P(j)}^{i+j-2} =
\nn \\
&= \sum_P {\rm Det}
\left[
{\begin{array}{ccrc}
1           &h_{P(2)}   &h^2_{P(3)}\ldots &h^{N_1-1}_{P(N_1)}    \\
h_{P(1)}    &h^2_{P(2)} &h^3_{P(3)}\ldots &h^{N_1}_{P(N_1)}    \\
h_{P(1)}^2  &h^3_{P(2)} &h^4_{P(3)}\ldots &h^{N_1+1}_{P(N_1)}    \\
            &           &\ldots           &                      \\
h_{P(1)}^{N_1-1}  &h_{P(2)}^{N_1} &h_{P(3)}^{N_1+1}\ldots
&h^{2N_1-2}_{P(N_1)}
\end{array}}
\right]
\label{detform1}
\ee
where the sum is over all the $N!$ permutations $P$ of $N$ elements
$1\ldots N$, in order to conclude that (\ref{mucomamo4}) reproduces
our old formula (\ref{todarepr}), (\ref{todachainmatr1}):
$Z_N = {\rm Det}_{ij} \partial^{i+j-2}\hat\Phi$.

For $p-1=2$ we need to use a more complicated analogue of
(\ref{detform1}):
\be
&\Delta(h)\Delta(h,h')\Delta(h') =
\label{detform2} \\
&= \sum_P\sum_{P'}{\rm Det}\left[
{\begin{array}{crccrc}
1           &h_{P(2)}
\ldots &h^{N_1-1}_{P(N_1)}   &1  &\bar h_{\bar P(2)}
\ldots &\bar h^{N_2-1}_{\bar P(N_2)} \\
h_{P(1)}    &h^2_{P(2)}
\ldots &h^{N_1}_{P(N_1)}  &\bar h_{\bar P(1)}  & \bar h_{\bar P(2)}^2
\ldots &\bar h^{N_2}_{\bar P(N_2)} \\
h_{P(1)}^2  &h^3_{P(2)}
\ldots &h^{N_1+1}_{P(N_1)}  &\bar h_{\bar P(1)}^2 &\bar h_{\bar P(2)}^3
\ldots &\bar h^{N_2+1}_{\bar P(N_2)}  \\
            &    &     \ldots &  & &                     \\
h_{P(1)}^{{\cal N}_1-1}  &h_{P(2)}^{{\cal N}_1}
\ldots &h^{{\cal N}+N_1-2}_{P(N_1)}
&\bar h_{\bar P(1)}^{{\cal N}_2-1}  &\bar h_{\bar P(2)}^{{\cal N}_2}
\ldots &\bar h^{{\cal N}+N_2-2}_{P(N_2)}
\end{array}}
\right]
\nn
\ee
where ${\cal N} = \sum_{\alpha =1}^{p-1} N_\alpha$. Making use of this
formula, we conclude that the r.h.s. of (\ref{mucomamo4}) for $p-1=2$ is also
representable in the form of determinant:
\be
{\rm Det}
 \left[
\begin{array}{crccrc}
\hat\Phi &\partial\hat\Phi \ldots &\partial^{N_1-1}\hat\Phi
  &\hat{\bar\Phi} &\partial\hat{\bar\Phi} \ldots
&\partial^{N_2-1}\hat{\bar\Phi} \\
\partial\hat\Phi &\partial^2\hat\Phi \ldots &\partial^{N_1}\hat\Phi
  &\partial\hat{\bar\Phi} &\partial^2\hat{\bar\Phi} \ldots
&\partial^{N_2}\hat{\bar\Phi}
\\
  &         & \ldots  & &    &                        \\
\partial^{{\cal N}-1}\hat\Phi &\partial^{\cal N}\hat\Phi \ldots
&\partial^{{\cal N}+N_1-2}\hat\Phi
  &\partial^{{\cal N}-1}\hat{\bar\Phi} &\partial^{\cal N}\hat{\bar\Phi} \ldots
&\partial^{{\cal N}+N_2-2}\hat{\bar\Phi}
\end{array}
\right]
\nn
\ee
here $\hat\Phi = \hat\Phi_1,\ \hat{\bar\Phi} = \hat\Phi_2$
and all arguments $ l^{(\alpha)} = 0$.
It is especially easy to check formula (\ref{detform2})
in the simplest case of $N_1=N_2=1$. Then it just says that
$\bar h - h = {\rm Det}\left[\begin{array}{cc}
1&1 \\ h&\bar h \end{array}\right]$.
Analogues expressions for $p-1>2$ are more involved, they are no longer just
determinants: this is obvious already
from consideration of the simplest case of $N_1=\ldots=N_{p-1}=1$,
the product $\prod_{\alpha=1}^{p-2}(h^{(\alpha)} - h^{(\alpha+1)})$
is no longer determinant of any nice matrix.

\subsection{Orthogonal polinomials}

Formalism of orthogonal polinomials was intensively used at the early days of
the theory of matrix models. It is applicable to scalar-product
eigenvalue models
and allows to further transform (diagonalize) the remaining determinants into
products. In variance with both reduction from
original $N^2$-fold matrix integrals to the eigenvalue problem, which (when
possible) reflects a physical phenomenon - decoupling of angular
(unitary-matrix) degrees of freedom (associated with $d$-dimensional gauge
bosons), - and with occurence of determinant formulas which reflects
integrability of the model,  orthogonal polinomials appear more as a technical
device. Essentially orthogonal polinomials are
necessary if wants to explicitly separate dependence on the the size $N$ of
the matrix in the matrix integral ("zero-time") from dependencies on all other
time-variables and to explicitly construct variables, which satisfy Toda-like
equations. However, modern description of integrable hierarchies in terms of
$\tau$-functions does not require explicit separation of the
zero-time and treats it more or less on the equal fooring with all other
variables, thus making the use of orthogonal polinomials unnecessary. Still
this technique remains in the arsenal of the matrix model theory\footnote{
Of course, one can also use this link just with the aim to put the
rich and beautifull mathematical theory of orthogonal polinomials into the
general context of string theory. Among interesting problems here is the
matrix-model description of $q$-orthogonal polinomials.
} and we now
briefly explain what it is about. At the end of this section two simple
applications will be also described: one to evaluation of the volume of the
unitary group, another - to direct proof of equivalence of the ordinary
1-matrix model and the Gaussian Kontsevich model. Both these examples make use
of explicitly known orthogonal Hermite polinomials and in this sense are not
quite representative: usually orthogonal polinomials are $not$ known
explicitly. Some applications of such "abstract" theory of orthogonal
polinomials to the study of matrix models will be mentioned in the following
sections.

In the context of the theory of scalar-product matrix models orthogonal
polinomials naturally arise when one notes that after
partition functions appears in a simple determinantal form of
eq.(\ref{dimamodet}),
any linear change of basises $h^i \rightarrow Q_i(h) = \sum_kA_{ik}h_k,
\ \bar h^j \rightarrow \bar Q_j(\bar h) = \sum_l B_{jl}\bar h^l$ can be easily
performed and
$Z \longrightarrow Z\cdot {\rm det} A\cdot {\rm det} B$. In particular, if $A$
and $B$
are triangular with units at diagonals, their determinants are just unities
and $Z$ does not change at all. This freedom is, however, enough, to
diagonalize the scalar product and choose polinomials $Q_i$ and $\bar Q_j$ so
that
\be
\langle Q_i(h) \mid \bar Q_j(\bar h) \rangle = e^{\phi_i}\delta_{ij}.
\label{orthopol}
\ee
$Q_i$ and $\bar Q_j$ defined in this way up to normalization are called
orthogonal polinomials. (Note that $\bar Q$ does not need to be a {\it
complex} conjugate of $Q$: "bar" does not mean complex conjugation.)
Because of above restriction on the form of matrices
$A$ and $B$ these polinomials are normalized so that
\be
Q_i(h) = h^i + \ldots;\ \ \ \bar Q_j(\bar h) = \bar h^j + \ldots
\nn
\ee
i.e. the leading power enters with the $unit$ coefficient. From
(\ref{dimamodet}) and (\ref{orthopol}) it follows that
\be
Z_N = \prod_{i=1}^N e^{\phi_{i-1}}.
\label{ZNprod}
\ee
This formula is essentially the main outcome of orthogonal polinomials theory
fro matrix models: it provides complete separation of the $N$-dependence of
$Z$ (on the size of the matrix) from that on all other parameters
(which specify the shape of
potential, i.e. the measure $d\mu_{h,\bar h}$), this information is encoded in
a rather complicated fashion in $\phi_i$.
As was already mentioned, any feature of matrix model can be examined already
at the level of eq.(\ref{dimamodet}), which does not refer to orthogonal
polinomials and thus they are not really relevant for the subject.

We can, however, reverse the problem and ask, what can matrix models provide
$for$ the theory of orthogonal polinomials.\footnote{Of course, we can hardly
get anything $new$ for that theory, but the purpose is to see, which features
are immediate consequences of the "physically-inspired" approach. As usually
this can help to somehow organize the existing knowledge in appropriate
system. This is, however, not our goal in these notes: only a very simple
example will be mentioned, which will be also of use in our futher
considerations.}
The first question to ask in the theory of orthogonal polinomials is:

Given the measure $d\mu_{h,\bar h}$, what are the corresponding orthogonal
polinomials?

Usually the answer to this type of questions is not at all straightforward.
Its complexity, however, depends on what one agrees to accept as a suitable
answer. Of particular interest for our purposes below would be integral
representations. It would be very helpfull to have just an integral
transformation, converting the set of orthogonal polinomials for given
$d\mu_{h,\bar h}$ into some standard set, like $Q_i^{(0)} = x^i$.
Unfortunately,
such transformation is rarely available, though there are important examples:
classical orthogonal polinomials and their $q$-analogues (expressed through
the ($q$-)hypergeometric functions, which usually possess integral
representation of a simple form, see \cite{movi} for an introductory review of
such integral formulas, which are in fact well known in CFT).
The simplest example of this kind, which will be used below is the set of
Hermite polinomials:
\be
{\rm He}_k(h) = \frac{1}{\sqrt{2\pi}} e^{\frac{h^2}{2}}
\int (ix)^ke^{-\frac{x^2}{2}-ixh} dx
= (h-\frac{d}{dh})^k\cdot 1 = \nn \\
= e^{\frac{h^2}{2}}(-\frac{d}{dh})^k e^{-\frac{h^2}{2}} =
\frac{1}{2^k}e^{\frac{h^2}{4}}(h - 2\frac{d}{dh})^ke^{-\frac{h^2}{4}} =
h^k + \ldots,
\label{hepo}
\ee
orthogonal with the $local$ measure $d\mu_h = e^{-\frac{h^2}{2}}$.

For generic measure the answer of this type does not exist in any universal
form. However, matrix models still provide a somewhat peculiar integral
representation for $any$ measure, with the number of integrations depending on
the number of polinomial.
In order to obtain this expression, let us consider a slight generalization of
formula (\ref{meas1})
\be
Z_N\{\lambda _\gamma\} \equiv
\frac{1}{N!} \prod_{k=1}^N \int d\mu_{h_k,\bar h_k}
\Delta(h)\Delta(\bar h)\prod_{k,\gamma} (\lambda _\gamma - h_k).
\label{meas2}
\ee
Then
$\Delta(h)\prod_{k,\gamma} (\lambda _\gamma - h_k) =
\Delta(h,\lambda )/\Delta(\lambda )$,
and $\lambda _\gamma$ can be just considered as $h_{N+\gamma}$,
which are $not$ integrated over in (\ref{meas2}). Then it is clear that
\be
\Delta(h,\lambda ) = {\rm Det} \left(
\begin{array}{cl}        Q_{i-1}(h_k)  & Q_{N+\gamma-1}(h_k)  \\
                     Q_{i-1}(\lambda _\delta)
& Q_{N+\gamma-1} (\lambda _\delta)
\end{array}
\right)
\ee
while $\Delta(\bar h) = {\rm Det}_{jk} \bar Q_{j-1}(\bar h_k)$. Since all the
$Q_{N+\gamma-1}(h_k)$ are orthogonal to all $\bar Q_{j-1}(\bar h_k)$
(because $N+\gamma-1 \neq j-1$), we obtain:
\be
Z_N\{\lambda _\delta\} =
\frac{{\rm det}_{\gamma\delta}Q_{N+\gamma-1}(\lambda _\delta)}
{\Delta(\lambda )} Z_N.
\label{meas3}
\ee
In particular,
\be
Q_N(\lambda ) = \frac{Z_N\{\lambda \}}{Z_N},
\ee
where both the numerator and denominator can be represented by $N\times
N$-matrix integrals.

Inverse "main question" of the theory of orthogonal polinomials is:

Given a set of polinomials
\be
Q_i(h) = h_i + \ldots,\nn \\
\bar Q_j(\bar h) = \bar h_j + \ldots,
\nn
\ee
 what is the measure
$d\mu_{h,\bar h}$ w.r.to which they form an orthogonal system?

We shall not discuss the complete answer to this question and consider only
the case of the {\it local} measure, when $\bar Q_i = Q_i$.
Then usually the answer does not exist at all: not {\it every} system of
polinomials is orthogonal w.r.to some local measure. It is easy to find the
necessary (and in fact sufficient) condition. As was mentioned above, the
local measure is distinguished by the property that multiplication by (any
function of) $h$ is Hermitean operator:
\be
\langle hf(h)\mid g(\bar h)\rangle =
\langle f(h) \mid \bar h g(\bar h)\rangle, \ \ {\rm if} \
d\mu_{h,\bar h} \sim \delta(h-\bar h).
\ee
This property implies, that the coefficients $c_{ij}$ in the recurrent
relation
\be
hQ_i(h) = Q_{i+1}(h) + \sum_{j=0}^i c_{ij}Q_j(h)
\ee
are almost all vanishing. Indeed: for $j<i$
\be
\new
\begin{array}{c}
c_{ij} = \frac{\langle hQ_i(h) \mid Q_j(\bar h)\rangle}{\langle Q_j(h) \mid
Q_j(\bar h)\rangle} =
\frac{\langle Q_i(h) \mid \bar hQ_j(\bar h)\rangle}{\langle Q_j(h) \mid
Q_j(\bar h)\rangle} =
 \\ =
\delta_{i,j+1}\frac{\langle Q_i(h) \mid Q_i(\bar h)\rangle}{\langle Q_j(h)
\mid Q_j(\bar h)\rangle} =
\delta_{j,i-1} e^{\phi_i -\phi_{i-1}}.
\end{array}
\ee
In other words, polinomials, orthogonal w.r.to a local measure are obliged to
satisfy the "3-term recurrent relation":
\be
hQ_i(h) = Q_{i+1}(h) + C_iQ_i(h) + R_iQ_{i-1}(h)
\label{3termrel}
\ee
(the coefficient in front of $Q_{i+1}$ can be of course changed by the change
of normalization). Parameter $C_i$ vanishes if the measure is even (symmetric
under the change $h \rightarrow -h$), then polinomials are split into two
orthogonal subsets: even and odd in $h$. Partition function (\ref{ZNprod})
of the $one$-component model can be expressed through parameters
$R_i =  e^{\phi_i -\phi_{i-1}}$ of the 3-term relation:
\be
Z_N = Z_1 \prod_{i=1}^{N-1} R_i^{N-i},
\ee
thus defining a one-component matrix model (i.e. particular shape of
potential), associated with any system of orthogonal polinomials.

Our "inverse main question" in the case of the local measure should be now
formulated as follows: Given a set of orthogonal polinomials $Q_i(h) = h^i
+\ldots$ {\it which satisfy the 3-term relation} (\ref{3termrel}), what is
the measure $d\mu_h$?

As every complete orthogonal system of functions, orthogonal polinomials
satisfy the completeness relation:
\be
\sum_{i=0}^{\infty} e^{-\phi_i} \bar Q_i(\bar h)Q_i(h) =
\delta^{\{d\mu\}}(\bar h,h),
\ee
where $\delta$-function, associated with the measure $d\mu_{h,\bar h}$ is
defined so that
\be
\int\int f(h)\delta^{\{d\mu\}}(\bar h,h') d\mu_{h,\bar h} = f(h')
\ee
for any function $f(h)$.
Since for the {\it local} measure $d\mu_h = e^{U(h)}dh$ the $\delta$-function
is just $\delta^{\{d\mu\}}(\bar h,h) = e^{-U(h)}\delta(\bar h - h)$,
as an answer to our question we can take a representation of $U(h)$
in terms of the corresponding orthogonal polinomials:
\be
e^{-U(h)}\delta(\bar h - h) = \sum_{k=0}^{\infty} \frac{Q_k(\bar h)Q_k(h)}
{\langle Q_k \mid Q_k\rangle}.
\label{orthopoltomeas}
\ee
As usually this relation should be understood as analytical continuation.
The squared norms $\mid\mid Q_k \mid\mid^2$ in denominator are expressed
through the coefficients $R_i$ of the 3-term relation up to an overall
constant: $\mid\mid Q_k \mid\mid^2 = \prod_{i=1}^k R_i\mid\mid Q_0 \mid\mid^2$.

For example, in the case of Hermite polinomials (\ref{hepo}) we have:
\be
{\rm He}_{k+1}(h) = (h - \frac{d}{dh}){\rm He}_k(h) =
h{\rm He}_k(h) - \frac{d}{dh}{\rm He}_k(h) =
\nn \\
= h{\rm He}_k(h) - k{\rm He}_{k-1}(h)
\ee
(the last equality holds because $\frac{d}{dh}$ and $h-\frac{d}{dh}$ play the
role of annihilation and creation operators respectively). This means that the
3-term relation is satisfied with $R_k = k$ and thus
$\mid \mid {\rm He}_k\mid\mid^2 = \mid \mid {\rm He}_0\mid\mid^2 k!$
We shall use the normalization condition
$\mid \mid He_0\mid\mid^2 = \sqrt{2\pi}$. Then for $e^{-U(h)}$ we get:
\be
\new
\begin{array}{c}
e^{-U(h)}\delta(\bar h-h) =  \\
\sum_{k=0}^{\infty} \frac{{\rm He}_k(\bar h){\rm He}_k(h)}
{\mid\mid He_k\mid\mid^2} =
\frac{1}{\sqrt{2\pi}}\sum_{k=0}^{\infty}\frac{1}{k!}
(h-\frac{d}{dh})^k(\bar h-\frac{d}{d\bar h})^k \cdot 1 =
 \\
=  \frac{1}{\sqrt{2\pi}} e^{\frac{h^2}{2}+\frac{\bar h^2}{2}}
  \sum_{k=0}^{\infty}\frac{1}{k!}(\frac{d^2}{dhd\bar h})^k
  e^{-\frac{h^2}{2}-\frac{\bar h^2}{2}}
= \frac{1}{\sqrt{2\pi}} e^{\frac{h^2}{2}+\frac{\bar h^2}{2}}
  e^{\frac{d^2}{dhd\bar h}} e^{-\frac{h^2}{2}-\frac{\bar h^2}{2}}
=  \\
= \frac{1}{\sqrt{2\pi}} {\rm Im} \int\int \frac{d\alpha d\bar\alpha}{2\pi}
e^{-\alpha\bar\alpha} e^{\frac{h^2}{2}+\frac{\bar h^2}{2}}
e^{\alpha\frac{d}{dh}+\bar\alpha\frac{d}{d\bar h}}
 e^{-\frac{h^2}{2}-\frac{\bar h^2}{2}} =
 \\ =
\frac{1}{\sqrt{2\pi}} {\rm Im} \int\int \frac{d\alpha d\bar\alpha}{2\pi}
e^{-\frac{1}{2}(\alpha +\bar\alpha)^2}
e^{-\frac{1}{2}(\alpha +\bar\alpha)(h + \bar h)}
e^{-\frac{1}{2}(\alpha -\bar\alpha)(h - \bar h)} =
\nn \\ = e^{\frac{h^2}{2}}\delta(h-\bar h).
\end{array}
\ee

\subsection{Scalar-product models in Miwa parametrization}

We shall now make the first step towards clarification of the interrelation
between the scalar-product and Kontsevich models.
We already know that in the latter
case the important role is played by representation of time-variables in the
form of
\be
T_k = \frac{1}{k} {\rm tr} \Lambda ^{-k},
\label{MiwaT}
\ee
with $n\times n$ matrix $\Lambda $,
which will be further refered to as Miwa parametrization (expressions of
some similar form were first introduced in \cite{Miwa}). Let us now perfrom
such transformation in the case of the scalar-product model.
Let us use eq.(\ref{tbartmeas}) to define the time-dependence of the measure,
only ignore the $\bar t$-variables. Namely, introduce
$d\mu_{h,\bar h} = e^{U(h)} d\hat\nu_{h,\bar h}$ (i.e. $d\hat\nu_{h,\bar h} =
e^{\bar U(\bar h)}d\hat\mu_{h,\bar h}$). Substitute
\be
t_k = \mp\left(\frac{1}{k}{\rm tr}\Lambda ^{-k} + r_k\right)
\ee
and obtain:
\be
\new
\begin{array}{c}
e^{U(h)} = e^{-\hat V(h)} e^{\mp{\rm tr} \sum_{k=1}^{\infty}\frac{1}{k}
\left(\frac{h}{\Lambda }\right)^k} =
e^{-\hat V(h)}\frac{{\rm det}^{\pm 1}(\Lambda -h\cdot I)}{{\rm
det} \Lambda } = \\
= \frac{ e^{-\hat V(h)}}{{\rm det} \Lambda }
\prod_{\gamma =1}^n (\lambda _\gamma-h)^{\pm 1},
\end{array}
\ee
where $\hat V(h) \equiv \pm \sum_k r_kh^k$. Let us choose {\it upper} signs in
these formulas. Then we can  use eqs.(\ref{meas2}) and
(\ref{meas3}) to conclude that in Miwa parametrization
\be
Z_N^{\{d\mu\}} =
\frac{1}{({\rm det} \Lambda )^N} Z_N^{\{d\hat\nu\}}\{ \lambda _\delta\} =
Z_N^{\{d\hat\nu\}}
\frac{{\rm det}_{\gamma\delta}\hat Q_{N+\gamma-1}(\lambda _\delta)}
{\Delta(\lambda )({\rm det} \Lambda )^N},
\nn
\ee
where $d\hat\nu_{h,\bar h} \equiv e^{-\hat V(\bar h)}d\nu_{h,\bar h}$ and
$\hat Q_k$ are the corresponding orthogonal polinomials.
In other words, we reduced the model with potential $U(h)$ to another model
with potential, $-\hat V(h)$, and expressed the difference in terms of
orthogonal polinomials $\hat Q_k$:
\be
\frac{Z_N^{\{d\mu\}}}{Z_N^{\{d\hat\nu\}}} =
\frac{1}{({\rm det} \Lambda )^N}\cdot
\frac{{\rm det}_{\gamma\delta}\hat Q_{N+\gamma-1}
(\lambda _\delta)}{\Delta(\lambda )}.
\label{reduhv}
\ee
If $\hat V(h)$ is adjusted to give rise to some simple orthogonal polinomials,
(i.e. if the new model $Z_N^{\{d\hat\nu\}}$ is easy to solve),
this representation can considerably simplify the original model.

Another interpretation of this formula is that we obtained a GKM-like
representation of the from of (\ref{KIev"N}) for the {\it discrete}
scalar-product model. The only difference is that $\varphi_\gamma^{\{V\}}$ in
(\ref{KIev"N}) are changed for $\hat Q_{\gamma-1}$ in (\ref{reduhv}). This is
an important difference, because $\varphi_\gamma^{\{V\}}$ in GKM are defined
to by integral formulas like (\ref{hatvarphi}),
$\varphi_\gamma^{\{V\}} =\ \langle\langle x^{\gamma -1} \rangle\rangle$
or, alternatively, satisfy the
recursive relations like (\ref{Kasch}). Moreover, generic
$\varphi_\gamma^{\{V\}}$ are infinite formal series in $\lambda^{-1}$, while
$Q_{\gamma-1}$ are orthogonal {\it polinomials}. This discreapancy is one of
important stimuli for further developement of the concept of Generalized
Kontsevich model, as well as for search for convenient integral
representations for orthogonal polinomials.

There is, however, at least one interesting situation when the two formulas
indeed coincide. This is the case of Gaussian potentials $V$ and $\hat V$,
when both $\varphi_\gamma^{\{V\}}$ and
$Q_{\gamma-1}$ are represented by orthogonal Hermite polinomials, which
possess  integral representation, exactly adequate in the context of
GKM. This is the subject of our consideration in the next subsection.

\subsection{Equivalence of the discrete 1-matrix and Gaussian Kontsevich
models}

Let us take the ordinary 1-matrix model with the $local$ measure $d\mu_h =
e^{U(h)}dh $ to be the scalar-product model, considered in
the previous subsection and take Miwa parametrization with upper signs and
with $r_k =
-\frac{1}{2}\delta_{k,2}$ (as we did in the section 2.6). Then $\hat V(h) =
\sum_k r_kh^k = -\frac{h^2}{2} = \frac{(ih)^2}{2}$.
The relevant orthogonal polinomials $\hat Q$ are just Hermite
polinomials of $imaginary$ argument:\footnote{
Note that this system of functions $\varphi_k = i^{-k}{\rm He}_k(ih)$
looks like $\varphi_0 = 1,\
\varphi_1 = h,\ \varphi_2 = h^2+1, \ldots$, and does not resemble any set of
orthogonal polinomials with a local measure (for example the product
$\varphi_0\cdot\varphi_2 = h^2+1$ may seem positive definite, this being
inconsistent with orthogonality requirement $\langle
\varphi_0\mid\varphi_2\rangle = 0$). The thing is that integration
at the l.h.s. of eq.(\ref{1mamoidgako}) is well defined only along the
imaginary axis, while integrals along the real axis are understood as
analytical continuation.
} $Q_k^{\{-\frac{h^2}{2}dh\}} = i^{-k}{\rm He}_k(ih) = h^k + \ldots$. These
polinomials possess an integral representation (\ref{hepo}):
\be
i^{1-k}{\rm He}_{k-1}(ih) =
\frac{1}{\sqrt{2\pi}} e^{-\frac{h^2}{2}}\int x^{k-1}
e^{-\frac{x^2}{2} + xh} dx  \stackrel{(\ref{varphiko})}{=}
\varphi_k^{\{\frac{x^2}{2}\}}(h).
\ee
Using (\ref{reduhv}) and (\ref{KIev"N}) we obtain a remarkable relation
between the two matrix models:
\be
\frac{Z_N\{t_0 = 0; t_k = -\frac{1}{k}{\rm tr}\Lambda ^{-k} +
\frac{1}{2}\delta_{k,2}\}}
{Z_N\{t_k = \frac{1}{2}\delta_{k,2}\}} =
\frac{\int_{N\times N} dH e^{\sum_{k=0}^{\infty} t_k {\rm Tr}
H^k}}{\int_{N\times N} dH e^{\frac{1}{2}H^2}} = \nn \\
= \frac{e^{-{\rm tr}\frac{\Lambda ^2}{2}}}
{(2\pi)^{\frac{n^2}{2}}({\rm det}\Lambda )^N}
\int_{n\times n}  dX ({\rm det}X)^N e^{-{\rm tr}\frac{X^2}{2} + \Lambda X}
= {\cal Z}_{\frac{X^2}{2}}\{N,t\},
\label{1mamoidgako}
\ee
where ${Z_N\{t_k = \frac{1}{2}\delta_{k,2}\}} = (-2\pi)^{\frac{N^2}{2}}c_N$.
This relation can be also regarded as an identity
\be
\frac{\int_{N\times N}dH e^{\frac{1}{2}{\rm Tr}H^2} Det(\Lambda \otimes I-
I\otimes
H)}{ \int_{N\times N}dH e^{\frac{1}{2}{\rm Tr}H^2}} =   \nn \\
= \frac{\int_{n\times n}dX e^{-\frac{1}{2}{\rm tr}X^2} {\rm
det}^N(X+\Lambda )}{\int_{n\times n}dX e^{-\frac{1}{2}{\rm tr}X^2}},
\ee
valid for any $\Lambda$.
Note that integrals are of differents sizes: $N\times N$ at the l.h.s. and
$n\times n$ at the r.h.s. While $N$-dependence is explicit at both sides of
the equation, the $n$-dependence at the l.h.s. enters only implicitly: through
the allowed domain of variation of variables
$t_k = -\frac{1}{k}{\rm tr}\Lambda ^{-k} + \frac{1}{2}\delta_{k,2}$.
(This can serve as an illustration to the general statement that the shape of
Kontsevich partition function ${\cal Z}_V$, considered as a function of $T$'s
rather than $ L$ or $\Lambda $, is independent of the matrix size $n$.)
Identity (\ref{1mamoidgako}) was anticipated
from the study of Ward identites for the
Gaussian Kontsevich model in \cite{ChMa} (see eq.
(2.53)
in the section 2.6 above),
and it was derived in the present form in ref.\cite{Toda}.

Eq.(\ref{1mamoidgako}) can be used to perfrom analytical continuation in $N$
and define what is $Z_N$ for $N$, which are not positive integers. Since $c_N
= 0$ for all {\it negative} integers (see eq.(\ref{cnneg}) below), the same is
true for $Z_N$. In the next section 4 we shall see that it is characteristic
property of $\tau$-functions of {\it forced} hierarchies.

\subsection{Volume of unitary group}

Formalism of orthogonal polinomials provides also a simple derivation of
eq.(\ref{volun}) for the volume of unitary group. Consider
eq.(\ref{1mamoev}) with $U(H) = H^2$. Then Gaussian matrix integral can be
easily evaluated:
\be
c_N\int_{N\times N} dH e^{-\frac{1}{2}{\rm Tr}H^2} =
c_N\prod_{i=1}^N \int dH_{ii}e^{-\frac{1}{2}H_{ii}^2}
\prod_{i<j}^N \int d^2H_{ij} e^{-\mid H_{ij}\mid^2} =
(2\pi)^{N^2/2}, \nn
\ee
while according to eqs.(\ref{orthopol}) and (\ref{ZNprod})
the same integral is equal to
\be
\frac{1}{N!}\prod_{i=1}^N\int_{-\infty}^{+\infty}e^{-\frac{1}{2}h_i^2}
\prod_{i>j}^N (h_i-h_j)^2 =
\prod_{j=1}^N \mid\mid {\rm He}_{j-1} \mid\mid^2. \nn
\ee
Here $\mid\mid {\rm He}_{j-1} \mid\mid$
stand for the norms of orthogonal Hermite
polinomials (\ref{hepo}), $\mid\mid {\rm He}_k \mid\mid^2  =  \sqrt{2\pi} k!$.
Comparing the two expressions for the same integral we get:
\be
c_N^{-1} = (2\pi)^{\frac{N^2}{2}}\prod_{k=0}^{N-1} \frac{1}{\sqrt{2\pi}k!} =
\frac{(2\pi)^{\frac{N(N-1)}{2}}}{\prod_{k=0}^{N-1} k!}.
\label{cnform}
\ee
According to (\ref{cN})
\be
c_N^{-1} = N!\frac{{\rm Vol}_{U(N)}}{({\rm Vol}_{U(1)})^N}
\nn
\ee
and ${\rm Vol}_{U(1)} = 2\pi$. Thus we obtain eq.(\ref{volun}):
\be
{\rm Vol}_{U(N)} = \frac{(2\pi)^{\frac{N(N+1)}{2}}}{\prod_{k=0}^N k!}.
\nn
\ee

An example of somewhat more sophisticated (quantum) group-theoretical
quantity, arising from Gaussian matrix models, is provided by the following
formula for the $q$-factorial \cite{CAD} (see also \cite{KMSW2}):
\be
\frac{1}{(q,q)_N} \equiv  \prod_{n=1}^N \frac{1}{1-q^n} =
\frac{\int\int_{N\times N} dH[dU] e^{-m^2{\rm Tr}H^2 + {\rm
Tr}HUHU^{\dagger}}}{{\rm Vol}_{U(N)} \int_{N\times N} dH e^{-m^2{\rm Tr}H^2}}.
\ee
Integral in the numerator is over Hermitean ($H$) and unitary ($U$) $N\times
N$ matrices, and $q \equiv m^2 - \sqrt{m^4-1}$.

Explicit expression (\ref{cnform}) can be used to prove that
$c_N = 0$ for all {\it negative} integer $N$ \cite{Toda}.
Eq.(\ref{cnform}) defines $c_N$ only for positive integer $N$ as a finite
product. There is an obvious prescription for analytical continuation of such
products, provided continuation of the items is known (it can be considered as
implied by the similar formula for integrals with the varying upper limit):
Let
\be
F(N) = \sum_{k = -\infty}^N f(k).
\ee
Then
\be
S(N) \equiv \sum_{k=1}^N f(k) = F(N) - F(0)
\ee
and, obviously, $F(0) - F(-N) = \sum_{k = 1-N}^0 f(k)$,  so that
\be
S(-N) \equiv F(-N) - F(0) = -\sum_{k=0}^{N-1} f(-k).
\ee
Exponentiation of this
formula gives the rule for the products. In the case of $c_N$ one can
treat factorials in (\ref{cnform}) as Gamma-functions,
\be
(2\pi)^{\frac{N(N-1)}{2}} c_N = \prod_{k=1}^N \Gamma(k),
\ee
and obtain:
\be
(2\pi)^{-\frac{N(N+1)}{2}} c_{-N} = \left( \prod_{k=0}^{N-1}
\Gamma(-k)\right)^{-1} = 0,
\label{cnneg}
\ee
because of the poles of $\Gamma$-functions.

\






\bigskip

\section{Integrable structure of eigenvalue models}

\setcounter{equation}{0}

\subsection{The concept of integrability}

Integrable structure of dynamical system implies that all the dynamical
characteristics - solutions of equations of motion for a classical system and
functional integrals for a quantum one - can be found exactly. According to
this description the notion of integrability is not very concrete, and in fact
it evolves with time, including more and more classes of theories into the
class of integable systems. Nowadays we consider the following types of
theories as clearly belonging to this class:

- Free motion (classical or quantum) on group manifolds and homogeneous
spaces;

- 2-dimensional conformal theories and their "integrable massive deformations";

- Integrable hierarchies of the (multicomponent) KP and Toda type and their
reductions;

- Functional integrals, subjected to conditions of (generalized)
Duistermaat-Heckman theorem;

- (Eigenvalue) matrix models;

- Topological theories;

- Many supersymmetric models (at least those allowing for Nicolai
transformation and/or Duistermaat-Heckman-like description);

- Systems with (infinitely) many local integrals of motion.

This list (nothing to say about the order of items) is rather arbitrary.
Also different items are not really different and (as it should be) can be
considered as different descriptions of the same reality. Now we discuss
very briefly at least some of the most important views on the concept of
integrability.

Often the notion of integrability is related to occurence of "many enough"
integrals of motion ("many enough" means equal to the number of degrees of
freedom). This is, however, not such a rigid definition as one can think.
In fact, in classical mechanics there is usually a complete set of integrals
of motion available: just initial conditions in the phase space (or, to be more
sophisticated, angle-action variables). The problem is, however, that

a) these obvious integrals are very complicated (non-local and multi-valued)
functionals of the $current$ coordinates, and

b) in general situation they are very "unstable" under a small change of
current coordinates ("divergency of trajectories").

In order to avoid these problems one usually imposes a "locality" condition on
equations of motion. While this is a reasonable thing to do for particular
classes of theories (e.g. possessing a well defined kinetic term, which is
quadratic in momenta), this is $not$ a nice decription in general situation,
since "locality" is not invariant under arbitrary (including non-local) change
of variables. In practice, when approached from this side, integrability
implies a kind of "regular" behaviour of trajectories and some more or less
nicel defined transformation from "natural" (or, better to say, "original")
coordinates to the action-angle variables.

Situation becomes even less clear when quantum theory is considered, since
"chaotic behaviour" no longer implies anything really "chaotic" for the quantum
system. Again, very much depends on what kind of observables one wants to
consider, and any notion of "regularity" is not enough under arbitrary
change of variables.

This can be made even more transparent, if one recalls the idea of
universality classes, so important in the modern theory. The idea is
that even in the cases when behaviour of the system seems absolutely haotic
from any naive point of view (like in the cases of turbulence or quantum
gravity), one can and should introduce new variables (which can be very
complicated functions of original ones), which have smooth and well defined
correlation functions. In most cases one is not attempting to find a
$complete$ set of such variables (and thus some information is lost), but this
reflects nothing but the current state of knowledge, and in fact in studies of
$2d$ quantum gravity the $goal$ of $complete$ descrption is already clearly
fromulated.

Despite these comments, the "definition" of integrability in terms of "many
enough" $local$ integrals of motion should be put at the first place in our
discussion, because most of the systems which were so far considered as
integrable more or less naturally get into this class, allowing for some
prefered choice of dynamical variables ("more or less" appears because some
"minor" non-locality is usually present in any interesting examples, where
angle-action variables are not obvious from the very beginning).

This "definition" is so unclear because we attempted to look for a generic
description of integrability. Most interesting approaches, however, go from
another direction. One starts from some simple system and then
perform a change of variables, which makes it looking much more complicated
(being still simple in its essense). This appears to be a much more fruitfull
view on the problem and in fact all the other items of our list above are
describable in this kind of terms.

A trivial, but surprisingly representative example of this approach is
provided by a free particle, moving in flat $D$-dimensional space. The
eigenfunctions of Laplace operator are just plain waves or, equivalently,
spherical harmonics. The radial part of the $j$-th harmonic is already a
not very simple function, satisfying the equation
\be
\left( -\frac{d^2}{dr^2} + \frac{D-1}{r}\frac{d}{dr} +
\frac{C_2(j)}{r^2}\right) \psi(r) = E\psi(r).
\label{ralap}
\ee
This equation is of course less trivial than the original Laplace equation,
but their solutions are related in a simple way. In order to find a solution
of (\ref{ralap}), say, for $j=0$, one should just take an angular average of a
plane wave:
\be
\phi_k(r) = \int e^{ik\vec r\vec\nu}d^{D-1}\vec\nu; \ \ \
\mid\vec\nu\mid = 1.
\ee
This integral representation expresses the solutions of (\ref{ralap}) through
Bessel functions, and this is in fact the proper way to derive the well-known
formula:
\be
\phi_k(r) = 2^{\frac{D}{2}-1}\Gamma\left(\frac{D}{2}\right)
(kr)^{1-\frac{D}{2}} J_{\frac{D}{2}-1}(kr).
\ee
If one expands the exponent in the integral in a series, the standard
expansion for the Bessel function arises.

A slightly more involved example is the quantum mechanical model of a particle
in the potential $e^{-q}$, i.e. the theory of equation
\be
\left(-\frac{d^2}{dq^2} + e^{-q}\right)\psi(q) = 0
\label{exppot}
\ee
(one of course recognizes a simplified version of Toda models).
It can be solved by projection of the simple Sshr\"odinger equation for a
particle, moving on the upper part of the hyperboloid
$x_0^2-x_1^2-x_2^2 = 1;\ \ x_0 > 0$ \cite{OlPe}. If
\be
x_0 = \cosh\frac{q}{2} + \frac{1}{2}z^2e^{q/2}; \ \
x_1 = \sinh \frac{q}{2} - \frac{1}{2}z^2e^{q/2}; \ \
x_2 = ze^{q/2},
\nn
\ee
then $q = \log(x_0+x_1)$, Laplace operator on hyperboloid is
\be
L = \frac{\partial^2}{\partial q^2} - \frac{1}{2}\frac{\partial}{\partial q}
+\frac{1}{4}e^{-q}\frac{\partial^2}{\partial z^2}
\ee
and average of the wave function $\psi_ \lambda(q,z)$
provides the following expression for solutions of (\ref{exppot}):
\be
\psi_ \lambda (q) = e^{i \lambda q}\int_0^{\infty}
t^{2i \lambda -1}e^{-(t+e^q/t)}dt.
\ee

This idea, which is sometimes refered to as "projection method" (see
\cite{OlPe} for a broad review), reveals hidden symmetries of some complicated
systems (which do not possess any symmetry at all in the usual, Noether-like,
sence of the word), by considering them as embedded into wider theories with
more degrees of freedom. Quantum mechanical examples of applicability of the
method are by no means exhausted by the two systems above, one can consider
various projections, starting from (exactly solvable problem of) the free
motion on any group manifold, and in general this gives rise to a very
important
theory of "zonal spherical functions", which nowadays is increasingly
attracting attention because of its obvious links to integrability theory and
quantum geometry (see \cite{Zab} for discussion of the latter relation and
\cite{ZabFKe}, where  also relations with orthogonal polinomials and
Generalized Kontsevich model are partly revealed). An extremely important
example of free motion on a group manifold (in the infinite-dimensional -
Kac-Moody - case) is provided by 2-dimensional WZNW
(Wess-\-Zumino-\-Novikov-\-Witten) model and the corresponding version of
projection method is known as  Hamiltonian reductions in conformal
field theory. Again the resulting theories (like minimal conformal models) do
not possess any kind of symmetry in the usual sense of the word, but still
they are very simple and exactly solvable, remembering their origin in the
theory of free fields.

In principle the theory which is reduced, i.e. complemented by constraints
(initial conditions), does not need to be absolutely symmetric, i.e. to have
Casimir operator or even zero (as in the WZNW case) as its Hamiltonian. It can
be in fact possible to use the projection method to gain a lot of information
about reductions of theories with more sophisticated Hamiltonians, which are
non-trivial functions of group generators. The simplest example is provided by
the theory of quantum-mechanical "quasi-exactly-solvable models"
\cite{QES},\cite{Turb1} and its CFT-generalizations \cite{Turb1},\cite{Halp}.
A more elaborated technique has the name of "localization theory",\footnote{
For various views and approaches to this theory see \cite{KhCh},\cite{STS},
\cite{AFS},\cite{AS},
\cite{DH}.\cite{NiDH},\cite{Wit2YM}, \cite{NT},\cite{NP}.
(So far there are no connections with Andersson localization in the solid state
physics.)
}
(known also as geometrical quantization, Fourier analysis on group manifolds
and Duistermaat-Heckman theory), it provides a very wide generalization of the
above averaging procedure, which maped plain waves into Bessel functions. The
classical sample system to illustrate all the aspects of integrability,
starting from free motion and ending with anyonic statistics,
$W_\infty$-algebras and $2d$ Yang-Mills theory, is the Calogero-Sutherland
system, which can be associated in a uniform way with any simple Lie algebras
and in an "intermediately involved" form looks like a multiparticle theory
in 1+1 dimension with interaction potential
$g^2\sin^{-2}\epsilon (x_i-x_j)$. (See \cite{OlPe} for the introduction to the
theory of Calogero-type models and \cite{Vas},\cite{GoNe} for the new
developements.)

All this discussion was necessary just to illustrate a very simple idea: the
theory of free particles, though trivial, is in fact unexhaustively deep. It is
enough to impose sophisticated initial conditions or perform a sophisticated
change of variables in order to obtain very complicated dynamical systems,
which after they are studied $per\ se$ appear to be surprisingly system, the
reason for this simplicity being that the real underlying dynamics is just
trivial - that of the free particles,- though it may be very hard probelm to
reveal this simplicity when the system is given.
It is advantage of the $general$ theory, that one can begin from the proper
side: just from the theory of free particles and just start making it more and
more complicated, by introducing different kind of variables, considering
correlators of sophisticated operators and so on. Everything what can be
obtained in this way is by definition trivially integrable, though it may be
not so simple to guess for somebody who did not know where the particular
system at the end of this procedure appeared from.

We now proceed to discussion of particularly important realization of this
idea: the theory of $\bar\partial$-operators in $1_C$ dimension (i.e. the
theory of free holomorphic fields in 2 real dimensions). When considered as
functions of moduli of the bundles over Riemann surfaces (i.e. of boundary
conditions, imposed on $2d$ free fields), these simple objects (known as
"$\tau$=functions") start looking a little involved and after all appear
related to sophisticated non-linear equations (but of course integrable)
in 2 and 3 dimensions (like KdV or Kadomtsev-Petvishvili (KP) equation).
We do not attempt to present an exhaustive theory of $\tau$-functions and
integrable hierarchies (besides being still uncompleted, this is a very big
field), but instead concentrate on the very core of it, which is just simple
determinant formulas for the simplest $\tau$-functions (namely, associated
with free-fermion theory and level $k=1$ Kac-Moody algebras). This issue will
be discussed in some details, because besides being the basis of integrable
hierarchies theory, it is also exactly the place, where the links with the
matrix models are found.

\subsection{The notion of $\tau$-function}

There are several different definitions of $\tau$-functions, but all of them
are particular realizations of the following idea:
$\tau$-function is a generating functional of all the correlation functions in
the theory of free particles in 1+1 dimensions. This basic quantity is a kind
of "$det \ D$", where "$D$" is a time-evolution operator (continuous or
discrete) and "$det$" is a sort of a product over eigenvalues of "$D$",
which is usually expressed in the form of a functional integral, associated
with free particles (it is not $a\ priori$ Gaussian in original variables).
This quantity is the most general definition of $\tau$-function.

In practice one is usually more specific. The mostly well studied version of
$\tau$-function arises if one thinks about free particles of a peculiar type:
free fermions with quadratic Hamiltonian and continuous time evolution, i.e.
one considers a theory of spin-1/2 $b,c$-system (fermions) $\tilde \psi(\bar
 z, z), \psi(\bar z, z)$, described by the functional
integral
\be
\new
\begin{array}{c}
\tau\{A\} \sim \ Det(\bar\partial + {\cal A}) \sim \nn \\
\sim \int D\tilde\psi D\psi
\exp\left({ \int_{d^2 z} \tilde\psi\bar\partial\psi}\right)
\exp\left({\int_{d^2 z}\int_{d^2\tilde z} A( z,\tilde z)
\delta(\bar{\tilde z}-\bar z)\psi( z)\tilde\psi(\tilde z)}
\right)
\end{array}
\label{fftf}
\ee
where $\bar z$ plays the role of time and
${\cal A} = A( z,\tilde z)\delta(\bar{\tilde z}-\bar z)
d\bar{\tilde z}d\bar z$ is some $(\frac{1}{2},1; \frac{1}{2},
1)$-bidifferential (i.e. contains a factor of
$d\tilde z^{1/2}d\bar{\tilde z}d z^{1/2}d\bar z$).

Of course, one can think about more general $\tau$-functions, involving many
fermions (this is often done), and more general $b,c$- and $\beta,
\gamma$-systems, in particular, arising in the context of WZNW model,
associated to any Kac-Moody algebra of any level.\footnote{The main technical
difference between generic and "free-fermion" cases is that Lagrangian of
generic free field theory is not just quadratic in scalar fields $\phi$, but
can also contain particular combinations of exponents $e^\phi$. It also
deserves noting that the most general expresiion, {\it quadratic} in {\it
scalar} fields, if rewritten in terms of fermions is in fact {\it quartic}
(but of course $not$ a generic quartic interaction arises in this way).
Integrable nature of certain quartic-fermion interactions is well known from
the theory of Thirring models (in this class of models interactions are usually
local).
}
Also of
interset is consideration of $discrete$ time evolution (described by
difference rather than differential equations), though, as usual in the $2d$
theories, this is not a really independent problem.

In the language of matrix models the restriction to $free-fermion$
$\tau$-functions is essentially equivalent to restriction to $eigenvalue$
models. Serious consideration of non-eigenvalue models, aimed at revealing
their integrable (solvable) structure will certainly involve the theory of
generic $\tau$-functions, but both these things are matters of the future
research, and we"ll not go into details about them in these notes.

\subsection{$\tau$-function, associated with the free fermions}

Because of specific form of the Lagrangian in (\ref{fftf}) the functional
integral can be easily represented in Hamiltonian form, provided topology of
the 2-surface on which $\bar z, z$ are coordinates, is trivial
(genus 0: sphere or annulus). Namely, consider $\tilde\psi$ and $\psi$ as
operator-valued functions of $ z$ only (not of the time $\bar z$).
Then the only reminiscent of kinetic term $\int_{d^2 z}
\tilde\psi\bar\partial\psi$ is canonical commutation relation:
\be
\phantom. [\tilde\psi(\tilde z),\psi( z)]_+ =
\delta(\tilde z -  z) d\tilde z^{1/2}d z^{1/2}.
\ee
Then
\be
\tau\{A\} \sim
\langle 0 \mid \exp\left({\oint_{d\tilde z}\oint_{d z}
A( z,\tilde z)\psi( z)\tilde\psi(\tilde z)}\right)
\mid 0 \rangle.
\label{hrtf}
\ee

Now it is usual to expand around $ z = 0$:
\be
&\psi( z) = \sum_{n\in Z}\psi_n z^n d z^{1/2}; \ \ \
\tilde\psi( z) = \sum_{n\in Z}\tilde\psi_n z^{-n-1}d z^{1/2};
\nn \\
&\phantom. [\tilde\psi_m,\psi_n]_+ = \delta_{m,n}; \nn  \\
&\psi_m\mid 0 \rangle = 0 \ \ {\rm for}\ m<0; \ \ \
\tilde\psi_m \mid 0 \rangle = 0 \ \ {\rm for}\ m\geq 0; \nn \\
&A( z,\tilde z) = \sum_{m,n\in Z}  z^{-m-1}\tilde z^n
A_{mn}d z^{1/2}d\tilde z^{1/2}; \nn
\ee
so that
\be
\oint_{d\tilde z}\oint_{d z}
A( z,\tilde z)\psi( z)\tilde\psi(\tilde z) =
\sum_{m,n\in Z}  A_{mn}\psi_m\tilde\psi_n.
\nn
\ee
In fact this expansion could be around $any$ point $ z_0$ and on a
2-surface of any topology: topological effects can be easily included as
specific shifts of the functional $A( z,\tilde z)$ - by combinations
of the "hadle-gluing operators". Analogous shifts can imitate the change of
basic functions $ z^n$ for $ z^{n+\alpha}$ and more complicated
expressions (holomorphic 1/2-differentials with various boundary conditions on
surfaces of various topologies).

One can now wonder, whether $local$ functionals $A( z,\tilde z) =
U( z)\delta(\tilde z- z)d z^{1/2}d\tilde z^{1/2}$
play any special role. The corresponding contribution to the Hamiltonian looks
like\footnote{
Note that normalization factor here is different by a factor of
$\frac{1}{\sqrt{2}}$ from that in discussion of discrete models in sections
2.3, 2.7 and 2.8. This is not just a change of {\it notations}, since Miwa
transformation can lead to different results when this normalization is
changed. See a footnote in section 4.6 below for more detailed discussion.
}
\be
H_{Cartan} =
\oint_{d z}U( z)\psi( z)\tilde\psi( z) =
\oint_{d z} U( z)J( z),
\ee
where
\be
J( z) = \psi( z)\tilde\psi( z) =
\sum_{n\in Z}J_n z^{-n-1}d z
\ee
is the $U(1)_{k=1}$ Kac-Moody current;
\be
J_n = \sum_{m\in Z}\psi_m\tilde\psi_{m+n}; \ \ \ [J_m,J_n] = m\delta_{m+n,0}.
\ee
If scalar function (potential) $U( z)$ is expanded as
$U( z) = \sum_{k\in Z} t_k z^k$, then
\be
H_{Cartan} =
\sum_{n\in Z} t_kJ_k.
\ee
This contribution to the whole Hamiltonian can be considered distinguished for
the following reason. Let us return to original expression (\ref{hrtf}) and
try to consider it as a generating functional for all the correlation
functions of $\tilde\psi$ and $\psi$.
Naively, variation w.r.to $A( z,\tilde z)$ should produce bilinear
combination $\psi( z)\tilde\psi(\tilde z)$ and this would solve the
problem. However, things are not just so trivial, because operators involved
do not commute (and in particular, the exponential operator in (\ref{hrtf})
should still be defined less symbolically, see next subsection). Things would
be much simpler, if we can consider $commuting$ set of operators: this is
where abelian $\hat{U(1)}_{k=1}$ subgroup of the entire $GL(\infty)_{k=1}$
(and even its purely commuting Borel subalgebra) enters the game. Remarkably,
it is sufficient to deal with this abelian subgroup in order to reproduce all
the correlation functions.\footnote{
We once again emphasize that this trick is specific for the free fermions and
for the level $k=1$ Kac-Moody algebras, which can be expressed entirely in
terms of free fields, associated with Cartan generators (modulo some
unpleasant details, related to "cocycle factors" in the Frenkel-Kac
representations \cite{FK},
which are in fact reminiscents of free fields associated with the non-Cartan
generators (parafermions) \cite{Turb}, - but can, however, be put under the
carpet or/and taken into account "by hands" as "unpleasant but
non-essential(?) sophistications).
}
The crucial point is the identity for free fermions (generalizable to any $b,
c$-systems):
\be
:\psi( \lambda)\tilde\psi(\tilde \lambda):\ = \ :\exp\left({
\int_{ \lambda}^{\tilde \lambda} J}\right):
\label{bosid}
\ee
which is widely known in the form of bosonization formulas:\footnote{
Formulas in brackets are indeed correct, before them the usual symbolic
relations are written. Using these formulas we get:
\be
:\psi(\lambda)\tilde\psi(\tilde \lambda):\  = \  :e^{
\phi(\tilde \lambda)-\phi( \lambda)}:\ = \ :e^{
\int_{ \lambda}^{\tilde \lambda} \partial\phi}: \ = \ :e^{
\int_{ \lambda}^{\tilde \lambda} J}:
\nn
\ee
This identity can be of course obtained within fermionic theory, one should
only take into account that $\psi$-operators are nilpotent, so that exponent
of a single $\psi$-operator would be just a sum of two terms (polinomial).
}
if $J( z) = \partial\phi( z)$,
\be
\tilde\psi(\tilde \lambda) \sim \ :e^{
\phi(\tilde \lambda)}:
\ \ \ \left(\ :\psi(\infty)\tilde\psi(\tilde \lambda):\ =\ :e^{
(\phi(\tilde \lambda) - \phi(\infty))}:\ \right); \nn \\
\psi( \lambda) \sim \ :e^{-
\phi( \lambda)}:
\ \ \ \left(\ :\psi( \lambda)\tilde\psi(\infty):\ =\ :e^{
(\phi(\infty) - \phi( \lambda))}:\ \right).
\nn
\ee
This identity implies that one can generate any bilinear combinations of
$\psi$-operators by variation of potential $U( z)$ only, moreover this
variation should be of specific form:
\be
\Delta\oint UJ = \Delta\left( \sum_{k\in Z} t_kJ_k \right) =
\int_{ z}^{\tilde z} J = \sum_{k \in Z}
\int_{ z}^{\tilde z}   z^{-k-1} d z = \nn \\
= \sum_{k \in Z} \frac{1}{k}J_k \left(\frac{1}{ z^k} -
\frac{1}{\tilde z^k}\right),
\nn
\ee
i.e.
\be
\Delta t_k = \frac{1}{k}\left(\frac{1}{ z^k} -
\frac{1}{\tilde z^k}\right)
\ee
Note that this is {\it not} an infinitesimal variation and that it has
exactly the
form, consistent with Miwa parametrization used in the previous section 3.

Since any bilinear combination can be generated in this way from $U( z)$,
it is clear that the entire Hamiltonian $\sum A_{mn}\tilde\psi_m\psi_n$ can be
also considered as resulting from some transformation of $V$ (i.e. of
"time-variables" $t_k$). In other words,
\be
\tau\{A\} = {\cal O}_A[t] \tau\{A = U\}.
\nn
\ee
These operators ${\cal O}_A$ are naturally interpreted as elements of the
group $GL(\infty)$, acting on the "Universal Grassmannian" \cite{SeWi},
\cite{Kac},\cite{Orlov}, parametrized by
the matrices $A_{mn}$ modulo changes of coordinates $ z \rightarrow
f( z)$. This representation for $\tau\{A\}$ is, however, not very
convenient, and usually one considers {\it infinitesimal} version of the
transformation, which just shifts $A$
\be
\tau\{t\mid A+\delta A\} = \hat{\cal O}_{\delta A}[t] \tau\{t\mid A\},
\label{trtf}
\ee
note that this transformation clearly distinguishes between the dependencies
of $\tau$ on $t$ and on all other components of $A$. The possibility of such
representation with the privileged role of Cartan generators is the origin of
all simplifications, arising in the case of free-fermion
$\tau$-functions.\footnote{
It is also the reason, why these are free-fermion $tau$-functions, that appear
in the study of ordinary integrable hierarchies: the Hamiltonian flows,
which describe evolution in different $t$-directions just commute, because
$t$'s are associated with the commuting Cartan generators of $GL(\infty)$. In
the more general situation the flows would form closed but $non-abelian$
algebra.
}

Relation (\ref{trtf}) is the basis of the orbit interpretation of
$\tau$-functions \cite{Kac}. It is also important to understand the role of
the "string equation" and othert constraints, imposed on $\tau$-functions in
the theory of matrix models. These arise as some particular subalgebras in the
set of $\hat{\cal O}$-operators, and their role is to specify particular
points $A$ in the Grassmannian, of which this subalgebra is a
stabilizer.\footnote{
This relation is straightforward in the case of Virasoro constraints, since
Virasoro algebra is just a subalgebra of the $GL(\infty)$ acting on
$\tau$-functions, and thus is a symmetry (covariance) of associated integrable
hierarchies \cite{Orlov}. $W$-constraints do not form Lie-subalgebra of this
$GL(\infty)$, they arise after certain reduction, which in turn exists in a
simple form $not$ everywhere on the Grassmannian
(in particular $W$ is {\it not} a symmetry
of entire KP hierarchy \cite{KodeVo}): here we deal with a more
sophisticated self-consistency relation, which remains to be understood in
full details (e.g., it is unknown, whether reduction exists at least at {\it
any}
Virasoro-stable point, which would significantly simplify this kind of
consideration). In fact, the entire relation between the constraints and
$\tau$-functions is not exhaustively worked out: for example, there is still
no clear and satisfactory proof, that the full set of Virasoro and/or
$W$-constraints implies that partition function is a $\tau$-function, which
would be pure algebraic and not refer to the uniqueness of solutions to the
constraints. Result, widely discussed in the literature, see \cite{FKN},
is that string equation (the lowest Virasoro constraint $L_{-1}Z = 0$), if
imposed on $Z$, which is somehow known to be the properly reduced
$\tau$-function, implies the entire set of Virasoro and $W$-constraints
(though even this proof can still have some loopholes).
}
The simplest examples are in fact provided by formulas from the section 2.3
above, where combinations of the screening charges describe $A$'s, which are
stable points of {\it discrete} Virasoro- and $W$-constraints (in the latter
case the {\it multi}-fermion system is used).

The fact that the $\tau$-function at all the points $A$ of Grassmannian can be
obtained by the group action from $\tau\{0\}$, has an implication, known as
Hirota equation. The idea \cite{Kac} is just that there are Casimir operators
in the group, which commute with the group action and thus the eigenvalue of
the Casimir operator is the same for $\tau\{A\}$ at all $A$. In the
free-fermion case the simplest example of Casimir operator is given by
\be
J_0 = \oint J = \oint\psi\tilde\psi = \sum_{n\in Z} \psi_n\tilde\psi_n.
\ee
The eigenvalue of this operator for the vacuum state $\mid 0 \rangle$ is an
infinite subtraction constant, and this makes equation
$J_0 {\cal O}_A\mid 0 \rangle = {\cal O}_A J_0 \mid 0 \rangle =
{\rm const} \cdot{\cal O}_A\mid 0 \rangle$, or
$J_0\tau\{A\} = {\rm const}\cdot\tau\{A\}$ not very interesting. However, this
operator is represented in bilinear form and in such cases the following trick
is usually usefull.

If operator, which is bilinear in generators of the algebra, $T^aT^a$,
commutes with the action of the group, so does $T^a\otimes T^a$, if the group
action on tensor product of representations is defined as
$\mid\ \rangle\otimes\mid\ \rangle \rightarrow {\cal O}_A\mid\
\rangle\otimes {\cal O}_A\mid\ \rangle$. (Indeed, then
$(T^a\otimes I + I\otimes T^a)^2$ commutes with the group action and so does
$T^a\otimes T^a = \frac{1}{2}\left( (T^a\otimes I + I\otimes T^a)^2 -
T^aT^a\otimes I - I\otimes T^aT^a\right)$.)
If further, $T^a\otimes T^a$ annihilates the product of two vacuum states:
\be
\left(T^a\otimes T^a\right)\ \mid 0 \rangle\otimes\mid 0 \rangle  = 0,
\label{annvac}
\ee
then the same equation holds for all $A$:
\be
\left(T^a\otimes T^a\right)\
\mid {\cal O}_A \rangle\otimes\mid {\cal O}_A \rangle  = 0.
\ee
Condition (\ref{annvac}) is trivially valid in our case:
\be
\sum_{n\in Z} \psi_n \mid 0 \rangle\otimes\tilde\psi_n\mid 0 \rangle  = 0,
\ee
since in every term in the sum one of the vacuum states is annihilated: the
first one of $n\geq 0$ and the second one if $n<0$.\footnote{
It is easy to verify directly that $\sum_n \psi_n \otimes \tilde\psi_n$ is
indeed a Casimir operator in the tensor product:
\be
&\phantom. [\sum_n \psi_n \otimes \tilde\psi_n, \
I\otimes \sum_{l,m} A_{lm}\psi_l\tilde\psi_m +  \sum_{l,m}
A_{lm}\psi_l\tilde\psi_m \otimes I] = \nn \\
&=
\sum_n\left( \psi_n \otimes \sum_m A_{nm}\tilde\psi_m -
\sum_l A_{ln}\psi_l\otimes \tilde\psi_n \right) =
\sum_l\sum_m A_{lm} (\psi_l\otimes \tilde\psi_m - \psi_l\otimes\tilde\psi_m)
= 0. \nn
\ee
}
Thus we obtain the relation
\be
\sum_{n\in Z} \psi_n \mid {\cal O}_A \rangle \otimes
\tilde\psi_n \mid {\cal O}_A \rangle = 0,
\ee
which can be now multiplied from the left by
\be
\langle 0 \mid \psi(\infty) e^{H_{Cartan}(t)} \otimes
\langle 0 \mid\tilde\psi(\infty)  e^{H_{Cartan}(t')}
\nn
\ee
($t'_k$ do not need to coincide with $t_k$) and after insertions of
$\psi$-operators are expressed as the shifts of times, we obtain:
\be
\sum_{n\in Z}  D_n^- \tau\{t\mid A\} \otimes D_n^+ \tau\{t\mid A\} = 0,
\label{prehirota}
\ee
where $\sum_{n\geq 0}D^{\pm}_n z^{-n} =
\exp\left(\pm \sum_{k>0} \frac{1}{k z^k}
\frac{\partial}{\partial t_k}\right)$.
This is particular form of Hirota equation \cite{Hirota},
which is often used to {\it define} $\tau$-functions, associated
with integrable hierarchies. If one takes (\ref{hrtf})
for the definition, as it is more natural to do in the general "theory of
everything" and as we did above, eq.(\ref{prehirota}) is the starting
point for the path, leading to hierarchies in conventional form of
differential equations, the Lax and pseudodifferential representations
{\it naturally} appearing on the way. We do not go along this path in these
notes.

The last remark to be made, before we proceed to more detailed formulas,
is that $\tau$-functions can be considered as determinants $Det \bar\partial$
of the $\bar\partial$-operators acting on fields with some complicated
boundary conditions (like
$\psi( z) \sim \exp\left(\sum_{k>0} t_k z^{-k}\right)$
in the simplest case of $t$-dependencies). Entire A-dependence is usually
described in this context as that on the point of the "universal module space",
which once appeared in the study of string models on Riemann surface of
arbitrary genus \cite{UMS}. From this point of view more general
$\tau$-functions are sections of the bundles over universal module space,
associated with conformal models, more sophisticated then just the theory of
free fermions (and $b,c$-systems). The WZNW model is, of course, the most
important example to be studied in this context.

The crucial feature of all the quantites, associated in this way to conformal
models is applicability of Wick theorem, reducing multipoint correlation
functions to pair correlators. In the free fermion case this is just a
consequence of quadratic form of Lagrangian, in generic situation this follows
from existence of holomorphic operator algebra, which allows to define the
correlators by fixation of monodromy properties, dictated by pairwise
collision of points. Wick theorem is the concrete source of determinant
formulas for $tau$-functions, which are used in order to establish their
relations with matrix models and other branches of string theory.

After this discussion of the context where free-fermion $\tau$-functions
can and do appear, we turn now to more detailed and exact formulas, relevant
in this particular free-fermion case.
They are mostly due to Japanese school \cite{DJKM}, though many other people
contributed to this field after it was established. The only sophisticated
part of the work with these  formulas is accurate accounting for the normal
ordering routine which will be mostly unnecessary for our purposes.
We shall mostly follow the
presentation of papers \cite{KMMOZ},\cite{GKM},\cite{Toda}.

\subsection{Basic determinant formula for the free-fermion correlator}

Let us consider the following matrix element:
\be
\tau_N\{t,\bar t\mid G\} =
\langle N \mid e^H \ G \ e^{\bar H} \mid N \rangle
\label{pretaf}
\ee
where
\be
\new
\begin{array}{c}
\psi( z) = \sum_{n\in Z}\psi_n z^n d z^{1/2}; \ \ \
\tilde\psi( z) = \sum_{n\in Z}\tilde\psi_n z^{-n-1}d z^{1/2};\\
G = \exp\left( \sum_{m,n\in Z}{A}_{mn}\psi_m\tilde\psi_n \right);  \\
H = \sum_{k>0} t_kJ_k, \ \ \ \bar H = \sum_{k>0} \bar t_kJ_{-k} \\
J( z) = \psi( z)\tilde\psi( z) =
\sum_{n\in Z}J_n z^{-n-1}d z; \ \ \ J_n =
\sum_k\psi_k\tilde\psi_{k+n};   \\
\phantom. [\tilde\psi_m,\psi_n]_+ = \delta_{m,n}; \ \ \
[J_m,J_n] = m\delta_{m+n,0};  \\
\psi_m\mid N \rangle = 0, \ \ m<N; \ \ \ \
\langle N \mid \psi_m = 0, \ \ m\geq N;  \\
\tilde\psi_m\mid N \rangle = 0, \ \ m\geq N; \ \ \ \
\langle N \mid \tilde\psi_m = 0, \ \ m< N;  \\
J_m\mid N \rangle = 0,\ \ m>0; \ \ \ \
\langle N \mid J_m = 0, \ \ m<0.
\end{array}
\label{notfermcor}
\ee
The "$N$-th vacuum" $\mid N \rangle$ is defined as the Dirac sea, filled up to
the level $N$:
\be
\mid N \rangle =
\prod_{i=N}^\infty \tilde\psi_i\mid\infty\rangle =
\prod_{i=-\infty}^{N-1} \psi_i \mid-\infty\rangle \ ; \nn \\
\langle N \mid = \langle \infty\mid \prod_{i=N}^\infty \psi_i =
\langle -\infty\mid \prod_{i=-\infty}^{N-1} \tilde\psi_i,
\ee
where the "empty" (bare) and "completely filled" vacua  are defined so that:
\be
&\tilde\psi_m \mid -\infty \rangle = 0, \ \ \ \langle -\infty \mid \psi_m = 0,
\nn \\
&\psi_m \mid \infty \rangle = 0, \ \ \ \langle \infty \mid \tilde\psi_m = 0
\ee
for $any$ $m\in Z$. For the only reason that operators $J$, $H$, $\bar H$ and
$G$ are defined so that they have usually $\tilde\psi$ at the very right and
$\psi$ at the very left, we have also:
\be
&J_m \mid -\infty \rangle = 0, \ \ \ \langle -\infty\mid  J_m = 0,
\nn \\
&G^{\pm 1} \mid-\infty \rangle = \mid-\infty \rangle; \ \ \
\langle -\infty \mid G^{\pm 1} = \langle -\infty \mid; \nn \\
&e^{\pm \bar H}  \mid-\infty \rangle = \mid-\infty \rangle; \ \ \
\langle -\infty \mid e^{\pm H} = \langle -\infty \mid.
\ee
Now we can use all these formulas to rewrite our original correlator
(\ref{pretaf}) as:
\be
\new
\begin{array}{c}
\langle N \mid e^H \ G\ e^{\bar H}\mid N \rangle =  \\
=  \langle  -\infty\mid \left(\prod_{i=-\infty}^{N-1} \tilde\psi_i\right)
                  e^H \ G\ e^{\bar H}
\left(\prod_{i=-\infty}^{N-1} \psi_i\right) \mid-\infty\rangle =  \\
= \langle  -\infty\mid e^{-H} \left(\prod_{i=-\infty}^{N-1}\tilde\psi_i\right)
                  e^H \ G\ e^{\bar H}
\left(\prod_{i=-\infty}^{N-1} \psi_i\right) e^{-\bar H}\mid-\infty\rangle =
 \\
= \langle  -\infty\mid \prod_{i=-\infty}^{N-1} \tilde\Psi_i[t]
\prod_{j=-\infty}^{N-1} \Psi_j^G[\bar t] \mid-\infty\rangle =
 \\
= {\rm Det}_{-\infty < i,j < N}
\langle  -\infty\mid \tilde\Psi_i[t] \Psi_j^G[\bar t] \mid-\infty\rangle =
 \\
= {\rm Det}_{i,j<0} {\cal H}_{i+N,j+N}.
\end{array}
\label{pretafdet}
\ee
The last two steps here were introduction of "$GL(\infty)$-rotated" fermions,
\be
\tilde\Psi_i[t] \equiv e^{-H}\psi_i e^H; \ \ \
\Psi_j[\bar t] \equiv e^{\bar H} \psi_j e^{-\bar H};\ \ \
\Psi_j^G[\bar t] \equiv G\Psi_j[\bar t] G^{-1},
\label{bigpsi}
\ee
and application of the Wick theorem to express multifermion correlation
function through pair correlators
\be
\new
\begin{array}{c}
{\cal H}_{ij}(t,\bar t) \equiv
\langle  -\infty\mid \tilde\Psi_i[t] \Psi_j^G[\bar t] \mid-\infty\rangle =
 \\ =
\langle  -\infty\mid \tilde\Psi_i[t]\ G \ \Psi_j[\bar t] \mid-\infty\rangle ,
\end{array}
\label{Hmatrcor}
\ee
(once again the fact that $G^{-1}\mid-\infty\rangle = \mid-\infty\rangle$
was used). The only non-trivial dynamical information entered through
applicability of the Wick theorem, and for that  it was crucial that all the
operators $e^H,\ e^{\bar H},\ G$ are $quadratic$ exponents, i.e. can only
modify the shape of the propagator, but do not destroy the quadratic form of
the action (fields remain $free$).
This is exactly equivalent to the statement that "heisenberg" operators
$\Psi[t]$ are just "rotations"
of $\psi$, i.e. that transformations (\ref{bigpsi}) are $linear$.

We shall now describe these transformations in  a little more explicit form.
Namely, their entire time-dependence can be encoded in terms of "Shur
polinomials" $P_n(t)$.
These are defined to have a very simple generating
function (which we already encountered many times in the theory of matrix
models):
\be
\sum_{n\geq 0} P_n(t)z^n = \exp\left({\sum_{k=1}^{\infty} t_kz^k}\right)
\ee
(i.e. $P_0 = 1,\ \ P_1 = t_1,\ \ P_2 = \frac{t_1^2}{2} + t^2$ etc.), and
satisfy the relation
\be
\frac{\partial P_n}{\partial t_k} = P_{n-k}.
\ee
Since
\be
\exp\left({\sum_{k=1}^{\infty} t_kz^k}\right) =
\prod_{k>0} \left(\sum_{n_k\geq 0}\frac{1}{n_k!}\ t_k^{n_k}z^{kn_k}\right),
\nn
\ee
Shur polinomials can be also represented as
\be
P_n(t) = \sum_{\stackrel{\{n_k\}}{\sum_{k>0} kn_k = n}}
\left(\prod_{k> 0}\frac{1}{n_k!}\ t_k^{n_k}\right).
\label{Shupoex}
\ee
Now, since
\be
e^{-B}Ae^B = A + [A,B] + \frac{1}{2!}[[A,B],B] +
\frac{1}{3!}[[[A,B],B],B] + \ldots
\nn
\ee
and
\be
\phantom. [\tilde\psi_i,J_k] = \tilde\psi_{i+k}, \ \
[ [\tilde\psi_i,J_{k_1}], J_{k_2}] = \tilde\psi_{i+k_1+k_2}, \ \ldots,
\nn
\ee
we have for every fixed $k$:
\be
e^{-t_kJ_k} \tilde\psi_i e^{t_kJ_k} =
\sum_{n_k\geq 0} \frac{t_k^{n_k}}{n_k!}\ \tilde\psi_{i+kn_k}.
\nn
\ee
It remains to note that all the harmonics of $J$ in $H = \sum_{k>0} t_kJ_k$
commute with each other, to obtain:
\be
\new
\begin{array}{c}
\tilde\Psi_i(t) = e^{-H} \tilde\psi_i e^H =
\left( \prod_{k>0} e^{-t_kJ_k}\right) \tilde\psi_i
\left( \prod_{k>0}e^{t_kJ_k}\right) =  \\
= \sum_{n\geq 0}\tilde\psi_{i+n} \left(
 \sum_{\stackrel{\{n_k\}}{\sum_{k>0} kn_k = n}}
\left(\prod_{k> 0}\frac{1}{n_k!}\  t_k^{n_k}\right)\right)
\stackrel{(\ref{Shupoex})}{=} \\
= \sum_{n\geq 0}\tilde\psi_{i+n} P_n(t) = \sum_{l\geq i}\tilde\psi_{l}
P_{l-i}(t).
\end{array}
\ee
Similarly, relation $[J_k,\psi_j] = \psi_{k+j}$ implies, that
\be
\Psi_j(\bar t) = e^{\bar H}\psi_j e^{-\bar H} =
\sum_{n\geq 0}\psi_{j+n} P_n(\bar t) = \sum_{m\geq j}\psi_{m}P_{m-j}(\bar t)
\label{bigpsishur}
\ee
and finally\footnote{
Eqs.(\ref{bigpsishur}) can be  also interpreted as representations of Shur
polinomials in terms of fermionic correlators in the bare vacuum:
\be
&P_m(\bar t) = \langle -\infty \mid \tilde\psi_{j+m}e^{\bar H} \psi_j
\mid-\infty \rangle\ ; \nn \\
&P_m( t) = \langle -\infty \mid \tilde\psi_i e^{H} \psi_{i+m}
\mid-\infty \rangle\  \nn
\ee
}
\be
&{\cal H}_{ij} = \sum_{\stackrel{l\geq i}{m\geq j}}
\langle -\infty \mid \tilde\psi_l\ G\ \psi_m \mid-\infty \rangle\
P_{l-i}(t)P_{m-j}(\bar t) = \nn \\         & =
\sum_{\stackrel{l\geq i}{m\geq j}} T_{lm} P_{l-i}(t)P_{m-j}(\bar t),
\label{HversTlm}
\ee
which implies also that
\be
\frac{\partial{\cal H}_{ij}}{\partial t_k} = {\cal H}_{i+k,j}; \nn \\
\frac{\partial{\cal H}_{ij}}{\partial\bar t_k} = {\cal H}_{i,j+k} .
\label{todaeqforH}
\ee
Matrix
\be
T_{lm} \equiv \langle -\infty \mid \tilde\psi_l\ G\ \psi_m \mid-\infty \rangle
\ee
is the one which defines fermion rotations under the action of
$GL(\infty)$-group element $G$:
\be
&G\psi_mG^{-1} = \sum_{l\in Z} \psi_lT_{lm}; \nn \\
&G^{-1}\tilde\psi_l G = \sum_{m\in Z} T_{lm}\tilde\psi_m,\ {\rm or}\
G\tilde\psi_l G^{-1} = \sum_{m\in Z} (T^{-1})_{lm}\tilde\psi_m.
\ee
If $G=1$, $T_{lm} = \delta_{lm}$. If all $t_k = \bar t_k = 0$,
${\cal H}_{ij} = T_{ij}$.

\subsection{Toda-lattice $\tau$-function and linear reductions of
Toda hierarchy}

In the previous subsection we derived a formula
\be
\tau_N\{t,\bar t\mid G\} = \left.{\rm Det}\right._{i,j<0} {\cal H}_{i+N,j+N}
\label{todatau}
\ee
for the basic correlator, which defines
"Toda-lattice $\tau$-function". For obvious reasons $\bar t$ are often refered
to as negative-times. $\tau$-function
can be normalized by dividing over the same
quantity for all time-variables vanishing, but this is not always convenient.
Eq.(\ref{todatau}) has generalizations - when similar
matrix elements in a multifermion system is considered - this leads to
"multicomponent Toda" (or AKNS) $\tau$-functions. Generalizations to
arbitrary conformal models should be considered as well. It has also particular
"reductions", of which the most important are: KP (Kadomtsev-Petviashvili),
forced (semi-infinite) and Toda-chain $\tau$-functions.
This is the subject to be discussed in this subsection.

Idea of linear reduction is that the form of operator $G$, or, what is
the same, of
the matrix $T_{lm} $ in eq.(\ref{HversTlm}), can be adjusted in such a way,
that $\tau_N\{t,\bar t\mid G\}$ becomes independent of some variables, i.e.
equation(s)
\be
\left(\sum_k \alpha\frac{\partial}{\partial t_k} +
 \sum_k \bar\alpha\frac{\partial}{\partial \bar t_k} +
\sum_k \beta_k D_N(k) + \gamma \right) \tau_N\{t,\bar t\mid G\} = 0
\label{redtotau1}
\ee
can be solved as equations for $G$ for all the values of $t,\bar t$ and $N$ at
once. (In (\ref{redtotau1}) $D_N(k)f_N \equiv f_{N+k} - f_N$.)
In this case the system of integrable equations (hierarchy), arising from
Hirota equation for $\tau$, gets reduced and one usually speaks about "reduced
hierarchy". Usually equation (\ref{redtotau1}) is imposed directly on matrix
${\cal H}_{ij}$, of course than (\ref{redtotau1}) is just a corollary.

We shall refer to the situation when (\ref{redtotau1}) is fulfilled for {\it
any} $t, \bar t, N$  as to "strong reduction". It is often reasonable to
consider also "weak reductions", when (\ref{redtotau1}) is satisfied on
particular infinite-dimensional hyperplanes in the space of time-variables.
Weak reduction is usually a property of entire $\tau$-function as well, but
not expressible in the from of a local linear equation, satisfied identicall
for {\it all} values of $t, \bar t, N$.
Now we proceed to concrete examples:

{\it Toda-chain hierarchy}. This is a {\it strong} reduction. The
corresponding constraint (\ref{redtotau1}) is just
\be
\frac{\partial {\cal H}_{ij}}{\partial t_k} = \frac{\partial {\cal
H}_{ij}}{\partial \bar t_k},
\ee
or, because of (\ref{todaeqforH}), ${\cal H}_{i+k,j} = {\cal H}_{i,j+k}$.
It has an obvious solution:
\be
{\cal H}_{i,j} = \hat{\cal H}_{i+j},
\ee
i.e. ${\cal H}_{ij}$ is expressed in terms of a one-index quantity $\hat{\cal
H}_i$. It is, however, not enough to say, what are restrictions on ${\cal
H}_{ij}$ - they should be fulfiled for all $t$ and $ \bar t$ at once, i.e.
should be resolvable as equations for $T_{lm}$. In the case under
consideration this is simple: $T_{lm}$ should be such that
\be
T_{lm} = \hat T_{l+m}.
\ee
Indeed, then
\be
&{\cal H}_{ij} = \sum_{l,m} T_{lm} P_{l-i}(t) P_{m-j}(\bar t) =
\sum_{l,m} \hat T_{l+m} P_{l-i}(t)P_{m-j}(\bar t) = \nn \\
&= \sum_{n\geq 0} \hat T_{n+i+j} \left(\sum_{k=0}^n P_k(t)P_{n-k}(\bar
t)\right),
\nn
\ee
and
\be
\hat{\cal H}_i = \sum_{n\geq 0} \hat T_{n+i} \left(\sum_{k=0}^n
P_k(t)P_{n-k}(\bar t)\right).
\label{todachainH}
\ee

{\it Volterra hierarchy}. Toda-chain $\tau$-function can be further {\it
weakly}
reduced to satisfy the identity
\be
\left.\frac{\partial \tau_{2N}}{\partial t_{2k+1}}\right|
_{\{t_{2l+1}=0\}}  = 0, \ \ \ {\rm for\ all}\ k,
\label{volterrared}
\ee
i.e. $\tau_{2N}$ is requested to be even function of all odd-times $t_{2l+1}$
(this is an example of "global characterization" of the weak reduction).
Note that (\ref{volterrared}) is imposed only on {\it Toda-chain}
$\tau$-function with {\it even} values of zero-time. Then (\ref{volterrared})
will hold whenever $\hat{\cal H}_i$ in (\ref{todachainH}) are even (odd)
functions of $t_{\rm odd}$ for even (odd) values of $i$. Since Shur
polinomials $P_k(t)$ are even (odd) functions of odd-times for even (odd) $k$,
it is enough that the sum in (\ref{todachainH}) goes over even (odd) $n$
when $i$ is even (odd). In other words, the restriction on $T_{lm}$ is that
\be
T_{lm} = \hat T_{l+m}, \ \ \ {\rm and} \ \ \ \hat T_{2k+1} = 0 \ \ {\rm for\
all}\ k.
\ee

{\it Forced hierarchies}. This is another important example of strong
reduction. It also provides an example of {\it singular}
$\tau$-functions,
arising when $G = \exp \left(\sum A_{mn}\psi_m\tilde\psi_n\right)$
blows up and normal ordered operators should be used to define
regularized
$\tau$-functions. Forced hierarchy appears when $G$ can be
represented in the form \cite{KMMOZ}  $G = G_0P_+$, where
projection operator $P_+$ is such that
\be
P_+ \mid N \rangle = \mid N \rangle \ \ {\rm for}\ N\geq N_0,
\nn \\
P_+ \mid N \rangle = 0 \ \ {\rm for}\ N < N_0.
\label{Pplus}
\ee
Explicit expression for this operator is\footnote{
Normal ordering sign $\ : \ \ :\ $ means that all operators
$\tilde\psi$ stand to the {\it left} of all operators $\psi$.
The product at the r.h.s. obviously implies both the property
(\ref{Pplus}) and projection property $P_+^2 = P_+$.
}
\be
P_+ = \ :\exp \left( - \sum_{l<N_0} \tilde\psi_l\psi_l\right): \ =
\prod_{l<N_0} (1 - \tilde\psi_l\psi_l) =
\prod_{l < N_0} \psi_l\tilde\psi_l.
\nn
\ee
Because of (\ref{Pplus}), $P_+\mid -\infty \rangle = )$, and
the identity $G\mid -\infty \rangle = \mid -\infty \rangle$,
which was essentially used in the derivation in (4.27), can be
satisfied only if $G_0$ is singular and $T_{lm} = \infty$.
In order to avoid this problem one usually introduces in the
vicinity of such
singular points in the universal module space a sort of normalized
(forced) $\tau$-function $\tau_N^f \equiv \frac{\tau_N}{\tau_{N_0}}$.
One can check that now $T^f_{lm} = \infty$ for all $l,m < N_0$, and
$\tau^f$ can be represented as determinant of a final-dimensional
matrix \cite{UT},\cite{KMMOZ}:
\be
&\tau_N^f = {\rm Det}_{N_0\leq i,j < N} {\cal H}^f_{ij}\ \ \ {\rm for}\ \
N>N_0;
\nn \\
&\tau_{N_0}^f = 1; \nn \\
&\tau_{N}^f = 0\ \ \ {\rm for }\ \ N < N_0.
\ee
For $N>N_0$ we have now determinant of a {\it finite}-dimensional
$(N-N_0)\times (N-N_0)$ matrix.
The choice of $N_0$ is not really essential, therefore it is better to put
$N_0 = 0$ in order to
simplify formulas, phraising  and relation with the discrete matrix models
($N_0$ is easily restored if everywhere $N$ is substituted by $N-N_0$).
For forced hierarchies one can also represent $\hat\tau$ as
\be
\tau_N^f = {\rm Det}_{0\leq i,j < N}
\partial_1^i\bar\partial_1^j {\cal H}^f,
\ee
where ${\cal H}^f = {\cal H}^f_{00}$ and $\partial_1 = \frac{\partial}{\partial
t_1}$, $\bar \partial_1 = \frac{\partial}{\partial \bar t_1}$. For
{\it forced Toda-chain} hierarchy this turns into even simpler expression:
\be
\tau_N^f = {\rm Det}_{0\leq i,j < N} \partial_1^{i+j}\hat{\cal H}^f,
\ee
while for the {\it forced Volterra} case we get a product of two Toda-chain
$\tau$-functions with twice as small value of $N$ \cite{Bowick}:
\be
\tau_{2N}^f &= \left({\rm Det}_{0\leq i,j < N} \partial_2^{i+j}\hat{\cal
H}^f\right) \cdot
       \left( {\rm Det}_{0\leq i,j < N} \partial_2^{i+j} (\partial_2
\hat{\cal H}^f)\right) = \nn \\
&= \tau_N^f[\hat{\cal H}^f]\cdot \tau_N^f[\partial_2\hat{\cal H}^f].
\ee

Forced $\tau^f_N$ can be {\it always} represented in the form of a
scalar-product matrix model. Indeed,
\be
{\cal H}_{ij} = \sum T_{lm}P_{l-i}(t)P_{m-j}(\bar t) =
\oint\oint e^{U(h)+\bar U(\bar h)} h^i\bar h^j T(h,\bar h) dhd\bar h,
\ee
where $T(h,\bar h) \equiv \sum_{lm} T_{lm} h^{-l-1}\bar h^{-m-1}$, and
$e^{U(h)} = e^{\sum_{k>0}t_kh^k} = \sum_{l\geq 0} h^lP_l(t)$.
Then, since ${\rm Det}_{0\leq i,j < N}h^i = \Delta_N(h)$ - this is where it
is essential that the hierarchy is forced -
\be
{\rm Det}_{0\leq i,j < N} {\cal H}_{ij} = \prod_i
\oint\oint e^{U(h_i)+\bar U(\bar h_i)}
T(h_i,\bar h_i)dh_id\bar h_i
\cdot \Delta_N(h)\Delta_N(\bar h),
\ee
i.e. we obtain a scalar-product model with
\be
d\mu_{h,\bar h} = e^{U(h)+\bar U(\bar h)}T(h,\bar h) dhd\bar h.
\ee
Inverse is also true: partition function of every scalar-product model is
forced Toda-lattice $\tau$-function - see section 4.7 for more details.

{\it KP hierarchy}.  In this case we just ignore the  dependence of
$\tau$-function on times $\bar t$. Every Toda-lattice $\tau$-function can be
considered also as KP $\tau$-function: just operator $G^{KP} \equiv Ge^{\bar
H}$ (a point of Grassmannian) becomes $\bar t$-dependent. Usually
$N$-dependence is also eliminated - this can be considered as a little more
sophisticated change of $G$. When $N$ is fixed, extra changes of
field-variables are allowed, including transformation from Ramond to
Neveu-Schwarz sector etc. Often KP hierarchy is from the very beginning
formulated in terms of Neveu-Schwarz (antiperiodic) fermionic fields
(associated with  principal
representations of Kac-Moody algebras), i.e. expansions in the first line of
(\ref{notfermcor}) are in semi-integer powers of $z$:
$\psi_{NS}( z) = \sum_{n\in Z}\psi_n z^{n-\frac{1}{2}} d z^{1/2}$.

Given a KP $\tau$-function one can usually construct a Toda-lattice one with
the {\it same} G, by introducing in appropriate way dependencies on $\bar t$
and $N$. For this purpose $\tau^{KP}$ should be represented in the form of
(\ref{todatau}):
\be
\tau^{KP}\{t\mid G\} = {\rm Det}_{i,j<0} {\cal H}_{ij}^{KP},
\label{addkptau1}
\ee
where ${\cal H}_{ij}^{KP} = \sum_l T_{lj} P_{l-i}(t)$. Since $T_{lm} $ is a
function of $G$ only, it does not change when we built up a Toda-lattice
$\tau$-function:
\be
&\tau_N\{t,\bar t\mid G\} = {\rm Det}_{i,j<0} {\cal H}_{i+N,j+N}; \nn \\
&{\cal H}_{ij} = \sum_{l,m}T_{lm} P_{l-i}(t)P_{m-j}(\bar t) =
\sum_m {\cal H}_{im}^{KP} P_{m-j}(\bar t).
\ee
Then
\be
\tau^{KP}\{ t\mid G\} = \tau_0\{t,0\mid G\}.
\ee
If we go in the opposite direction, when Toda-lattice $\tau$-function is
considered as KP $\tau$-function,
\be
\tau_0\{t,\bar t\mid G\} = \tau^{KP}\{t \mid \tilde G(\bar t)\}; \nn \\
\tilde {\cal H}^{KP}_{ij} = \sum_m {\cal H}_{im}P_{m-j}(\bar t)\ \
{\rm and} \nn \\
\tilde T_{lj}\{\tilde G(\bar t)\} = \sum_m T_{lm}\{G\}P_{m-j}(\bar t).
\label{addkptau2}
\ee

KP reduction in its turn has many further weak reductions (KdV and  Boussinesq
being the simplest examples). We shall mention them again in section 4.9
below, after Miwa transformation of representation (\ref{todatau}) will
be considered in the next subsection.

\subsection{Fermion correlator in Miwa coordinates}

Let us now return to original correlator (\ref{pretaf}) and
discuss in a little more details the implications of bosonization
identity (\ref{bosid}). In order not to write down integrals of $J$, we
introduce scalar field:\footnote{
One can consider $\phi $ as introduced for simplicity of notation,
but it should be kept in mind that the scalar-field
representation is in fact more fundamental for {\it generic} $\tau$-functions,
not related to the level $k=1$ Kac-Moody algebras (this phenomenon is well
known in conformal filed theory, see \cite{GMMOS} for more details).
}
\be
\phi(z) = \sum_{\stackrel{k\neq 0}{k\in Z-0}}\frac{J_{-k}}{k}z^k
+ \phi_0 + J_0\log z,
\ee
such that $\partial\phi(z) = J(z)$. Then (\ref{bosid}) states that:
\be
:\psi( \lambda)\tilde\psi(\tilde \lambda):\ = \
:e^{\phi(\tilde \lambda)-\phi( \lambda)}:
\label{bosid'}
\ee
"Normal ordering" here means nothing more but the requirement to neglect all
mutual contractions (or correlators) of operators in between $:\ \ :$ when
Wick theorem is applied to evaluate corrletion functions.
One can also get rid of the normal ordering sign at the l.h.s. of
(\ref{bosid'}), then
\be
\psi( \lambda)\tilde\psi(\tilde \lambda) = \
:e^{\phi(\tilde \lambda)}:\  :e^{-\phi( \lambda)}:
\label{bosid"'}
\ee
In distinguished coordinates on a sphere, when the free field propagator is
just $\log( z-\tilde z)$, one also has:
\be
\psi( z)\tilde\psi(\tilde z) =
\frac{1}{ z-\tilde z}\ :\psi( z)\tilde\psi(\tilde z):
\nn
\ee

Our task now is to express operators $e^H$ and $e^{\bar H}$ through the field
$\phi$. This is simple:
\be
H = \oint_0 U( z)J( z) = \oint_0 U( z)\partial\phi( z) =
- \oint_0 \phi( z)\partial U( z).
\ee
Here as usual $U( z) = \sum_{k>0} t_k z^k$ and integral is around
$ z = 0$. This is very similar to generic linear functional of
$\phi_-(\lambda) \equiv -\sum_{k>0}\frac{1}{k}J_k\lambda^{-k}$,
\be
H = \int \phi_-(\lambda)f(\lambda) d\lambda,
\label{hamneg}
\ee
one should only require that\footnote{
As it is usual nowadays, a factor of $2\pi i$ is assumed to be included into
the definition of contour integral $\oint$.
}
\be
\partial U( z) = \int \frac{f(\lambda)}{ z - \lambda}d\lambda,
\nn
\ee
i.e.
\be
U( z) = \int\log\left(1-\frac{ z}{\lambda}\right) f(\lambda)d\lambda.
\ee
In terms of time-variables this means that
\be
t_k = -\frac{1}{k}\int \lambda^{-k}f(\lambda)d\lambda.
\label{miwatransint}
\ee
Here we required that $U( z = 0) = 0$, sometimes it can be more natural
to introduce also
\be
t_0 = \int \log \lambda\ f(\lambda)d\lambda.
\ee
This change from the time-variables to "time density" $f(\lambda)$ is known as
Miwa
transformation. In order to establish relation with fermionic representation
and also with matrix models we shall need it in "discretized" form:
\be
t_k &= \frac{\xi}{k}\left(\sum_{\gamma} \lambda_\gamma^{-k} -
\sum_\gamma \tilde \lambda_\gamma^{-k} \right), \nn \\
t_0 &= -\xi\left(\sum_\gamma \log \lambda_\gamma  - \sum_\gamma\log \tilde
\lambda_\gamma\right).
\label{miwatrans}
\ee
We changed integral over $\lambda$ for a discrete sum
(i.e. the density function $f(\lambda)$ is
a combination of $\delta$-functions, picked at some points
$\lambda_\gamma,\ \tilde \lambda_\gamma$. This is
of course just another basis in the space of the linear functionals, but the
change from one basis to another one is highly non-trivial. The thing is, that
we selected the basis where amplitudes of different $\delta$-functions are the
{\it same}: parameter $\xi$ in (\ref{miwatrans}) is {\it independent} of
$\gamma$. Thus the real parameters are just positions of the points
$\lambda_\gamma, \ \tilde \lambda_\gamma$,
while the amplitude is defined by the density
of these points in the integration (summation) domain. This domain does
not need to be {\it a priori} specified:
it can be real line, any other contour or - better -
some  Riemann surface.)
Parameter $\xi$ is also unnecessary to introduce, because basises with
different $\xi$ are essentially equivalent. We shall soon put it equal to {\it
one}, but not before Miwa transformation will be discussed in a little more
detail.

Our next steps will be as follows.
Substitution of (\ref{hamneg}) into (\ref{miwatrans}), gives:
\be
H = -\xi\sum_\gamma \phi_-(\lambda_\gamma) +
\xi\sum_\gamma\phi_-(\tilde\lambda_\gamma).
\ee
In fact, what we need is not operator $H$ itself, but the state which is
created when $e^H$ acts on the vacuum state $\langle N\mid$. Then, since
$\langle N\mid J_m = 0$ for $m<0$, $\langle N\mid e^{-\xi\phi_-(\lambda)}$ is
essentially equivalent to $\langle N\mid e^{-\xi\phi(\lambda)}$ with
$\phi_-(\lambda)$ substituted by entire $\phi(\lambda)$. If $\xi = 1 $,
$e^{-\phi(\lambda)}$ can be further changed for $\psi(\lambda)$ and we
obtain an expression for the correlator (\ref{pretaf}) an expression where
$e^H$ is substituted by a product of operators $\psi(\lambda_\gamma)$. The
same is of course true for $e^{\bar H}$. Then Wick therem can be applied and a
new type of determinant formulas arises like, for example,
\be
\tau \sim \frac{\Delta(\lambda,
\tilde\lambda)}{\Delta^2(\lambda)\Delta^2(\tilde\lambda)}
{\rm det}_{\gamma\delta} \langle N\mid \psi(\lambda_\gamma)
\tilde\psi(\tilde\lambda_\delta)\ G \ \mid N\rangle
\ee
It can be also obtained directly from (\ref{pretafdet}), (\ref{Hmatrcor}) and
(\ref{HversTlm}) by Miwa transformation.
The rest of this subsection describes this derivation in somewhat more
details.

The first task is to substitute $\phi_-$ by $\phi$. For this purpose we
introduce operator
\be
\sum_{k=-\infty}^{\infty} t_kJ_k = H_+ + H_-,
\ee
where $H_+ = \sum_{k>0} t_kJ_k$ is just our old $H$, $H_- = \sum_{k\geq 0}
t_{-k}J_k$, and "negative times" $t_{-k}$ are defined by "analytical
continuation" of the same formulas (\ref{miwatransint}) and (\ref{miwatrans}):
\be
t_{-k} = \frac{1}{k} \int \lambda^k f(\lambda)d\lambda =
-\frac{\xi}{k} \left(\sum_\gamma \lambda_\gamma^k - \sum_\gamma
\tilde\lambda^k_\gamma \right).
\ee
Then
\be
\sum_{k=-\infty}^{\infty} t_kJ_k = H_+ + H_- =
-\xi\left(\sum_\gamma \phi(\lambda_\gamma) - \sum_\gamma
\phi(\tilde\lambda_\gamma) \right).
\label{miwadetder1}
\ee
Further,
\be
e^{H_+ + H_-} = e^{-\frac{1}{2}s(t)} e^{H_+}e^{H_-} =
      e^{\frac{1}{2}s(t)} e^{H_-}e^{H_+},
\ee
where
\be
\new
\begin{array}{c}
s(t) \equiv \sum_{k>0} kt_kt_{-k} =
-\xi^2 \sum_{k>0}\frac{1}{k} \left(\sum_\gamma
\left(\lambda_\gamma^{-k} - \tilde\lambda_\gamma^{-k}\right)
\sum_\delta \left(\lambda_\delta^k - \tilde\lambda_\delta^k \right)\right) = \\
= \xi^2 \log \left(\left. \prod_{\gamma,\delta}\right.^\prime\
\frac{(1-\frac{\lambda_\delta}{\lambda_\gamma})
(1-\frac{\tilde\lambda_\delta}{\tilde\lambda_\gamma})}
{(1-\frac{\tilde\lambda_\delta}{\lambda_\gamma})
(1-\frac{\lambda_\delta}{\tilde\lambda_\gamma})}\right) \ + \ {\rm const},
\end{array}
\ee
where prime means that the terms with $\gamma = \delta$ are excluded from the
product in the numerator and accounted for in the infinite "constant", added
at the r.h.s. In other words,
\be
e^{\frac{1}{2}s(t)} &= {\rm const}\cdot
\left(
\frac{\prod_{\gamma > \delta} (\lambda_\gamma - \lambda_\delta)
(\tilde\lambda_\gamma - \tilde\lambda_\delta)}
{\prod_\gamma \prod_\delta (\lambda_\gamma - \tilde\lambda_\delta)}
\right)^{\xi^2} =  \nn \\ &=
{\rm const}\cdot
\left(
\frac{\Delta^2(\lambda) \Delta^2(\tilde\lambda)}{\Delta(\lambda,
\tilde\lambda)}
\right)^{\xi^2}.
\label{miwadetder2}
\ee
Since $\ \langle N\mid J_m = 0\ $  for all $m<0$, we have
$\langle N\mid e^{H_-} = \langle N\mid$, and therefore
\be
\langle N\mid e^H \equiv \langle N\mid e^{H_+} =
\langle N\mid e^{H_-}e^{H_+} =
e^{-\frac{1}{2}s(t)}\langle N\mid e^{H_+ + H_-}.
\ee
{}From eq.(\ref{miwadetder1}),
\be
e^{H_+ + H_-} = {\rm const}\cdot \prod_\gamma \ :e^{-\xi\phi(\lambda_\gamma)}:
\ :e^{\xi\phi(\tilde\lambda_\gamma)}:
\ee
where "const" is exactly the same as in (\ref{miwadetder2}).
If $\xi = 1$, eq.(\ref{bosid"'}) can be used to write:\footnote{
The choice of $\xi$ can be dictated by particular purposes. Here we impose the
requirement on Miwa transform to represent $e^H = e^{H_{Cartan}}$ as a product
of dimension-1/2 operators - this is
most natural from the point of view of Hirota equations and simplifies the
relation with integrable hierarchies. However, in section 2.7 and 2.8 we used
another requirement (and there $\xi = \frac{1}{\sqrt{2}}$ rather than $\xi =
1$). There the 1-matrix model was considered, which is characterized by
especially simple form of the $full$ hamiltonian (product of dimension-zero
operators), and it was more  important to adjust
operators, which arise from $e^{H_{Cartan}}$ after Miwa transform to have
simple correlators with $e^{A\psi\tilde\psi}$. When analyzing the 1-matrix
model from this point of view one should also keep in mind that it was
actually represented in s.2.3 in terms of {\it two} complex fermions.
Screening charges are $Q^{(+)} = \oint e^{\sqrt{2}\phi} =
\oint\tilde\psi_1\psi_2 = \oint e^{\phi_1-\phi_2},\ \
Q^{(-)} = \oint e^{-\sqrt{2}\phi} = \oint\tilde\psi_2\psi_1 =
\oint e^{\phi_2 - \phi_1}$, while $\phi = \frac{1}{\sqrt{2}}(\phi_1 - \phi_2)$.
The Hamiltonian is $H_{Cartan} = \frac{1}{\sqrt{2}}\sum_k t_kJ_k =
\frac{1}{2} \sum_k t_k(J_k^1 - J_k^2)$, and Miwa transformation generators
insertions of operators $\chi_1\tilde\chi_2$, where
$\chi_1$ and $\tilde\chi_2$
have dimension 1/8 (rather than 1/2 as in the one (complex)-fermion system,
considered in $this$ section).
}
\be
\langle N\mid e^H \ = \
\frac{\Delta(\lambda,\tilde\lambda)}
{\Delta^2(\lambda) \Delta^2(\tilde\lambda)}
\langle N \mid \prod_\gamma \psi(\lambda_\gamma)
\prod_\gamma \tilde\psi(\tilde\lambda_\gamma)
\ee
Similarly,
\be
e^{\bar H} \mid N \rangle =
\prod_\delta \psi(\bar\lambda_\delta)
\prod_\delta \tilde\psi(\tilde{\bar\lambda}_\delta)
\mid N \rangle
\frac{\Delta(\bar\lambda,\tilde{\bar\lambda})}
{\Delta^2(\bar\lambda) \Delta^2(\tilde{\bar\lambda})},
\ee
where
\be
\bar t_k = - \frac{1}{k} \sum_\delta \left(\bar\lambda_\delta^k -
                  \tilde{\bar\lambda}_\delta^k\right)
\ee
and we used the fact that $J_m \mid N \rangle = 0 $ for all $m>0$.
Finaly,
\be
\new
\begin{array}{c}
\tau_N\{t, \bar t \mid G\} =
\langle N \mid e^H \ G \ e^{\bar H} \mid N \rangle =
\frac{\Delta(\lambda,\tilde\lambda)}
{\Delta^2(\lambda) \Delta^2(\tilde\lambda)}
\frac{\Delta(\bar\lambda,\tilde{\bar\lambda})}
{\Delta^2(\bar\lambda) \Delta^2(\tilde{\bar\lambda})}
\cdot   \\ \cdot
\langle N \mid  \prod_\gamma \psi(\lambda_\gamma)
\prod_\gamma \tilde\psi(\tilde\lambda_\gamma)\ G\
\prod_\delta \psi(\bar\lambda_\delta)
\prod_\delta \tilde\psi(\tilde{\bar\lambda}_\delta)
\mid N \rangle .
\end{array}
\label{tautodamiwdet}
\ee
Singularities at the coinciding points are completely eliminated from this
expression, since poles and zeroes of the correlator are canceled by those
coming from the Van-der-Monde determinants.

Let us now put $N=0$ and define normalized $\tau$-function
\be
\hat \tau_0\{t,\bar t \mid G\} \equiv
\frac{\tau_0\{t,\bar t\mid G\}}{\tau_0\{0,0\mid G\}},
\ee
i.e. divide r.h.s. of (\ref{tautodamiwdet}) by $\langle 0 \mid  G  \mid
0 \rangle$.
Wick theorem now allows to rewrite the correlator at the r.h.s.
as a determinant of the block matrix:
\be
\new
{\rm det}\left(\begin{array}{cc}
\frac{\langle 0 \mid \psi(\lambda_\gamma)
\tilde\psi(\tilde\lambda_\delta)\ G \
\mid 0 \rangle}{\langle 0 \mid \ G \ \mid 0 \rangle}
  & \frac{\langle 0 \mid \psi(\lambda_\gamma)\ G \
\tilde\psi(\tilde{\bar\lambda}_\delta) \mid 0 \rangle}{\langle 0 \mid \ G \
\mid 0 \rangle} \\
-\frac{\langle 0 \mid \tilde\psi(\tilde\lambda_\delta)\ G \
\psi(\bar\lambda_\gamma) \mid 0 \rangle}{\langle 0 \mid \ G \ \mid 0 \rangle}
  & \frac{\langle 0 \mid \ G\ \psi(\bar\lambda_\gamma)
\tilde\psi(\tilde{\bar\lambda}_\delta) \mid 0 \rangle}{\langle 0 \mid \ G \
\mid 0 \rangle}
\end{array} \right)
\label{blockform}
\ee

Special choices of points $\lambda_\gamma, \ldots,
\tilde{\bar\lambda}_\delta$ can lead to simpler formulas.
If $\tilde{\bar\lambda}_\gamma \rightarrow {\bar\lambda}_\gamma$, so that
$\bar t_k \rightarrow 0$, the matrix elements at the right lower block in
(\ref{blockform}) blow up, so that the off-diagonal blocks can be neglected.
Then
\be
\new
\begin{array}{c}
\tau_0\{t,\bar t \mid G\} \rightarrow \tau^{KP}\{t \mid G\} =
\frac{\langle 0 \mid e^H \ G \ \mid 0 \rangle}{\langle 0 \mid \ G \ \mid 0
\rangle} =   \\
= \frac{\Delta(\lambda,\tilde\lambda)}
{\Delta^2(\lambda) \Delta^2(\tilde\lambda)}
{\rm det}_{\gamma\delta} \frac{\langle 0 \mid   \psi(\lambda_\gamma)
\tilde\psi(\tilde\lambda_\delta)\ G\ \mid 0 \rangle}{\langle 0 \mid \ G \ \mid
0 \rangle} .
\end{array}
\label{kpdetrep}
\ee
This function no longer depends on $\bar t$-times and  is
just a KP $\tau$-function.

Matrix element
\be
\varphi(\lambda,\tilde\lambda) = \frac{\langle 0 \mid \psi(\lambda)\tilde\psi(
\tilde\lambda)\  G\ \mid 0 \rangle}{\langle 0 \mid G \mid 0 \rangle}
\ee
is singular, when $\lambda \rightarrow \tilde\lambda$:
$\varphi(\lambda,\tilde\lambda) \rightarrow \frac{1}{\lambda - \tilde\lambda}$.
If now in (\ref{kpdetrep}) all $\tilde\lambda \rightarrow \infty$,
\be
\tau^{KP}\{t \mid G\} =
\frac{{\rm
det}_{\gamma\delta}\varphi_\delta(\lambda_\gamma)}{\Delta(\lambda)},
\label{KPdetmain}
\ee
where
\be
\varphi_\delta(\lambda) \equiv \langle 0 \mid \psi(\lambda)
\left(\partial^{\delta-1}\tilde\psi\right)(\infty)\
G \ \mid 0 \rangle \sim \lambda^{\delta
-1} \left(1 + {\cal O}\left(\frac{1}{\lambda}\right)\right).
\ee
This is the main determinant representation of KP $\tau$-function in Miwa
parametrization.

Starting from representation (\ref{KPdetmain}) one can restore the
corresponding matrix ${\cal H}^{KP}_{ij}$ in eq.(\ref{addkptau1}) \cite{Toda}:
\be
{\cal H}^{KP}_{ij}\{t\} = \oint z^i\varphi_{-j}(z) e^{\sum_k t_kz^k} dz,
\label{addkptau3}
\ee
i.e.
\be
T_{lj}^{KP} = \oint z^l\varphi_{-j}(z).
\ee
Then obviously $\displaystyle{\frac{\partial {\cal H}^{KP}_{ij}}{\partial t_k}
= {\cal H}^{KP}_{i+k,j}}$. Now we need to prove that the $\tau$-function is
given at once by $\frac{{\rm det}\
\varphi_\gamma(\lambda_\delta)}{\Delta(\lambda)}$ and ${\rm Det} {\cal
H}^{KP}_{ij}\{t\}$. In order to compare these two expressions one should take
$t_k = \frac{1}{k}\sum_\gamma^n \lambda_\gamma^{-k}$, so that
\be
\exp\left({\sum_{k>0} t_kz^k}\right) = \prod_{\gamma =1}^n
\frac{\lambda_\gamma}{\lambda_\gamma - z} =
\left(\prod_\gamma^n \lambda_\gamma\right)
\sum_\gamma \frac{(-)^\gamma}{z-\lambda_\gamma}
\frac{\Delta_\gamma(\lambda)}{\Delta(\lambda)},
\ee
where
\be
\Delta_\gamma(\lambda) = \prod_{\stackrel{\alpha>\beta}{\alpha,\beta\neq
\gamma}}(\lambda_\alpha - \lambda_\beta) =
\frac{\Delta(\lambda)}{\prod_{\alpha\neq \gamma}(\lambda_\alpha -
\lambda_\gamma)},
\ee
and
\be
\left.{\cal H}^{KP}_{ij}\right|_
{t_k = \frac{1}{k}\sum_\gamma^n \lambda_\gamma^{-k}} =
\left(\prod_\gamma^n \lambda_\gamma\right)
\sum_\gamma \frac{(-)^{\gamma+1}\Delta_\gamma(\lambda)}{\Delta(\lambda)}
\lambda_\gamma^i\varphi_{-j}(\lambda_\gamma).
\label{derivkpvertod1}
\ee
As far as $n$ is kept finite, determinant of the infinite-size matrix
(\ref{derivkpvertod1}),
$\displaystyle{\left.{\rm Det}_{i,j<0}{\cal H}^{KP}_{ij}\right|_{t_k =
\frac{1}{k}\sum_\gamma^n \lambda_\gamma^{-k}} = 0}$
since it is obvious from (\ref{derivkpvertod1}) that
the rank of the matrix is equal to $n$. Therefore let us consider the maximal
non-vanishing determinant,
\be
\new
\begin{array}{c}
\left.{\rm Det}_{-n\leq i,j<0}{\cal H}^{KP}_{ij}\right|_{t_k =
\frac{1}{k}\sum_\gamma^n \lambda_\gamma^{-k}} = \\ =
\left(\prod_\gamma^n \lambda_\gamma\right)^n
{\rm det}_{i\gamma} \left(
\frac{(-)^{\gamma+1}\Delta_\gamma(\lambda)}{\lambda_\gamma^i\Delta(\lambda)}
\right)
\cdot {\rm det}_{\gamma j}\varphi_{j}(\lambda_\gamma)
=  \\ =
\frac{{\rm det}_{\gamma j}\varphi_{j}(\lambda_\gamma)}{\Delta(\lambda)}.
\end{array}
\label{derivkpvertod2}
\ee
We used here the fact that determinant of a matrix is a product of
determinants and reversed the signs of $i$ and $j$. Also used were some simple
relations:
\be
\new
\begin{array}{lc}
&\prod_{\gamma=1}^n \frac{\Delta_\gamma(\lambda)}{\Delta(\lambda)} =
\frac{1}{\Delta^2(\lambda)}, \nn \\
&{\rm det}_{i\gamma}\frac{1}{\lambda_\gamma^i} = \left(\prod_\gamma^n
\lambda_\gamma\right)^{-1}\Delta(1/\lambda), \\
&\Delta(1/\lambda)  = \prod_{\alpha>\beta}\left(\frac{1}{\lambda_\alpha} -
\frac{1}{\lambda_\beta}\right) = (-)^{n(n-1)/2} \Delta(\lambda)
\left(\prod_\gamma^n\lambda_\gamma\right)^{-(n-1)},  \\
{\rm thus}& \\
&\left(\prod_\gamma^n\lambda_\gamma\right)(-)^{n(n-1)/2}
\prod_{\gamma=1}^n \frac{\Delta_\gamma(\lambda)}{\Delta(\lambda)}
{\rm det}_{i\gamma}\frac{1}{\lambda_\gamma^i} = \frac{1}{\Delta(\lambda)}.
\end{array}
\ee
Since (\ref{derivkpvertod2}) is true for any $n$, one can claim that in the
limit $n \rightarrow \infty$ we recover the statement, that
$\tau^{KP}\{t\} = {\rm Det}_{i,j<0} {\cal H}_{ij}^{KP}$ with
${\cal H}_{ij}^{KP}$ given by eq.(\ref{derivkpvertod1}) (that formula does
not refer directly to Miwa parametrization and is defined for any $t$ and any
$j<0$ and $i$). This relation between $\varphi_\gamma$'s and ${\cal
H}_{ij}^{KP}$ can now be used to introduce negative times $\bar t_k$ according
to the rule (\ref{addkptau2}). Especially simple is the prescription for
zero-time: ${\cal H}_{ij} \rightarrow {\cal H}_{i+N,j+N}$, when expressed in
terms of $\varphi$ just implies that
\be
\frac{{\rm det}\ \varphi_\gamma(\lambda_\delta)}{\Delta(\lambda)} \rightarrow
\frac{{\rm det}\ \varphi_{\gamma +N}(\lambda_\delta)}
{({\rm det}\Lambda)^N\Delta(\lambda)}.
\ee
Generalizations of  (\ref{addkptau3}), like
\be
{\cal H}_{ij}\{t,\bar t\} = \oint \oint z^i\bar z^j
\langle 0 \mid \psi(z) \ G\ \tilde\psi(\bar z)\mid 0 \rangle
e^{\sum_{k} (t_kz^k + \bar t_k\bar z^k)} dzd\bar z,
\ee
also can be considered.

\subsection{Matrix models versus $\tau$-functions.}

We are now prepared to return to our main subject and discuss integrability
properties of eigenvalue matrix models. The claim is that partition functions
of all these models, when considered as functions of time-variables
(parametrizing the shapes of  potentials) are in fact $\tau$-functions of
(perhaps, multicomponent) Toda-lattice and/or KP type. (Interesting
non-eigenvalue models are believed to be related to integrable systems of more
general type, not restricted to level $k=1$ Kac-Moody algebras.)

Partition functions are, however, not generic Toda or KP $\tau$-functions:
first, they usually belong to some reduced hierarchies, second, the relevent
operators $G$ (points of Grassmannian) are restricted to stay in peculiar
domains of the universal module space, specified by "string equations". String
equation is in fact nothing but the set of Ward-identities (Virasoro or
$W$-constraints in the examples under investigation), which are now
interpreted as equations on $G$. The very possibility of such interpretation
is highly non-trivial and reflects some deep relation between the constraints
and integrable structure. In the case of Virasoro constraints this is not a
puzzle, because Virasoro algebra is a symmetry (covariance) of the hierarchy,
the situation with other constraints is less clear (see the footnote at
section 4.3). In fact, when applied to a $\tau$-function of appropriately
reduced hierarchy, the infinitely many constraints usually become dependent
and it is enough to impose only the lowest Virasoro constraint $L_{-1}\tau =
0$
(or ${\cal L}_{-p}\tau = 0$, where $p$ is the degree of reduction), in order
to recover the entire set \cite{FKN}. It is this lowest constraint (or rather
its $t_1$-derivative, $\frac{\partial}{\partial t_1}(L_{-1}\tau) = 0$) that
traditionally carries the name of "string equation". It is often much simpler
to deduce than the entire set of identites, what is important in
practical applications (especially because determinant formulas, which imply
integrability, can  be also simpler to find in some situations than
the Ward identities).

In order to give a complete description of some sort of (matrix) models
from the point of view of integrability theory
it is enough to specify the hierarchy, to which it belongs (if partition
function is interpreted as $\tau$-function,
\be
Z_{\rm model}\{t\} = \tau \{t \mid G_{\rm model}\}
\ee
and the string equation which serves to fix operator $G$ - the point in the
universal module space.\footnote{
As we argued in the Introduction and in section 2.1, the word "matrix" can be
probably omited if generic Lagrangians are considered in other models of
quantum field theory. Also the universal module space (where moduli are - of
bundles over {\it spectral} Riemann surfaces)  can (and should) be
treated as a "space of theories". It is one of the great puzzles (and
beauties) of string theory, that Riemann surfaces appear both in the
world-sheet and spectral "dimensions". See \cite{GLM} for more discussion on
this issue.} After that it becomes an internal (yet unsolved) problem of
integrability theory to explain, what is so special about the set of points
$\{G_{\rm model}\}$ in this space. (We shall touch this problem in the next
subsection, devoted to Kac-Schwarz operators.) Alternatively, if there is
nothing special, it is an (unsolved) problem of matrix model theory to find
models, associated with {\it any} points $G$ in the universal module space (or
explain what is an obstacle, if any).

We proceed now to description of particular matrix models from this point of
view. As everywhere in these notes we consider only the most important classes
of scalar-product, conformal (multicomponent) and Generalized Kontsevich
models (GKM). All other examples (like models of complex, orthogonal, unitary
etc matrices) can be included into consideration with more or less effort
(see \cite{MMMM} and \cite{Bowick} for the cases of complex and unitary models
respectively), but they do not add much for the general theory that we are now
considering. String equations will be discussed in the next subsection.

{\it Scalar-product models}. These were exhaustively discussed in sections
3.5-3.7. We remind that all conventional multimatrix models (with inter-matrix
interaction of the form $\exp ({\rm Tr} H^{(\alpha)}H^{(\alpha+1)})$ belong to
this class. The crucial formulas are:
\be
Z_N &= {\rm Det}_N {\cal H}_{ij}^f = {\rm Det}_{0 \leq i,j \leq N-1}
{\cal H}_{ij}^f = \nn \\
&= {\rm Det}_{-N \leq i,j < 0}{\cal H}_{i+N,j+N}^f; \nn \\
{\cal H}_{ij}^f &= \frac{\partial^2}{\partial t_i\partial \bar t_j}
{\cal H}^f = \left(\frac{\partial}{\partial t_1}\right)^i
\left(\frac{\partial}{\partial \bar t_1}\right)^j {\cal H}^f.
\label{scapro1}
\ee
Here
\be
{\cal H}_{ij}^f = \langle h^i \mid \bar h^j\rangle =
\int d\hat\mu_{h,\bar h}e^{U(h)+\bar U(\bar h)} h^i\bar h^j.
\ee
Further,
\be
e^{U(h)} = e^{\sum_{k\geq 0} t_kh^k} = \sum_l h^lP_l(t); \nn \\
e^{\bar U(\bar h)} = e^{\sum_{k\geq 0}\bar t_k\bar h^k} =
\sum_m \bar h^mP_m(\bar t)
\ee
and thus
\be
{\cal H}_{ij}^f &= \sum_{l,m} \langle\langle h^{i+l}\mid \bar h^{j+m}\rangle
\rangle P_l(t)P_m(\bar t) =
\sum_{l,m} T_{lm}^f P_{l-i}(t)P_{m-j}(\bar t), \nn \\
T_{lm}^f &= \langle\langle h^{l}\mid \bar h^{m}\rangle\rangle,
\ee
where the
scalar product $\langle\langle \ \mid \ \rangle\rangle$ is w.r.to the measure
$d\hat\mu_{h,\bar h}$ (while $\langle \ \mid\ \rangle$ is w.r.to
$d\mu_{h,\bar h} = e^{U(h)+\bar U(\bar h)} d\hat\mu_{h,\bar h}$).

One would immediately recognize in these formulas
representation (\ref{todatau}) of Toda-lattice $\tau$-function,
be there no additional restriction that determinant in (\ref{scapro1}) is over
finite-dimensional $N\times N$ matrix (indices are constrained: $i,j\geq -N$).
This can be automatically taken into account if we require that
\be
T_{lm}^f = \infty \ \ {\rm for\ all}\ l,m<0,
\ee
and identify $Z_N$ as a  $\tau$-function $\tau^f$ of {\it forced}
Toda-lattice hierarchy (thus the supescript $f$ carried by ${\cal H}$ and $T$).
We conclude that partition functions of any scalar-product model is a
$\tau^f$-function of forced Toda-lattice hierarchy.

Let us now consider them as KP $\tau$-functions. This means that the $\bar
t$-dependence is simply ignored. However, $N$ will be preserved
explicitly as a parameter, labeling KP $\tau$-function.
After Miwa transformation $t_k = -\frac{1}{k}\sum_\gamma\lambda_\gamma^{-k} -
r_k$, described in section 3.7, we get:
\be
Z_N = \hat Z_N \frac{{\rm det}_{\gamma\delta}\hat
Q_{N+\gamma-1}(\lambda_\delta)}{\Delta(\lambda)},
\ee
where $\hat Q$ are orthogonal polinomials w.r.to the measure $d\hat\nu_{h,\bar
h} = e^{-\sum_k r_kh^k}d\hat\mu_{h,\bar h}$.
We conclude that in the framework of KP hierarchy the scalar-product models
are distinguished by the fact, that the corresponding
$\varphi_\gamma(\lambda)$ in (\ref{KPdetmain}) are {\it
polinomials} rather than infinite series in powers of $\lambda^{-1}$.

{\it 1-Matrix model}. This is particular example of scalar product model with
a {\it local} measure $d\mu_{h,\bar h} = e^{U(h)+\bar U(\bar h)}\delta(h-\bar
h)dhd\bar h$. In this case
\be
{\cal H}^f_{ij} = \langle h^i \mid \bar h^j \rangle =
\langle h^{i+j} \rangle = \frac{\partial}{\partial t_{i+j}}{\cal H}^f =
\left(\frac{\partial}{\partial t_1}\right)^{i+j}{\cal H}^f.
\ee
Thus in this case we deal with the (forced) Toda-chain reduction of
Toda-lattice hierarchy. In the end of this section 4 we use orthogonal
polinomials to present a detailed description of 1-matrix models  as
a Toda-chain $\tau$-function.

This model can be alternatively defined as Gaussian Kontsevich model: see
section 3.8. The fact that partition function is a $\tau$-function follows
then from the general statement for GKM, see below. The fact that it is
{\it forced} $\tau$-function is related to the property $c_{-N} = 0$,
mentioned at the end of s.3.8 (and proved in s.3.9). Also reduction to
Toda-chain hierarchy can be observed directly
in terms of GKM: see ref.\cite{Toda} for more details.

{\it Multicomponent (conformal) matrix models}. These are related to
multicomponent hierarchies, with $\tau$-functions representable as correlators
in multi-fermion systems. An example of determinant formula which substitutes
(\ref{todatau}) in the 2-component case, is given at the end of section 3.5,
where it is derived from consideration of the relevant matrix model
\cite{comamo}. For
derivation of the same determinant formula in the theory of $\tau$-functions
see ref.\cite{HOS}. Generic theory of multicomponent hierarchies is now making
its first steps and we do not review it in these notes. See \cite{Kavdl} for
the group-theory approach to the problem.

{\it Generalized Kontsevich model}. Determinant formulas for this case are
derived in Section 3.3. The most important expression is
\be
Z_V\{N,T\} = \frac{1}{({\rm det}\Lambda)^N}\frac{{\rm
det}_{\gamma\delta}\varphi_{\gamma+N}(\lambda_\delta)}{\Delta(\lambda)},
\label{gkmdet}
\ee
where
\be
\varphi_\gamma(\lambda) &= \frac{1}{\sqrt{2\pi}} e^{-\lambda V'(\lambda) +
V(\lambda)} \sqrt{V''(\lambda)} \int x^{\gamma-1}e^{-V(x) + V'(\lambda)x}dx =
\nn \\
&= \lambda^{\gamma-1}(1 + {\cal O}(\lambda^{-1}));
\label{gkmdet1}
\ee
and
\be
\varphi_\gamma(\lambda) = {\cal A}\varphi_{\gamma-1}(\lambda) =
{\cal A}^{\gamma-1}\Phi(\lambda).
\label{gkmKasch}
\ee
For $N=0$ this is just the representation, peculiar for KP $\tau$-function in
Miwa parametrization $T_k = \frac{1}{k}{\rm tr}\Lambda^{-k}$,
see eq.(\ref{KPdetmain}) above. Thus
\be
Z_V\{T\} = \tau^{KP}\{T \mid G_V\},
\ee
where it is operator $G$ (the point in Grassmannian) which depends on the
shape of potential $V(X)$. We also remind that the only way in which $Z$
depends on the size of the
matrix $n$ is through the {\it domain of variation} of the time variables $T$.
If (\ref{gkmdet}) is extended to full Toda-lattice $\tau$-function, by
introduction of negative times, we get \cite{Toda}:
\be
\new
\begin{array}{c}
Z_V\{T,N,\bar T\} =  \frac{{\cal C}_V^{-1}(\Lambda)}{({\rm det}\Lambda)^N}
e^{-\sum_{k>0}\bar T_k{\rm tr}\Lambda^{-k}} \times  \\
\times\int_{n\times n} dX ({\rm det}X)^N\exp\left( - {\rm tr}V(X) + {\rm
tr}\Lambda X +
\sum_{k>0} \bar T_k {\rm tr}X^{-k} \right)
\end{array}
\ee
When {\it this} extended partition function is considered as KP
$\tau$-function, we have instead of (\ref{gkmdet}):
\be
Z_V\{T,N,\bar T\} = \frac{1}{({\rm det}\Lambda)^N}\frac{{\rm
det}_{\gamma\delta}\varphi_{\gamma+N}^{\{\hat V\}}
(\lambda_\delta)}{\Delta(\lambda)},
\label{gkmdet2}
\ee
and relevant $\varphi$-functions are
\be
\varphi_{\gamma+N}^{\{\hat V\}}(\lambda)
&= \frac{1}{\sqrt{2\pi}} e^{-\lambda V'(\lambda) +
\hat V(\lambda)} \sqrt{V''(\lambda)} \int x^{\gamma-1}
e^{-\hat V(x) + V'(\lambda)x}dx =
\nn \\
&= \lambda^{N+\gamma-1}(1 + {\cal O}(\lambda^{-1})),
\ee
with
\be
\hat V(x) &\equiv V(x) - N\log x - \sum_{k>0}\bar T_kx^{-k}, \nn \\
V(x) &= \hat V_+(x)
\ee
(where $\hat V_+(x)$ is the positive-power fragment of Laurent series $\hat
V(x)$).
Functions $\varphi_\gamma(\lambda)$ in (\ref{gkmdet1}) are just equal to
$\displaystyle{
\left.\varphi_\gamma^{\{\hat V\}}(\lambda)\right|_{\bar T = 0}}$.

\subsection{String equations and general concept of reduction}

The role of string equation is to fix the point $G$ in the universal module
space (UMS), associated with the particular matrix model, so that partition
function, considered as a function of time variables, will appear as  the
corresponding $\tau$-function of a fixed shape. In this sense the idea behind
the string equation is exatly the same as reduction of integrable hierarchies.
The difference is that {\it linear} reductions, as defined in the section 4.5
above, are not enough to fix $G$ unambiguously: they just specify certain
subsets in the Grassmannian, which are still infinite-dimensional. The reason,
why  these are usually linear reductions that are considered in the
conventional theory of integrable hierarchies, is that they are associated
with the simplest possible - Kac-Moody - subalgebras in the entire
$GL(\infty)$. String equations, even their simplest examples, are usually
fragments of more complicated - Virasoro and $W$-algebras, and are in fact
considerably more restrictive. Moreover, string equation is usually a
{\it distinguished} fragment,
because it usually belongs to the Virasoro component
of Ward identities, and Virasoro is still a {\it Lie} subalgebra in
$GL(\infty)$. This is what makes the problem of string equations very similar
to the "classical" one with linear reduction.

More specifically, in order to include string equations (and in fact the
entire set of Virasoro - but not $W$ - constraints) into consideration of
reduction it is enough to allow the coefficients in (\ref{redtotau1})
to depend on $t$
and $\bar t$, without changing the order of time-derivatives. Of course, there
are no obvious reasons to think that {\it any} point $G$ in the UMS can be
selected by imposing this kind of linear-derivative constraints on
$\tau$-function, - and further investigation can require essential
generalization of such restricted notion of string equation. However, some of
the eigenvalue matrix models are already known to possess string equations
of such simple type, associated with Virasoro subalgebras of $GL(\infty)$.
We do not go into details of the general theory - it is far not completed yet,
- but instead present several examples of how string equations arise in
particular matrix models. These examples can illustrate also the
simplifications arising when only  string equations and not the entire sets of
Ward identities need be derived. In particular, it is clear that in cases when
$\tau$ is represented as ${\rm Det}_{ij} {\cal H}_{ij}$, a linear differential
equation imposed on ${\cal H}_{ij}$ will give rise to a similar equation on
$\tau$ itself. Most of known string equations can be derived with the help of
this technical idea. They are usually associated with invariance of integrals
under {\it constant} shifts of integration variables $\delta h =\ const$ in
scalar-product and other discrete models, and with the action of operator
${\rm tr} \frac{\partial}{\partial L_{tr}}$ in GKM. For somewhat more involved
ideas, associated with string equations see \cite{GMoo}.

{\it Scalar-product models}. String equation can be easily deduced for very
specific type of measures $d\hat\mu_{h,\bar h}$. Since integral
\be
{\cal H}_{ij} = \int h^i\bar h^j e^{U(h)+\bar U(\bar h)}
d\mu_{h,\bar h}
\ee
is invariant under the shift of integration variable $\delta h =\ const$,
\be
\int h^i\bar h^j e^{U(h)+\bar U(\bar h)}
d\hat\mu_{h,\bar h} \left[ ih^{-1} + \frac{\partial U(h)}{\partial h} +
\frac{\partial}{\partial h} \log (d\hat\mu_{h,\bar h}) \right] = 0,
\ee
or
\be
i{\cal H}_{i-1,j} + \sum_{k>0} kt_k\frac{\partial}{\partial t_{k-1}}
{\cal H}_{ij} +
\left[ S\left(\frac{\partial}{\partial t},\frac{\partial}{\partial \bar t}
\right)\right]_{ij} = 0.
\ee
String equation arises straightforwardly, when operator $S$ is linear. This is
true, if $\log (d\hat\mu_{h,\bar h}) \sim hf(\hat h)$  with any function
$f(h)$. If the measure $d\hat\mu_{h,\bar h}$ is also required to be symmetric
in $h$ and $\bar h$,  we obtain conventional 2-matrix model as the only
example:
\be
d\hat\mu_{h,\bar h} = e^{ch\bar h}dhd\bar h.
\ee
Equation for ${\cal H}_{ij}$ is:
\be
\phantom.\left(\sum_{k>0} kt_k\frac{\partial}{\partial t_{k-1}} +
c\frac{\partial}{\partial \bar t_1}\right) {\cal H}_{ij} =
- i{\cal H}_{i-1,j}.
\label{hfor2mamo}
\ee
Its implication for $\hat \tau_N$ is:
\be
\left(\sum_{k>0} kt_k\frac{\partial}{\partial t_{k-1}} +
c\frac{\partial}{\partial \bar t_1}\right) \hat\tau_N = 0,
\label{sefor2mamo}
\ee
since the r.h.s. of (\ref{hfor2mamo}) does not contribute to determinant (the
entries in the i-th row are proportional to those in the $i-1$-th row).

In particular case of {\it 1-matrix model} $\ c=0$, and we recognize the lowest
Virasoro constraint $L_{-1}\hat\tau_N = 0$.
Traditionally the name of string equation is
given not to $L_{-1}$-constraint itself, but to its $t_1$-derivative:
$\frac{\partial}{\partial t_1}(L_{-1}\hat\tau_N) = 0$.
For 2-matrix model
(\ref{sefor2mamo}) is the lowest ($m=1,\ n=0$) component of the Ward
identities $\left(\tilde W_{n-m}^{(m+1)}(t) - (-)^{m+n}c^{n+1}
\tilde W_{m-n}^{(n+1)}(\bar t)\right)\hat\tau_N = 0$.
Of course, there is also a similar equation with
$t \leftrightarrow \bar t$.

{\it Multicomponent (conformal) models}. The crucial feature of these models
is that intermatrix interaction, when rewritten in terms of eigenvalues,
usually contains only differences $\ h_i^{(\balpha)} - h_j^{(\bbeta)}\ $. Thus
there is usually covariance under {\it simultaneous} shift of all eigenvalues
$\ \delta h_i^{(\balpha)} = const\ $ by a {\it same} constant.
This gives rise to a string equation of the form
\be
\left( \sum_{\balpha} L_{-1}^{(\balpha)} \right) \tau_{\bf N} = 0.
\ee
See \cite{comamo} for details.

{\it Generalized Kontsevich Model}.

In order to derive string equation one should act on partition function
$Z_V\{T_k = \frac{1}{k}{\rm tr}\Lambda^{-k}\} =
{\cal C}_V^{-1} {\cal F}_V\{L = V'(\Lambda)\}$ with operator ${\rm
tr}\frac{\partial}{\partial L_{tr}} = {\rm tr}\frac{1}{V''(\Lambda)}
\frac{\partial}{\partial \Lambda_{tr}}$. We can rewrite the result of this
action in terms of time-derivatives,
\be
{\rm tr}\frac{\partial}{\partial L_{tr}} \log Z_V\{T\} =
-\sum_{k>0} \left({\rm tr}\frac{1}{V''(\Lambda)\Lambda^{k+1}}\right)
\frac{\partial}{\partial T_k}\log Z_V\{T\}.
\ee
Alternatively we can use the fact that $\ \displaystyle{
{\rm tr}\frac{\partial}{\partial L_{tr}} = \sum_\gamma
\frac{1}{V''(\lambda_\gamma)}\frac{\partial}{\partial\lambda_\gamma}}\ $,
$l = V'(\lambda)$, and
explicit expression for $Z_V$ in terms of eigenvalues (Miwa coordinates),
\be
\new
\begin{array}{c}
Z_V \sim e^{{\rm tr}V(\Lambda) - {\rm tr}\Lambda V'(\Lambda)}
\sqrt{\prod_\gamma V''(\lambda_\gamma)}\
\frac{{\rm det}\ \hat \varphi_\gamma(\lambda_\delta)}{\Delta(\lambda)} \sim \\
{\sim} \frac{{\rm det}\
\varphi_\gamma(\lambda_\delta)}{\Delta(\lambda)},
\end{array}
\label{detformzv}
\ee
to get:
\be
\new
\begin{array}{c}
\left({\rm tr}\frac{\partial}{\partial L_{tr}}\right) \log Z_V\{T\} =   \\
= \frac{1}{2} {\rm
tr}\frac{V'''(\Lambda)}{(V''(\lambda))^2}
+ \frac{1}{2}\sum_{\gamma > \delta} \frac{V''(\lambda_\gamma) -
V''(\lambda_\delta)}{\lambda_\gamma - \lambda_\delta} \cdot
\frac{1}{V''(\lambda_\gamma)V''(\lambda_\delta)} -  \\
- {\rm tr} \Lambda  + \sum_\beta
\frac{\partial}{\partial l_\beta}
\log{\rm det}_{\gamma\delta} \hat \varphi_\gamma(l_\delta).
\end{array}
\ee
Comparison of these two expressions gives:
\be
\new
\begin{array}{c}
\frac{{\cal L}_{-1}^{(V)} Z_V}{Z_V} \equiv
\frac{1}{Z_V}\left[ \sum_{k>0} \left({\rm
tr}\frac{1}{V''(\Lambda)\Lambda^{k+1}}\right)
\frac{\partial}{\partial T_k} +  \right. \\ \left. +
\frac{1}{2}\sum_{\gamma > \delta} \frac{V''(\lambda_\gamma) -
V''(\lambda_\delta)}{\lambda_\gamma - \lambda_\delta} \cdot
\frac{1}{V''(\lambda_\gamma)V''(\lambda_\delta)} - \frac{\partial}{\partial
T_1}\right] Z_V = \\
= - \frac{\partial}{\partial T_1}\log Z_V + {\rm tr} \Lambda -
\sum_\beta
\frac{\partial}{\partial l_\beta}
\log{\rm det}_{\gamma\delta} \hat \varphi_\gamma(l_\delta).
\end{array}
\label{steqend}
\ee
One can show that the r.h.s. is equal to zero, and thus the string
equation arises in the form
\be
{\cal L}_{-1}^{(V)} Z_V = 0.
\ee
If potential is monomial, $V_p = \frac{X^{p+1}}{p+1}$, then
$r_k = -\frac{p}{p+1}\delta_{k,p+1}$ and
\be
\new
\begin{array}{c}
{\cal L}_{-1}^{V_p} \rightarrow {\cal L}_{-p} \equiv \\
\equiv
\frac{1}{p} \left[ \sum_{k>0} (k+p)(T_{k+p} + r_{k+p})\frac{\partial}{\partial
T_k} + \frac{1}{2} \sum_{k=1}^{p-1} k(p-k)T_kT_{p-k}\right].
\end{array}
\label{asa}
\ee

The technical idea behind the proof \cite{GKM} is to represent
\be
\frac{\partial}{\partial T_1}\log Z_V = {\rm Res}
\frac{Z_V\{T_k + \frac{1}{k\lambda^k}\}d\lambda}{Z_V\{T_k\}},
\ee
and make use of the second determinant representation in (\ref{detformzv})
both in the denominator and the numerator:
\be
\frac{\partial}{\partial T_1}\log Z_V =
{\rm Res}\frac{d\lambda}{\prod_{\gamma=1}^n(\lambda - \lambda_\gamma)}
\cdot\frac{ {\rm det}\left( \begin{array}{cc}
\varphi_\delta(\lambda_\gamma) & \varphi_{n+1}(\lambda_\gamma) \\
\varphi_\delta(\lambda) & \varphi_{n+1}(\lambda)
\end{array}\right)}
{{\rm det}\ \varphi_\delta(\lambda_\gamma)}.
\label{as0}
\ee
Now we recall that
\be
\varphi_\gamma(\lambda) \sim \lambda^{\gamma -1} \left( 1
+ {\cal O}(\lambda^{-1})\right).
\label{as1}
\ee
At some moment we shall need even more:  in fact
\be
&\varphi_\gamma(\lambda) \sim \lambda^{\gamma -1} \left( 1
+ {\cal O}(\lambda^{-2})\right), \ \ {\rm i.e.} \nn \\
&\varphi_\gamma(\lambda) = \lambda^{\gamma -1} + c_\gamma\lambda^{\gamma-2} +
\ldots, \ \ {\rm and}\ \ c_\gamma = 0 \ \ {\rm for\ any} \ \gamma.
\label{as2}
\ee
This is a rather delicate property of GKM, it follows from two facts:
first, $\displaystyle{\varphi_1 = 1 + {\cal O}\left(\frac{V''''}{(V'')^2},
\frac{(V''')^2}{(V'')^3}\right)}$, thus $c_1 = 0$, and
second, Kac-Scwarz operator ${\cal A}$, defined in eq.(\ref{gkmKasch}) below,
does not have contributions with zero-th power of $\lambda$, thus $c_{\gamma
+1} = c_\gamma$. (For example, if $V(x) = \frac{x^2}{2} + ax$,
$\varphi_\gamma(x) = \frac{1}{\sqrt{2\pi}} \int x^{\gamma -1}
e^{-\frac{1}{2}(x-\lambda)^2}dx = \lambda^{\gamma-1} + 0\cdot \lambda^{\gamma
-2} + \ldots$: the dangerous terms with $a$ simply do not show up in the
expression for $\varphi_\gamma$.)

After this comment we can come back to evaluation of (\ref{as0}).
The product in denominator, which arised from the Van-der-Monde determinant,
is already proportional to $\lambda^n$:
$\prod_{\gamma =1}^n(\lambda - \lambda_\gamma) = \lambda^n\left(1 + {\cal
O}(\lambda^{-1})\right)$.
Because of this and the asymptotic formulas (\ref{as1}), it is clear that
if determinant in the numerator of (\ref{as0}) is rewritten as linear
combination of $n\times n$ determinants with the coefficients
$\varphi_\gamma(\lambda)$ from the last row, only items
with $\gamma \geq n$ can contribute. There are two such items: $\gamma = n$
and $\gamma = n+1$. In the expansion of $(n+1)\times (n+1)$ determinant
$\varphi_{n+1}(\lambda) $ is multiplied by ${\rm det}\
\varphi_{\gamma}(\lambda_\delta)$, which exactly cancels with determinant in
denominator, and the relevant contribution is
\be
{\rm Res}\ \frac{\varphi_{n+1}(\lambda)d\lambda}{\prod_{\gamma
=1}^n(\lambda - \lambda_\gamma)} = c_{n+1} + \sum_\gamma \lambda_\gamma =
\nn \\ = c_{n+1} + {\rm tr}\Lambda.
\ee
The item with $\varphi_n(\lambda)$ is
\be
\frac{{\rm det} \left(\varphi_1(\lambda_\gamma) \ldots
\varphi_{n-1}(\lambda_\gamma)\ \varphi_{n+1}(\lambda_\gamma)\right)}
{{\rm det} \left(\varphi_1(\lambda_\gamma) \ldots
\varphi_{n-1}(\lambda_\gamma)\ \varphi_{n}(\lambda_\gamma)\right)}
{\rm Res}\ \frac{\varphi_n(\lambda)d\lambda}{\prod_{\gamma
=1}^n(\lambda - \lambda_\gamma)}.
\label{detratvpn1}
\ee
The remaining residue is just unity. Determinant in the numerator differs from
the one in denominator by substitution of the colomn with entries
$\varphi_{n}(\lambda_\gamma)$ for that with $\varphi_{n+1}(\lambda_\gamma)$.

At last we can return to eq.(\ref{steqend}) and recall that
$\frac{\partial}{\partial l}\hat\varphi_\gamma(l) = \varphi_{\gamma +1}(l)$,
thus
\be
\new
\begin{array}{c}
\sum_\beta \frac{\partial}{\partial l_\beta} \log{\rm
det}_{\gamma\delta}\hat\varphi_\delta(l_\gamma) = \\
\frac{{\rm det} \left(\hat\varphi_1(l_\gamma) \ldots
\hat\varphi_{n-1}(l_\gamma)\ \hat\varphi_{n+1}(l_\gamma)\right)}
{{\rm det} \left(\hat\varphi_1(l_\gamma) \ldots
\hat\varphi_{n-1}(l_\gamma)\ \hat\varphi_{n}(l_\gamma)\right)},
\end{array}
\label{detvarphil}
\ee
what is just the same as (\ref{detratvpn1}), since $\hat\varphi_\delta$
differ from $\varphi_\delta$ by $\delta$-independent factor of
$e^{V(\lambda)-\lambda V'(\lambda)}\sqrt{V''(\lambda)}$. Thus we conclude that
the r.h.s. of (ref{steqend}) is equal to $-c_{n+1}$, which actually vanishes,
as was explained several lines above.

Two things deserve paying attention in this derivation. First, it was
absolutely crucial that we had $\frac{\partial}{\partial T_1}\log Z_V$ at the
r.h.s. of (\ref{steqend}) to make it vanishing, and therefore
$\frac{\partial}{\partial T_1}$ immedeately appears in the expression for the
${\cal L}_{-1}^{(V)}$ operator at the l.h.s. (this is the origin of
$r_k$-corrections in (\ref{asa}). Second, the result is both simple and
natural, but the proof is full of technical details and looks somewhat
artificial. It becomes even more involved, when the general formula
(\ref{tkderiv})
for $T_k$-derivatives of $Z_V$ with $1\leq k\leq p$ \cite{Krich} is derived,
which plays an important role in the theory of GKM and its applications to the
theory of quantum gravity. The proof of the string equation is just a
particular case of that formula, since using integral representation of
$\hat\varphi(l)$ one can represent the r.h.s. of (\ref{detvarphil}) as
$\frac{1}{Z_V}\langle {\rm tr} X \rangle$, where $\langle \ \ \rangle$ now
stand for the average, defined by Kontsevich integral. Thus
\be
{\cal L}_{-1}^{(V)} Z_V \stackrel{(\ref{steqend})}{=}
-\frac{\partial}{\partial T_1}Z_V + \langle {\rm tr}\Lambda - {\rm tr}X
\rangle \stackrel{(\ref{tkderiv})}{=} 0.
\ee

\subsection{On the theory of GKM}

We remind that GKM is abbreviation for the Generalized Kontsevich model,
This theory is the naturally broad collection of topics for a separate big
section in these notes. However, we decided not to include such detailed
presentation. This is because GKM theory seems too incomplete now. First, we
believe that the natural invariant formulation - of which existing matrix
integral is
only a specific realization - is still lacking. Second, GKM is not yet
generalized enough to fulfil its main purpose of incorporating infromation
about {\it all} the models of $2d$ gravity (in fact it should include even
more: the entire theory of integrable hierarchies and geometrical
quantization). Third, though the whole approach is very conceptual and deep,
many {\it proofs}, as available nowadays, are still very technical and long.
All this implies, that the proper view on the subject of GKM still needs to be
found. At the moment we could describe two complementary approaches: one,
starting from integral representations, another - from the Duistermaat-Heckman
(localization) theory and Fourier analysis on group manifolds. Though
intimately related, these two approaches are still technically different in
too many respects. The second one is more fundamental (since ordinary
integrals arise from discrete sums either in special limits or in the cases of
inifinite-dimensional algebras, and, more important, since integral
representation is only one of many possible ways to define the quantities of
interest). However, many of the most important results, obtained in the first
approach yet do not have their proper names and exact counterparts in the
second one. We believe that this whole issue will be very much clarified in
the near future and decided to postpone a detailed review till that time.
What we can not avoid in these notes, is giving at least a {\it list} of
topics,
already included in the theory of GKM, and this is the purpose of the present
subsection.

Kontsevich model with $V=\frac{X^3}{3}$ was derived by Maxim
Kontsevich \cite{Ko} from the original definition of topological $2d$ gravity,
given by E.Witten \cite{WitTG} in terms of generating functional for Chern
classes of certain bundles over Riemann surfaces. Generalization of this
reasoning (when more bundles are taken into consideration) leads to the theory
of Landau-Ginzburg Gravity (LGG), which is believed to be the same as GKM,
though not all the proofs are already avaliable.\footnote{
Intermidiate results include
the study of spherical approximation to LGG, which exhibits the structures,
peculiar for "quasiclassical integrable hierarchies" (of which Bateman
hierarchy to be briefly mentioned in section 5.2 below is an example), and
which also arise in "quasiclassical approximation" to GKM. For some results in
this direction see \cite{WittenN2},\cite{KriDu},\cite{TakTak},\cite{LP},
\cite{Krich},\cite{KhMa} and
references therein.
}

The crucial feature of non-perturbative partition functions, as we discussed at
the neginning of section 2, is their intrinsic integrability. For $2d$ gravity
this general idea acquires a very concrete formulation: partition functions
are usually just $\tau$-functions of conventional integrable hierarchies,
moreover - for LGG, associated with minimal models, these are just ordinary
multicomponent Toda hierarchies.\footnote{One can say that this is natural:
both such models and Toda hierarchies are associated with the level $k=1$
Kac-Moody algebras and corresponding simplified versions of the WZNW model.
However, too much still remains to be clarified about this "obvious"
connection.
}

M.Kontsevich found  representation for  generating
functional in the form of matrix integral, i.e. formulated a matrix model,
which later allowed to prove Witten's conjecture
that the functional is in fact a $\tau$-function.
The concept of GKM as a {\it universal} matrix model, including all the
information about generic (eigenvalue?) matrix models and thus all the models
of $2d(?)$ gravity was introduced in \cite{GKM}, and the analogue of
Kontsevich model with arbitrary potential $V(X)$, i.e. expression
\be
\new
\begin{array}{c}
\left.Z_V\{T\}\right|_{T_k = \frac{1}{k}{\rm tr}\Lambda^{-k}} =
C_V(\Lambda)^{-1}{\cal F}_V\left(V'(\Lambda)\right) \sim \\
\sim \frac{\sqrt{{\rm det} V''(\Lambda)}}
{(2\pi)^{n^2/2} e^{{\rm tr}\left(\Lambda V'(\Lambda) - V(\Lambda)\right)}}
\int_{n\times n} dX e^{- {\rm tr} V(X) + {\rm
tr} V'(\Lambda) X}
\end{array}
\label{GKoM}
\ee
was proposed as an intemediate step in this direction.\footnote{
We remind that ${\rm det} V''(\Lambda)$ is defined somewhat tricky, see s.2.5
above. The same matrix integral (\ref{GKoM}) was also considered
in refs.\cite{MaSe},\cite{Direv},\cite{Ko2},\cite{AvM}.
}
This (still restricted) version of GKM is already enough to unify all the $(p,
1)$-models of $2d$ gravity. In some sense, $(p,q)$-models with $q\neq 1$ are
also included, but in a very non-transparent way (using analytical
continuation), which does not even explicitly respect the $p \leftrightarrow
q$ symmetry. Partition function of such GKM, $Z_V\{T\}$ depends on two types
of variables: time-variables $\hat T_k$ and potential $V$. Formally these two
types of variables are absolutely different, $V$ being responsible for the
choice of particular LGG model or, what is essentially the same, of particular
reduction of Toda-lattice or KP hierarchy; while $\hat T_k$ are parameters of
the generating functional of all correlation functions in this particular
model. But of course, since we deal with exact (non-perturbative) approach,
there is almost no real difference between these types of
dependencies - on the model (vacuum state) and on the $\hat T$'s: the model
can be changed by non-infinitesimal shift of $\hat T$-variables.
Technically, in GKM this is reflected in the identity of the form
\cite{Krich}:
\be
{\cal Z}_{V_p}\{T\} = f_p(r\mid \hat T_k + r_k)\cdot
\tau\{\hat T_k + r_k\mid G_p\}
\label{Krich'}
\ee
where $r_k = \frac{p}{k(p-k)}{\rm
Res}\left(V'(\mu)\right)^{1-\frac{k}{p}}d\mu$
provide a specific parametrization of potentials $V$, which is here assumed to
be any polinomial of degree $p$, and $f_p$ is some simple function:
\be
f_p(r\mid \hat T_k + r_k) = \exp -\frac{1}{2} \sum_{i,j} A_{ij}(r)
(\hat T_i + r_i)(\hat T_j + r_j); \nn \\
A_{ij} = {\rm Res}\left(V'(\mu)\right)^{i/p}d\left(V'(\mu)\right)^{j/p} =
\frac{\partial^2\log \tau_0^{(p)}}{\partial t_i\partial t_j},
\ee
and $\tau_0^{(p)}$ is a $\tau$-function of "quasiclassical hierarchy".
What
is important, $G_p$ (which define the shape of the $\tau$-function
as function of $\hat T + r$) and  $f_p$  depend only on the degree $p$,
but not on the other details of the shape of the potential. This is a deep
formula. It accounts for two phenomena at once: First, it says that $Z$
depends on the {\it sum} of $\hat T$ and $r$.\footnote{
In Miwa parametrization
$\hat T_k = \frac{1}{k}{\rm tr}\left(V'_p(\Lambda)\right)^{-k/p}$. Throughout
these notes we used different time-variables $T_k =
\frac{1}{k}{\rm tr}\Lambda^{-k}$, which are independent of the potential $V$,
instead $V$-dependence of $Z_V$ - which we did not really study -
was rather nontrivial. If expressed in terms of $\hat T$, partition function
$\hat Z_V\{\hat T+r\} = Z_V\{T\}$ becomes almost independent of $V$: it changes
- abruptly - {\it only} when the degree $p$ of potential changes. This second
type of description is of course in better accordance with the symmetries of
particular model - which are different in different "vacua" (for different
$p$). Therefore these are  variables $\hat T + r$, rather than in
$T$ which arise naturally in Ward identites - as we saw in sections 2.5 and
2.6. $T$'s and $\hat T$'s are suited for different purposes: $T$'s are nice
when {\it universality} aspects of GKM are concerned, while $\hat T$'s arise
when specific features of particular models (orbits, vacua) are considered.
}
Second, dependence on $V$ is not {\it quite} smooth: when the degree of
potential changes, the shapes of the functions $f$ and $\tau$ also changes
abruptly. Another side of the same phenomenon is that partition function
$Z_V\{T\}$, which in principle is well defined as a matrix integral for all
choices of $V$ and $L$ (thus $\hat T$) at once, is in fact singular at some
points: there are phase transitions, manifesting the switch from one LGG model
to another. After a phase transition original integral expression becomes
sonewhat symbolical: it defines partition function only in the sense of
analytical continuation, and it is a separate problem to find an integral
representation, adequate in the new phases. In practice, what is nicely
decribed by the integral representation GKM in the form of eq.(\ref{GKoM}),
are $(p,1)$-models, with $p+1$ being just the power of potential $V(x)$. What
is not yet found, is analogous representation for $(p,q)$-models with $q\neq
1$ (it can involve multiple matrix integrals, and universal model is supposed
to be "matrix quantum mechanics in external fields").

Derivation of the crucial formula (\ref{Krich'})in any approach - starting from
GKM in the form of either LGG or matrix integrals - is still very tedious. In
matrix-model representation it relies upon identity \cite{Krich}
\be
\new
\begin{array}{c}
\frac{\partial Z_V}{\partial T_k} =
\langle {\rm tr}\Lambda^k - {\rm tr} X^k \rangle \equiv  \\
\equiv {\cal C}_V^{-1} \int \left({\rm tr}\Lambda^k - {\rm tr} X^k\right)
e^{-{\rm tr}V_p(X) +  {\rm tr}V_p'(\Lambda)X} dX \ \ \ {\rm for}\  1\leq k\leq
p,
\end{array}
\label{tkderiv}
\ee
which look trivial but are rather hard to derive. (A proof of the string
equation in GKM at the end of the previous subsection is the simplest example
of this kind of exersices.)
Certainly some simple derivation "in two lines" should exist, but it is not yet
found. Formulas of this kind are very important for all aspects of GKM theory.
Besides other things, they are just necessary to actually evaluate correlation
functions in $(p,1)$-models of $2d$ gravity, of which $Z_V\{T\}$ is a
generating functional. If instead of these "physical" questions, one asks
about integrability theory, identites of this sort also play important role.
For example, looking at (\ref{tkderiv}) for a special $k=p$ and special choice
of potential - monomial $V_p(X) = \frac{X^{p+1}}{p+1}$,- ane can note that the
r.h.s. vanishes: this is just a Ward identity, reflecting invariance under the
shift of the integration variable, $\delta X =\ const$.
This is a simplest version of a more general statement:\footnote{
This property was technically implicit in original Kontsevich's work \cite{Ko}
for $p+1=3$, where it was related with certain combinatorial identities. A
tricky proof, relying upon properties of $\tau$-functions, was given for any
$p$ in \cite{GKM}. An example of straightforward proof, again  for $p+1=3$, -
just in terms of Kontsevich matrix integrals - can be found in
ref.\cite{DFIZ}.
}
\be
{\rm if}\ V_p(X) = \frac{X^{p+1}}{p+1}, \ \ \ {\rm then} \ \ \ \
\frac{\partial Z_V}{\partial T_{pk}} = 0 \ \ {\rm for\ all\ } n\in Z_+.
\label{redcondk}
\ee
Looking from the point of view of integrable hierarchies, one immediately
recognizes (\ref{redcondk}) as an example of reduction condition
(\ref{redtotau1}). It corresponds to the so called $p$-reduction of
KP-hierarhy, of which KdV ($p=2$) and Boussinesq ($p=3$) are the most
celebrated examples. We refer to \cite{GKM} and \cite{Krich} for all details
and references, the only thing to mention here is that the slightly weaker
verison of the constraint (\ref{redcondk}),
\be
\frac{\partial Z_V}{\partial T_{pn}} = a_n = {\rm const},
\label{redcondk'}
\ee
where $a_n$ do not depend on any time variables, can be simply expressed in
Miwa parametrization: it is just the statement that $\varphi$-functions in
\be
Z_V = \frac{{\rm
det}_{\gamma\delta}\varphi_\gamma(\lambda_\delta)}{\Delta(\lambda)}
\nn
\ee
satisfy the $p$-reduction condition:
\be
\lambda^p \varphi_\gamma(\lambda) =
\sum_{\delta = 1}^{\gamma+p}
\hat{\cal V}_{\gamma\delta}\varphi_\delta(\lambda).
\label{predcogra}
\ee
This a restrictive relation, because $\varphi$'s are infinite seria in
$1/\lambda$, while at th r.h.s. in (\ref{predcogra}) there is only a finite
number of items. In GKM it is satisfied for monomial potential
just as a corollary of Gross-Newman
equation, or  more exact, of the Ward identity for the integral
\be
\varphi_\gamma(\lambda) \sim \int x^{\gamma-1} e^{-V(x)+V'(\lambda)x} dx.
\nn
\ee
Indeed, integral does not change under the shift $\delta x = \ const$, and
this implies:
\be
\int x^{\gamma-1} \left( V'(x) - V'(\lambda) - \frac{\gamma -1}{x}\right)
e^{-V(x)+V'(\lambda)x} dx = 0,
\nn
\ee
i.e.
\be
\sum_{k=1}^{p+1} kv_k \left(\varphi_{\gamma +k-1}(\lambda) -
\lambda^{k-1}\varphi_\gamma(\lambda)\right) - (\gamma-1)\varphi_{\gamma -1} =
0.
\label{calvdef}
\ee
If only $v_{p+1}\neq 0$, this leads to an identity of the required form of
(\ref{predcogra}). This description of reduction can be modified to allow for
non-monomial potentials, making use of the concept of "equivalent
hierarchies", see \cite{Tak},\cite{Krich}: in this framework the reduction
condition is
\be
V'(\lambda)\varphi_\gamma(\lambda) = \sum_\delta
{\cal V}_{\gamma\delta}\varphi_\delta,(\lambda),
\label{redcond'}
\ee
but classes of essentially different
reductions are labeled by the degree of potential only.

As we already discussed in the previous subsection, linear constraints like
(\ref{predcogra}) are not restrictive enough to fix the shape of the
$\tau$-function (the point $G$ in the universal module space) unambigously:
string equation should be also imposed. If expressed in terms of
$\varphi$'s, string equation is just the property
(\ref{gkmKasch}):
\be
\varphi_{\gamma +1} = {\cal A}\varphi_\gamma,
\label{recondigkm}
\ee
where the Kac-Schwarz operator
\be
{\cal A} = \frac{1}{V''(\lambda)}\frac{\partial}{\partial \lambda} -
\frac{1}{2}\frac{V'''(\lambda)}{(V''(\lambda))^2} + \lambda.
\ee
It has obvious generalization of the form
\be
{\cal A}_{p,q} = \frac{\partial}{\partial V'_p( \lambda)}
- \frac{1}{2}\frac{V'''_p(\lambda)}{(V''_p(\lambda))^2} + Q'_q(\lambda),
\label{KaschopQ}
\ee
where $Q_q(\lambda)$ is a polinomial of degree $q+1$ and (\ref{recondigkm}) is
substituted be
\be
\varphi_{\gamma +q} = {\cal A}_{p,q}\varphi_\gamma.
\ee
This generalization is
naturally related to the string equation in $(p.q)$ models, see \cite{KhMa}
and references therein.
Generic $(p,q)$ LGG model can be described by a system of constraints,
\be
(\lambda^p - {\cal V}_p)\{\varphi\} = 0,  \nn \\
{\cal A}_{p,q}\{\varphi\} = 0,
\label{constrset}
\ee
where both operators ${\cal V}_p$ and ${\cal A}_{p,q}$
are not uniquely fixed by
choosing $p$ and $q$, and there is also a freedom to change variables $\lambda
\rightarrow f(\lambda)$ and make a triangular transformation of basis
$\varphi_\gamma \rightarrow \varphi_\gamma + \sum_{\delta
<\gamma}C_{\gamma\delta}\varphi_\delta$. Altogether the set of equations
(\ref{constrset}) modulo these allowed transformations is finite -
$(p-1)(q-1)$ - dimensional, this is dimension of the module space of LGG
models with given $p$ and $q$. Kontsevich integral can be now used to
establish duality transformation from $(p,q)$ to $(q,p)$ model \cite{KhMa}:
\be
Z_{V,Q}(\Lambda) =
C_{V,Q}^{-1}(\Lambda) \int_{n\times n} dX
e^{-{\rm tr}S_{V,Q}(X,\Lambda) + {\rm
tr}V'(\Lambda)Q'(X)} Z_{Q,V}(X).
\label{VQGKM}
\ee
Here
\be
S_{V,Q}(x,\lambda) = \int^x V'(y)Q''(y)dy = \int^x V'(y)dQ'(y),
\ee
As usual, $C_{V,Q}(\Lambda)$ is  the quasiclassical approximation
to the integral, and
$Z_{V,Q}(\Lambda) \equiv \frac{{\rm det}_{\gamma\delta}
\varphi_\gamma(\lambda_\delta)}{\Delta(\lambda)}$, where $\varphi$ are
solutions to (\ref{constrset}) with ${\cal V}_p$ and ${\cal A}_{p,q}$
defined by eqs.(\ref{redcond'}) and (\ref{KaschopQ}) respectively.\footnote{
Also expression for  $r_k$-variables is now modified:
\be
r_k = \frac{p}{k(p-k)}{\rm Res} (V'_p(\mu))^{1-k/p}dQ'_p(\mu),
\nn
\ee
For monomial $V_p$ and $Q_q$  $\ r_k = -\frac{p}{p+q}\delta_{k,p+q}$.
}
This relation does {\it not} provide any
formula for $Z_{V_p,Q_q}(\Lambda)$ unless $q=1$. The case of $q=1$ is
distinguished because $Z_{Q_1,V_p}$ is trivial. Indeed, the 1-reduction
constraint $\lambda\varphi_\gamma = \varphi_{\gamma+1} + \sum_{\delta
\leq\gamma}{\cal V}_{\gamma\delta} \varphi_\delta$, implies that
${\rm det}_{\gamma\delta}\varphi_\gamma(\lambda_\delta) =
\Delta(\lambda)\prod_\delta \varphi_1(\lambda_\delta)$,
thus $Z_{Q_1,V_p} = \exp \sum_k a_kT_k$, what is essentially the same as
$Z_{Q_1,V_p} = 1$,\footnote{
Since $\varphi_1(\lambda) = 1 +\sum_{k>0}b_k\lambda^{-k}$, $\ \log
\varphi_1(\lambda) = \sum_{k>0}\frac{a_k}{k}\lambda^{-k}$, and the sum
$\sum_\delta \log\varphi_1(\lambda_\delta) =
\sum_{k>0}\frac{a_k}{k}\left(\sum_\delta \lambda_\delta^{-k}\right) =
\sum_{k>0} a_kT_k$. Addition of any {\it linear} combination of
time-variables to $\log\tau$ does not essentially change $\tau$-function.
For example,  the ordinary integrable equations
(like KdV or KP) are usually written in terms of variables like $u =
\frac{\partial^2}{\partial T_1}\log\tau$, which are {\it second} derivatives
of $\log\tau$.
}
and (\ref{VQGKM}) is just our old
formula (\ref{GKoM}) for the $(p,1)$ version of GKM.
(In fact $Q_1(X) \sim X^2$, and $Z_{Q_1,V_p}$ is
nothing but Gaussian Kontsevich model. It is trivial when the "zero-time"
$N=0$, as we assume here.) Matrix model realization of $Z_{V_p,Q_q}$ for
$q\neq 1$ is yet unknown.

This is not the only important further generalization of GKM (\ref{GKoM}).
Another one is implied by the formula
for ${\cal F}_V$ in terms of eigenvalues from the section 3.3,
\be
{\cal F}_V \sim \prod_{\gamma = 1}^n \int dx_\gamma e^{-V(x_\gamma)}
\Delta^2(x) I(x,l).
\label{KIverIZ}
\ee
As it was already mentioned in section 3.3, the Itzykson-Zuber integral
\be
I(x,l) \sim  \int [DU] e^{{\rm tr} UXU^\dagger L} \sim
\frac{{\rm det}_{\gamma\delta} e^{x_\gamma l\delta}}
{\Delta(x)\Delta(l)}
\ee
is in fact a coadjoint orbit integral and has group theoretical
interpretation: under certian conditions it turns into a character $\chi_R(g)
= {\rm Tr}_R g$ of the group $GL(n)$. Here $g \equiv e^L$ is considered as a
group element, representation $R$ is labeled by integer-valued parameters
$m_1,\ldots,m_n$ - essentially the lengths of rows in the Young diagramm.
Exact statement is:
\be
I(m,l)\cdot\frac{\Delta(l)}{\Delta(g)} = \frac{{\rm det}_{\gamma\delta}
g_\gamma^{m_\delta}}{\Delta(m)\Delta(g)} = \frac{\chi_R(g)}{d_R},
\ee
i.e. in order to get a character we should integrate over matrices $X$
with integer-valued eigenvalues.\footnote{
The ratio
\be
\frac{\Delta(l)}{\Delta(g)} = \prod_{\gamma >\delta} \frac{l_\gamma -
l_\delta}{e^{l_\gamma} - e^{l_\delta}}
\nn
\ee
is the usual correction factor, which is the price for the possibility to
reduce quantum-mechanical problem of motion on the orbit to a
{\it single} matrix integral. The full problem of
matrix quantum mechanics can and should be considered as a
multi-matrix (in fact, infinite-matrix) generalization of GKM (\ref{GKoM}),
which incorporates all the $(p,q)$ LGG models.
}
Dimension $d_R$ of representation can be also expressed in terms of
$m$-variables: $d_R = \Delta(m)$. As to the traces
${\rm tr} X^k = \sum_\gamma
x_\gamma^k \rightarrow \sum_\gamma m_\gamma^k$, which appear in the action of
GKM, they are very similar to the
$k$-th Casimir eigenvalue $C_k(R)$ (though is not exactly the same). Thus we
see that the integral in (\ref{KIverIZ})
is in fact very similar to
\be
{\cal F}_V^{qu}\{g,\bar g\} \equiv \sum_R \chi_R(\bar g)\chi_R(g)
e^{-\sum_{k=0}^{\infty}v_kC_k(R)},
\label{quanGKM}
\ee
evaluated at the point $\bar g = I$.
The only real difference is that instead of the integral we have a sum over
discrete values of $m$ (sum over all the representations, or a {\it model} of
$GL(n)$). This "discretized" (quantum?) GKM
is more general that the continuum one
which can be obtained by various limiting procedures. It is now obvious that
the theory of discretized GKM largely overlaps with that of $2d$ Yang-Mills
theory. The simplest ingredient of this theory is the classical result
\cite{chartau}, that
$GL(N)$ characters are in fact (singular) Toda-lattice and KP
$\tau$-functions. Moreover, the entire sum at the r.h.s. of (\ref{quanGKM}),
if considered as a function of $T_k = \frac{1}{k}{\rm tr} g^k$,
${\bar T}_k = \frac{1}{k}{\rm tr} \bar g^k$ is
in fact a Toda-lattice $\tau$-function. There are also features parallel to
(\ref{Krich'}).
We refer to \cite{ITEP12}
for a little more details about discretized GKM (see also
a recent paper \cite{GoNe}). This is one more very important direction of the
further investigation of GKM.

\subsection{1-Matrix model versus Toda-chain hierarchy}

At the end of this section we use an explicit example of dicrete 1-matrix
model \cite{GMMMO} to illustrate, how a more familiar Lax description of
integrable hierarchies arises from determinant formulas. This example will be
also usefull in Section 5.3 below, when one of the ways to take double-scaling
continuum limit of the 1-matrix model will be discussed.
Lax representation appears usually after some coordinate system is chosen
in the Grassmannian. In the example which we are now considering this
system is introduced by the use of orthogonal polinomials.

We already know from section 3.6, that partition function of 1-matrix model
(which is a {\it one}-component model) is given by
\be
Z_N = \ {\rm Det}_{0 < i,j \leq N} \langle h^i\mid h^j\rangle\  =
\prod_{i=0}^{N-1} e^{\phi_i} = Z_1 \prod_{i=1}^{N-1} R_i^{N-i},
\ee
where the last two representations are in terms of the norms of orthogonal
polinomials
\be
\langle Q_n \mid Q_m \rangle = e^{\phi_n}\delta_{nm}
\label{orthocond}
\ee
 and parameter of the
3-term relation
\be
\new
\begin{array}{c}
hQ_n(h) = Q_{n+1}(h) + c_nQ_n(h) + R_nQ_{n-1}(h), \\
Z_1 = e^{\phi_0} = \langle 1 \mid 1 \rangle , \ \ \ \ R_n =
e^{\phi_n-\phi_{n-1}}.
\end{array}
\nn
\ee
Of course all the information is contained in the determinant formula together
with the rule, which defines time-dependence of
${\cal H}_{ij}^f = \langle h^i\mid h^j\rangle  = \hat{\cal H}_{i+j}^f$:
\be
\new
\begin{array}{c}
\frac{\partial {\cal H}_{ij}^f}{\partial t_k} =
 {\cal H}_{i+k,j}^f = {\cal H}_{i,j+k}^f ,\ \ {\rm or} \\
\frac{\partial \hat{\cal H}_i^f}{\partial t_k} =
 \hat{\cal H}_{i+k}^f.
\end{array}
\ee
(The possibility to express everything in terms of ${\cal H}_i^f$ with a single
matrix index $i$ is the feature of Toda-chain reduction of generic
Toda-lattice hierarchy.)

However, in order to reveal the standard Lax representation we need to go into
somewhat more involved considerations. Namely, we consider representation of
two operators in the basis of orthogonal polinomials. First,
\be
h^k Q_n(h) = \sum_{m=0}^{n+k}
\frac{\langle n \mid h^k \mid m \rangle}{\langle m \mid m \rangle}
Q_m(h) =
\sum_{m=0}^{n+k} \gamma_{nm}^{(k)}Q_m(h)
\ee
(here the simplified notation is introduced for
$\langle n \mid f(h) \mid m \rangle \equiv
\langle Q_n \mid f(h) \mid Q_m \rangle$ and
$\displaystyle{\gamma_{nm}^{(k)} \equiv
\frac{\langle n \mid h^k \mid m \rangle}{\langle m \mid m \rangle}}$.)
Second,
\be
\new
\begin{array}{c}
\frac{\partial Q_n(h)}{\partial t_k} =
-\sum_{m=0}^{n-1}
\frac{\langle n \mid h^k \mid m \rangle}{\langle m \mid m \rangle}
Q_m(h) =
-\sum_{m=0}^{n-1}\gamma_{nm}^{(k)}Q_m(h), \\
\frac{\partial \phi_n}{\partial t_k} =
\frac{\langle n \mid h^k \mid n \rangle}{\langle n \mid n \rangle}
= \gamma_{nn}^{(k)}.
\end{array}
\ee
(These last relations arise from differentiation of orthogonality condition
(\ref{orthocond}):
\be
\new
\begin{array}{c}
e^{\phi_n}\frac{\partial \phi_n}{\partial t_k}\delta_{nm} =
\frac{\partial \langle Q_n \mid Q_m \rangle}{\partial t_k} =
\nn \\
= \langle \frac{\partial Q_n}{\partial t_k} \mid Q_m \rangle +
\langle Q_n \mid \frac{\partial Q_m}{\partial t_k} \rangle +
\langle Q_n\mid h^k \mid Q_m \rangle
\end{array}
\ee
by looking at the cases of $m<n$ and $m=n$ respectively.)

{}From these relations one immediately derives the Lax-like formula:
\be
\frac{\partial \gamma_{nm}^{(k)}}{\partial t_q} =
- \sum_{l=m-k}^{n-1} \gamma_{nl}^{(q)}\gamma_{lm}^{(k)} +
\sum_{l=m+1}^{n+k} \gamma_{nl}^{(k)}\gamma_{lm}^{(q)}
\label{laxrep1mamo}
\ee
or, in a matrix form,
\be
\frac{\partial \gamma^{(k)}}{\partial t_q}
= [ R\gamma^{(q)}, \gamma^{(k)}],
\ee
where
\be
R\gamma_{mn}^{(k)} \equiv \left\{
\begin{array}{c}
-\gamma_{mn}^{(k)} \ \ {\rm if} \ m>n, \\
\gamma_{mn}^{(k)} \ \ {\rm if} \ m<n
\end{array}
\right.
\ee
(We remind that usually $R$-matrix acts on a function
$f(h) = \sum_{n = -\infty}^{+\infty}f_nh^n$ according to the rule:
$Rf(h) = \sum_{n\geq l}f_nh^n - \sum_{n<l}f_nh^n$ with some "level" $l$.)
These $\gamma^{(k)}$ are not symmetric matrices, but one can also rewrite all
the formulas above in terms of symmetric ones:
\be
{\cal L}_{mn}^{(k)} \equiv e^{\frac{1}{2}(\phi_n - \phi_m)}\gamma_{mn}^{(k)}
= \frac{\langle m \mid h^k \mid n \rangle}{\sqrt{\langle m \mid m \rangle
\langle n \mid n \rangle}}
\ee

{}From eqs.(\ref{laxrep1mamo}) one can easily deduce Toda-equations for
$\phi_n$:
\be
\new
\begin{array}{c}
\frac{\partial^2\phi_n}
{\partial t_k\partial t_l} = \frac{\partial}{\partial t_k}
\frac{\langle n \mid h^l \mid n \rangle}{\langle n \mid n \rangle}
=  \\
= \left( \sum_{m>n} - \sum_{m<n}\right)
\frac{\langle n \mid h^k \mid m \rangle \langle m \mid h^l \mid n \rangle}
{\langle m \mid m \rangle\langle n \mid n \rangle},
\end{array}
\ee
where the r.h.s. can be expressed in terms of $R_m = e^{\phi_m - \phi_{m-1}}$.
In particular,
\be
\frac{\partial^2\phi_n}{\partial t_1\partial t_1} =
R_{n+1} - R_n = e^{\phi_{n+1} - \phi_n} - e^{\phi_n - \phi_{n-1}}.
\ee

Let  us also mention that in this formalism the Ward identities (Virasoro
constraints) follow essentially from the relation
\be
\left( \frac{\partial}{\partial h}\right)^\dagger =
- \frac{\partial}{\partial h} -
\sum_{k>0} kt_k h^{k-1},
\ee
where Hermitean conjugation is w.r.to the scalar product $\langle\ \mid\
\rangle$. For example, this relation implies, that
\be
\langle Q_n \mid \frac{\partial Q_n}{\partial h} \rangle =
- \langle \frac{\partial Q_n}{\partial h} \mid Q_n \rangle
- \sum_{k>0} kt_k \langle Q_n \mid h^{k-1} \mid Q_n \rangle.
\ee
Now we note that $\frac{\partial Q_n}{\partial h}$ is a polinomial of
degree $n-1$, thus
${\langle Q_n \mid \frac{\partial Q_n}{\partial
h} \rangle = 0}$. (In fact $$\displaystyle{
\frac{\partial Q_n}{\partial h} = -\sum_{k>0} kt_k
\left(\sum_{m=0}^{n-1} \gamma_{nm}^{(k-1)}Q_m \right) =
-\sum_{k>0} kt_k \frac{\partial Q_n}{\partial t_{k-1}}}.)$$
Also we recall that $\langle Q_n \mid h^{k-1} \mid Q_n \rangle =
\langle Q_n \mid Q_n \rangle \frac{\partial \phi_n}{\partial t_{k-1}}$,
and obtain:
\be
\sum_{k>0} kt_k\frac{\partial \phi_n}{\partial t_{k-1}} = 0
\ee
for any $n$. This should be supplemented by relation
$\frac{\partial \phi_n}{\partial t_0} = \phi_n$. In order to get the lowest
Virasoro constraint (string equation), $L_{-1}Z_N = 0$ or $L_{-1}\log Z_N = 0$
it is enough just to sum over $n$ from $0$ to $N-1$.

For more details about 1-matrix model, Toda-chain hierarchy and application of
the formalism of orthogonal polinomials in this context see \cite{GMMMO}.

\






\bigskip

\section{Continuum limits of discrete matrix models}

\setcounter{equation}{0}

\subsection{What is continuum limit}

Continuum limit of matrix models is, of course, the crucial issue for their
physical applications whenever these models are interpreted as discrete
(lattice) approximations to continuum theory. The very first thing to be kept
in mind is that it is $not$ the only possible view on matrix models. Another
approach considers them as describing $topological$ (and thus also in a
certain sense "discrete") properties of the theory. Such models, when appearing
in the field of, say, quantum gravity (which after all is a sort of a pure
topological theory) do $not$ require any continuum limit to be taken: their
discrete nature (occurence of {\it integer}-valued matrix indices)
reflects $not$ the discrete
approximation to the space-time (which does not really exist in quantum
gravity), but rather the essential discreteness of the underlying structures:
topology of the module spaces of geometries. Example of matrix models which
allow for this kind of interpretation - in terms of topology of module spaces
of bundles over Riemann surfaces - is provided by Kontsevich models, and this
is why they usually do not require any continuum limit and why we once called
them "continuous matrix models" in the Introduction to these notes. The models
which are usually interpreted in more traditional way - as lattice theories -
are represented by our "discrete" models, the 1-matrix, conventional and
"conformal" multimatrix models being included into this class. More
sophisticated examples are provided by "$c=1$"-theories, Kazakov-Migdal model
and, say, Wilson's QCD (and infinetely many other lattice theories). It is not
a surprise that continuum limits of some discrete models provide the theories
of Kontsevich type: this happens whenever continuum theory is supposed to have
a kind of topological nature. This is usually the case for quantum gravity
(which, as we said, is conceptually a topological theory in the "module space
of geometries" - the notion which is already made more or less explicit in the
$2d$ case), but in principle this can be also true for many other theories,
including exhaustive quantum theory of Yang-Mills fields (again there is
already considerable progress in this direction, as soon as $2d$ Yang-Mills
model is concerned). There should not be confusion about the presence of gauge
particles in dimensions greater than 2 (for Yang-Mills) and 3 (for gravity):
there is no reason to prevent generic topological theory from possessing
continuum spectrum of excitations, though explicit analogue of Kontsevich-like
description of such situations is not yet found (as we mentioned many times,
it should probably rely upon non-eigenvalue models).

We shall not discuss the non-trivial history of invention and understanding of
all these notions (the crucial steps being discovery of the "multiscaling
continuum limits" \cite{Kmamo},\cite{mamo}, which preserve integrable
structure of discrete models in continuum case; hypothesis of equivalence of
quantum and topological $2d$ gravities \cite{WitTG} and its proof
\cite{Witko},\cite{MMM}, provided by discovery of Kontsevich models \cite{Ko}
as a peculiar and powerful tool
for description of topology of the module spaces). Instead,
following the mean line of these notes,  we
shall concentrate on intrinsic relation between (multiscaling) continuum
limits and integrability: the notion of continuum limits is in fact built into
the theory of integrable hierarchies and the underlying representation theory
of Kac-Moody algebras.

In the case of the eigenvalue models the central issue here is the
interrelation between Toda-lattice and KP hierarchies, even its more narrow
aspect: elimination of the zero-time $N$, present in the Toda-lattice case.
{}From representation theory (or conformal field theory, what is essentially
the
same) point of view the thing is that the zero-time (which labels the filling
level
of Dirac sea in the fermionic picture) is associated with the zero-modes of
$scalar$ field and its elimination is just the change of boundary conditions
which eliminates zero-modes. The simplest example of this "twisting" procedure
is just transformation from periodic to antiperiodic scalars - it still
preserves possibility to have fermionic description (where it looks like a
switch from Ramond to Neveu-Schwarz sector), and thus does not take us out of
the field of conventional integrable hierarchies. In representation theory one
can interpret the same operation just as a switch from the
homogeneous to principal representation, which are associated with
the Toda-lattice and KP hierarchies respectively.

This remarkably simple description is of course far from obvious, if one
investigates continuum limit in naive way, without taking integrable structure
into account explicitly, but just sending the number of degrees of freedom in
discrete theory (i.e. the matrix size $N$) to infinity (together with the
inverse lattice spacing, if any). We refer to the classical review
\cite{Migrev}  for discussion of what are the naive continuum limits in
lattice gauge theories, i.e. what are the conditions for getting the
second-order phase transitions, which allow for a continuum-like scaling
behavious in the vicinity of the critical point, with critical exponents
defining all the continuum physics, from quantum dimesnion of the space-time
to spectrum of particles. The problem with naive continuum limits is that they
can easily destroy integrable structure of the theory (the underlying
hidden symmetries), unless special precaution is taken: the critical
point (which is in fact a low-codimensional hypersurface in the
infinite-dimensional space of parameters) should be approached from the
certain directions, so that Ward identities are not explicitly broken.

As soon as this word - Ward -dentities - is pronounced, we already get into
the field of integrable systems and the issue can be discussed inside this
field. The above-mentioned switch from periodic to antiperiodic fields is of
course apparent if the discrete and continuos Virasoro constraints
(represented by formulas
(1.2)
and
(1.3)
in the Introduction) are compared, but this is {\it a posteriori} information,
because so far we interpreted "continuous Virasoro constraints" as the
Ward-identities for the $V=X^3$-Kontsevich model, and it still remains to be
explained why Kontsevich model is indeed what arises after continuum limit is
taken. The simplest approach to {\it this} problem
is to make use of the identity
between discrete 1-matrix model and Gaussian Kontsevich model \cite{ChMa},
established in
section 3.8 above. Then the $X^3$-model arises in the large-$N$ limit just
when the matrix integral is evaluated by the steepest descent method
\cite{Toda}. We shall present this simple calculation in the last subsection
below, but before we take a somewhat more direct (and complicated)
approach in order to reveal at least some of ideas, underlying the entire
theory of continuum limits.

\subsection{From Toda-chain to KdV}

We begin with the simplest existing example: continuum limit, in which the
lowest  equation of the "Volterra hierarchy",
\be
\frac{\partial R_n}{\partial t} = - R_n(R_{n+1}-R_{n-1}),
\label{lowVolt}
\ee
turns into the lowest KdV equation:
\be
\frac{\partial r}{\partial T_3} = - \frac{1}{3}r''' - 2rr'.
\label{lowKdV}
\ee
Volterra hierarchy is a reduction of Toda-chain hierarchy, with
$R_n = e^{\phi_n-\phi_{n-1}}$, arising when all the odd-times $t_{2k+1} = 0$
and all $\phi_n$ are supposed to be independent of them. Therefore this
hierarchy is clearly related to the discrete 1-matrix model. We"ll turn to the
study of 1-matrix model in the next subsection, but here we just address the
transformation from (\ref{lowVolt}) to (\ref{lowKdV}) \cite{Nov},\cite{GMMMO}.

The basic idea of taking continuum limit is to change discrete "zero-time" $n$
for {\it continuum} variable $x$ (to be after all substituted by $T_1$ of the
continuous hierarchy). In other words, the idea is to consider a subset of
functions $R_n$, which satisfy Volterra equation and depend on $n$ very
smoothly, so that they can actually be substituted by a smooth function
$R(x)$. This is a very natural thing to do, of course, when one is interested
in the large-$n$ limit of the equation. Namely, one substitutes
(\ref{lowVolt}) by
\be
\frac{\partial R(x)}{\partial t} = - R(x)(R(x+\epsilon)-R(x-\epsilon)),
\label{lowVoltcon}
\ee
and take the limit $\epsilon \rightarrow 0$, which, after rescaling
$x \rightarrow \epsilon x$, gives rise to "Bateman equation",
\be
\frac{\partial R(x)}{\partial t} = - R(x)R'(x).
\label{Bateq}
\ee
This is a very interesting equation (see \cite{Fair} for description of the
amusements of the related theory, which is in fact intimately related to the
theory of jets). However, it is much simpler that KdV equation (for example,
it is {\it completely} integrable in the most trivial sense of the word:
entire set of solutions, satisfying {\it any} boundary conditions can be
immediately written down, see \cite{Fair}). KdV equation can be considered as
a sort of "quantization" of (\ref{Bateq}) (unfortunately this very interesting
subject did not yet attract enough attention and is not studied well enough).

Remarkably, Bateman equation is not the {\it only} possible limit of Volterra
equation: a fine tuning procedure ("double-scaling limit") exist, which can
provide less trivial - KdV - equation \cite{Nov}. Indeed, imagine, that in
continuum limit $R_n$ tends to a constant $R_0$, and the function $r(x)$
arises only as scaling approximation to this constant:
$R(x) = R_0(1 + \epsilon ^sr(x))$. Then the leading term at the r.h.s. of
(\ref{Bateq}) is $\epsilon RR'(x) =
-2\epsilon^s r(x) (1 + {\cal O}(\epsilon^2,\epsilon^s))$,
and instead of (\ref{Bateq}) we would get:
\be
\frac{\partial r}{\partial t} = - 2\epsilon R_0 r'(x) ((1 + {\cal
O}(\epsilon^2,\epsilon^s)).
\nn
\ee
This equation is even simpler that (\ref{Bateq}) - it is just linear, but in
fact it is too simple to preserve its form: by a simple change of
variables\footnote{
This change of variables is implied by the relation:
\be
\frac{\partial}{\partial t} + 2\epsilon^sR_0 \frac{\partial}{\partial x} =
\left( \frac{\partial \tilde t}{\partial t} +
2\epsilon^sR_0 \frac{\partial \tilde t}{\partial x}\right)
\frac{\partial}{\partial \tilde t} +
\left( \frac{\partial \tilde x}{\partial t} +
2\epsilon^sR_0 \frac{\partial \tilde x}{\partial x}\right)
\frac{\partial}{\partial \tilde x} =
\frac{\partial}{\partial \tilde t}.
\ee
}
\be
\tilde x = x - 2\epsilon R_0 t, \\
\tilde t = \epsilon^3 R_0 t
\label{chavacont}
\ee
it can be transformed into
\be
\frac{\partial r}{\partial \tilde t} = \epsilon^{-2} {\cal
O}(\epsilon^2,\epsilon^s),
\nn
\ee
and terms at the r.h.s. also deserves beeing taken into account.
Then we get:
\be
\frac{\partial r(x)}{\partial t} = -2\epsilon R_0 \left(1 + \epsilon^sr(x))
(r'(x) + \frac{1}{6}\epsilon^2r'''(x) + {\cal O}(\epsilon^4)\right) = \nn \\
= -2\epsilon R_0 \left( r'(x) + \frac{1}{6}\epsilon^2r'''(x)
+ \epsilon^s rr'(x) + \epsilon^2{\cal O}(\epsilon^2, \epsilon^s)\right)
\nn
\ee
and, after the change of variables (\ref{chavacont}),
\be
\frac{\partial r(\tilde x)}{\partial \tilde t} = - \frac{1}{3} r'''(\tilde x)
- 2\epsilon^{s-2}rr'(\tilde x) + {\cal O}(\epsilon^2, \epsilon^s).
\nn
\ee
It is now clear, that the choice $s=2$  is distinguished (a critical point)
and at this point we get:
\be
\frac{\partial r}{\partial T_3} =
-\frac{1}{3}\frac{\partial^3r}{\partial T_1^3} -
2r\frac{\partial r}{\partial T_1},
\ee
where new notation $T_1$ and $T_3$ is introduced for $\tilde x$ and $\tilde t$
respectively. This is already the KdV equation (\ref{lowKdV}), and our
conclusion is:

While the naive continuum limit of Volterra equation is just a
simple Bateman equation, the scaling limit can be fine tuned so that KdV
equation arises instead. The crucial ingredient of this adjustement is the
change of time-variables $\{t\} \longrightarrow \{T\}$, which involves
singular parameter $\epsilon$. The procedure can be easily generalized to the
entire Volterra hierarchy, and fine tuning allows to get the entire KdV
hierarchy in the limit of $\epsilon \rightarrow 0$. Usually transformation to
the "Kazakov variables" $\{T\}$ (they are a little different from those
originally introduced by V.Kazakov in \cite{Kmamo})
from $\{t\}$ is some linear {\it triangular} transformation.

An important detail is that this procedure
requires restriction to only {\it even} time-variables $t_{2m},\ m\geq 0$.
(If odd times are also involved, a {\it pair} of KdV hierarchies arises in the
continuum limit - this is not a "minimal" case.) Thus "irreducible"
realization of continuum limit requires {\it reduction} of original hierarchy.
This can be also seen from the fact that the lowest KdV equation arises from
the lowest Volterra equation, which is related to the {\it second} eqaution of
Toda-chain hierarchy.

Unfortunately this simple piece of theory (continuum limits in terms of
hierarchies) has never been worked out in full details (for the entire
Toda-lattice hierarchy, its multicomponent generalizations and their
reductions). As we already mentioned, this theory will involve the general
relation between homogeneous and principal representations of the (level
$k=1$) Kac-Moody algebras.

\subsection{Double-scaling limit of 1-matrix model}

Now we proceed to discussion of a slightly different approach to continuum
limits, which is directly adjusted to the needs of matrix models. The naive
idea \cite{mamo},\cite{FKN}
is to forget about integrability and just look at the Ward
identities (Virasoro constraints in the 1-matrix case) and take a continuum
limit of these identities. This approach makes close contact with the standard
technique of "loop equations" (Makeenko-Migdal equations \cite{MaMi})
in the theory of
matrix models, of which Virasoro and $W$-constraints are just particular
examples.\footnote{One of the puzzles in the theory of non-eigenvalue
models is to identify group-theoretical meaning of generic loop equations:
they are usually introduced as equations of motion rather than as Ward
identities (see discussion at the beginning of Section 2 above), and thus
their implications are more obscure and technical means to deal with them are
much more restricted. When group theory description will be found, it will
very soon reveal the (generalized) integrable structure of non-eigenvalue
models and it will be a big step forward in the whole theory.
}

However, carefull analysis of continuum limit of discrete Virasoro constraints
\cite{MMMM} makes it clear that the procedure is far less simple than one can
think in advance (usually derivations are not very carefull and details are
just "put under the carpet"). The crucial problem is that we want peculiar
(double scaling) rather than naive limit, and, as we mentioned in the previous
subsection, this also requires a certain reduction (elimination of the
odd-times $t_{2m+1}$).  If parity symmetry
(w.r.to the change of $H \rightarrow -H$ in the original matrix integral) is
taken into account, one
can easily throw away {\it first} derivatives w.r.to the odd-times $t_{2m+1}$,
just because
$\displaystyle{\left.\frac{\partial Z_N}{\partial t_{2m+1}}\right| _{t_{2k+1}
= 0} = 0}$, but this is no longer true as far as the {\it second} derivatives
$\displaystyle{\left.\frac{\partial^2 Z_N}{\partial t_{2m+1}\partial t_{2l-1}}
\right| _{t_{2k+1} = 0}}$ are concerned,
which appear in (the "quantum piece" of) the
Virasoro constraints (1.2). It is a highly non-trivial feature of loop
equations (having its origin in their integrable structure!), that in
continuum limit these terms can be in fact carefully eliminated.
The thing is that the second derivatives of $\log Z_N$ appear to be a {\it
local} objects, in the sense that they depend only on $Z_{\tilde N}$ with the
difference $\mid \tilde N-N\mid \leq m+l$, which does {\it not} blow up as $N
\rightarrow \infty$ in continuum limit. Moreover, the differences
$\displaystyle{\frac{\partial^2 \log Z_N}{\partial t_{2m+1}\partial t_{2l-1}} -
\frac{\partial^2 \log Z_N}{\partial t_{2m}\partial t_{2l}}}$ almost tend to
zero, leaving some simple (though vitaly important) correction to arising
continuous loop equations. This locality property allows one to
get rid of these dangerous odd-time derivatives,
substituting them just by second derivatives
w.r.to the even-times. Since such substitution is possible
only  for {\it logarithms} of $Z_N$, continuous constraints appear imposed on
the {\it square root} of original partition function (or on the
$\frac{1}{p}$-th power in the case of the $p-1$-component conformal models).
Another aspect of this trick to deal with the odd-time derivatives is that it
makes the entire derivation depending on the
fact that the theory is integrable - this is what guarantees the
above-mentioned
locality.  Since the way to reveal integrability,
by looking at the loop equations themselves is yet not very well
understood, the whole calculation becomes not quite self-contained (but of
course, if we know everything about integrable structure this is not a real
drawback, this is just a limitation of particular approach, starting from the
loop equations) . In particular, this is the only
loophole, which is still not filled in the description of continuum limit of
conformal (multi-component) matrix models, which in all other respects goes in
exactly in parallel with the 1-component (1-matrix) case.\footnote{
It transforms discrete $W$-constraints into continuum $W$-constraints, which
in their turn arise from the GKM with the appropriate potential \cite{GKM},
\cite{Mikh}. Unfortunately, since the GKM-inteprpretation of {\it discrete}
multicomponent models (like the one existing in the 1-matrix case, see s.3.8)
is yet unknown, the direct way to take their continuum limit - like the one
to be described in the next subsection for the 1-matrix case - is also yet
unavailable. For more details about conformal matrix models, their integrable
structure and continuum limits see refs.\cite{comamo}.
}

We shall now describe briefly the steps of this calculation for the 1-matrix
model, refering for all the details to refs.\cite{MMMM} and \cite{GMMMMO}.
Our previous discussion already contain motivations for the main steps, so we
do not need to go into detailed explanations. Manipulations below, involving
Kazakov variables can look a little artificial, but we repeat that they can be
interpreted as a switch from the Toda-type to KP-type hierarchies, which, as
we already saw in the previous subsection, is naturally associated with the
double-scaling continuum limit.

We start from the discrete Virasoro constraints (1.2), rewritten in terms of a
generating functional ("stress tensor" on the spectral plane):
\be
L_-(z) Z_N = 0,
\ee
where
\be L_-(z) = \sum_{n\geq -1}^{\infty} L_n z^{-n-2} = \frac{1}{2}\left(
J^2(z)\right)_-,
\ee
and
\be
\new
\begin{array}{c}
J(z) = \partial \phi(z) = \sum_{n= -\infty}^{\infty}J_n z^{-n-1}; \\
\phi(z) = \frac{1}{\sqrt{2}}\sum_{k\geq 0}t_kz^k -
\sqrt{2} \sum_{k>0} \frac{z^{-k}}{k}\frac{\partial }{\partial t_k};  \\
J_{-k} = \sqrt{2}\frac{\partial }{\partial t_k}; \ \ J_k =
\frac{1}{\sqrt{2}}kt_k, \ \ k\geq 0;  \\
\frac{\partial}{\partial t_0}Z_N = NZ_N.
\end{array}
\ee

Next, we need to reduce the original partition function:
\be
Z_N\{t\} \longrightarrow Z_N^{\rm red}\{t_{\rm even}\} \equiv
Z_N\{t_{\rm odd}=0, t_{\rm even}\}.
\ee
All odd virasoro generators $L_{2n+1}$
act trivially on $Z_N^{\rm red}$, since
$\displaystyle{\left.\frac{\partial Z_N}{\partial t_{2k+1}}\right|
_{t_{\rm odd}=0} = 0}$, and we need to consider only $L_{2n}$.
Introduce also\footnote{
Note that $\displaystyle{
\left.\phi^{\rm red}(z) \neq \phi(z) \right|_{t_{\rm odd}=0}}$ and similarly
$\displaystyle{
\left.L_{2n}^{\rm red}(z) \neq L_{2n} \right|_{t_{\rm odd}=0}}$: some factors
of $2$ in (\ref{redfields}) being responsible for this discreapancy. In fact
$L^{\rm red}$ are related to generators of the Virasoro constraints in the
{\it complex}-matrix model \cite{MMMM},
\be
Z_N^C = \int dM \exp \left(\sum_{k\geq 0}t_{2k} {\rm Tr} (MM^\dagger)^k\right)
\nn
\ee
and in continuum limit $\displaystyle{Z_N^C \sim \sqrt{Z_{2N}^{\rm red}}}$.
}
\be
\phi^{\rm red}(z) &\equiv \frac{1}{\sqrt{2}}\sum_{k\geq 0} t_{2k}z^{2k} -
\sqrt{2}\sum_{k>0} \frac{z^{-2k}}{k}\frac{\partial}{\partial
t_{2k}}; \nn \\
L^{\rm red}(z) &= \frac{1}{2}\left(\partial \phi^{\rm red}(z)\right)^2; \nn \\
L_{2n}^{\rm red} &\equiv \sum_{k>0} kt_{2k}\frac{\partial}{\partial t_{2k+2n}}
+
\sum_{k=0}^n \frac{\partial^2}{\partial t_{2k}\partial t_{2n-2k}}.
\label{redfields}
\ee

Now we have two issues to be discussed separately. The first one is the change
from $t_{2k}$ to Kazakov variables $T_{2m+1}$. The second is the difference
between constraints imposed on $Z^{\rm red}$ and $Z$.

The simplest way to {\it describe} Kazakov variables is to introduce one more
- antiperiodic - scalar field,
\be
\Phi(u) = \frac{1}{\sqrt{2}} \sum_{k\geq 0} T_{2k+1} u^{k+\frac{1}{2}}
- \sqrt{2} \sum_{k\geq 0} \frac{u^{-k-\frac{1}{2}}}{k+\frac{1}{2}}
\frac{\partial}{\partial \tilde T_{2k+1}}.
\ee
Here $\tilde T$ and $T$ are related by transfromation
\be
T_{2k+1} = \tilde T_{2k+1} + \epsilon^2 \frac{k}{k+\frac{1}{2}}\tilde
T_{2k-1} + 2\epsilon N\delta_{k,0}.
\ee
Impose now a relation:
\be
\partial\phi^{\rm red}(z) &= \frac{1}{\epsilon^2}U^{-1}
\partial\Phi(u) U; \nn \\
z^2 &= 1 + \epsilon^2 u,
\label{preKaz}
\ee
and in continuum limit $\epsilon$ is assumed to vanish.
This is a relation which maps homogeneous representations into principal, but
its invariant meaning (especially - from the point of view of conformal field
theory) does not seem to be enough understood. Anyhow, this relations
establishes a relation between $t_{\rm even}$ and $T$. Namely, comparing
the coefficients in front of the positive powers of $u$
at both sides of this equation, we get:
\be
T_{2k+1} &= \frac{1}{2}\epsilon^{2k+1}
\sum_{m\geq k}^{\infty} \frac{g_m
\Gamma(m+\frac{1}{2})}{(m-k)!\Gamma(k+\frac{3}{2})}, \ \ k\geq 0; \nn \\
g_m &= mt_{2m},\ \ m\geq 1; \ \ \ \ g_0 = 2N.
\label{Kazchava}
\ee
Inverse transformation looks like
\be
g_m = 2\sum_{k\geq m} (-)^{k-m}
\frac{T_{2k+1}\Gamma(k+\frac{3}{2})}
{\epsilon^{2k+1}(k-m)!\Gamma(m+\frac{1}{2})} .
\ee
Now,
\be
\frac{\partial}{\partial t_{2k}} =
\frac{1}{2}\sum_{m=0}^{k-1} \frac{\Gamma(k+\frac{1}{2})\epsilon^{2k+1}}
{(k-m-1)!\Gamma(m+\frac{3}{2})}\frac{\partial}{\partial\tilde T_{2m+1}},
\ee
and using this formula when comparing the negative powers of $u$ we find:
\be
U &= \exp\left(\sum_{m,n} A_{mn}\tilde T_{2m+1}\tilde T_{2n+1}\right), \nn \\
A_{mn} &= 2\frac{(-)^{m+n}}{\epsilon^{2(m+n+1)}}\cdot
\frac{\Gamma(m+\frac{3}{2})\Gamma(n+\frac{3}{2})}
{m!n!(m+n+1)(m+n+2)}.
\label{Umatrix}
\ee
The square of relation (\ref{preKaz}) is:
\be
\left(\partial\phi^{\rm red}\right)^2(z) = \frac{1}{\epsilon^4}U^{-1}
\left(\partial\Phi\right)^2(u) U,
\label{preKazVir}
\ee
or
\be
\sum_{p\geq 0}L_{2p}^{\rm red} z^{-2p-2} =
\frac{1}{\epsilon^4}U^{-1}
\left(\sum_{n\geq -1} \tilde{\cal L}_{2n} u^{-n-2}\right) U.
\ee
This equality implies that
\be
\new
\begin{array}{c}
U^{-1}\tilde{\cal L}_{2n} U = \epsilon^4 \sum_{p\geq 0}L_{2p}^{\rm red}
\oint_{\infty}\frac{u^{n+1}du}{z^{2p+2}} =  \\ =
\epsilon^{-2n} \sum_{p=0}^{n+1} (-)^{n+1-p} C^p_{n+1} L_{2p}^{\rm red},
\end{array}
\ee
since
\be
\new
\begin{array}{c}
\epsilon^4\oint_{\infty}\frac{u^{n+1}du}{z^{2p+2}} =
\oint_{\infty}\frac{u^{n+1}du}{(1+\epsilon^2 u)^{p+1}}
=  \frac{1}{\epsilon^{2n}}\frac{\Gamma(-p)}{(n+1-p)!\Gamma(-n-1)} = \nn \\ =
\frac{(-)^{n+p+1}}{\epsilon^{2n}}\frac{(n+1)!}{p!(n+1-p)!}
= \frac{(-)^{n+1-p}}{\epsilon^{2n}}C_{n+1}^p.
\end{array}
\ee
Explicit expressions for the generators $\tilde{\cal L}_{2n}$ (which are
harmonics
of the stress tensor
$\frac{1}{2}\left(\partial\Phi\right)^2(u)$ of antiperiodic field $\Phi(u)$),
are:
\be
\new
\begin{array}{c}
\tilde{\cal L}_{-2} = \sum_{k\geq 1} (k+\frac{1}{2}) T_{2k+1}
\frac{\partial}{\partial \tilde T_{2(k-1)+1}} + \frac{T_1^2}{4};  \\
\tilde{\cal L}_{0} = \sum_{k\geq 0} (k+\frac{1}{2}) T_{2k+1}
\frac{\partial}{\partial \tilde T_{2k+1}};  \\
\tilde{\cal L}_{2n} = \sum_{k\geq 0} (k+\frac{1}{2}) T_{2k+1}
\frac{\partial}{\partial \tilde T_{2(k+n)+1}} +  \\
+ \frac{1}{4}\sum_{k=0}^{n-1}
\frac{\partial^2}{\partial \tilde T_{2k+1}\partial \tilde T_{2(n-k-1)+1}} -
\frac{(-)^n}{16\epsilon^{2n}}; \ \ n>0.
\end{array}
\label{vircotil}
\ee
So far everything what was done was just change of variables and all relations
were exact for any $\epsilon$, no limits were taken.

Operators (\ref{vircotil}) are very similar to ${\cal L}_{2n}$, arising in the
"continuous Virasoro constraints" (1.3), imposed on partition function of
$X^3$-Kontsevich model. There are, however, two discreapancies.

First, $\frac{\partial}{\partial \tilde T}$ appear in (\ref{vircotil}) instead
of $\frac{\partial}{\partial T}$ in generators in (1.3). One can argue that
this difference is not really essential, since $\tilde T_{2k+1}$ and
$T_{2k+1}$ differ by terms, which are proportional to $\epsilon^2$ and thus
vanish in the continuum limit $\epsilon \rightarrow 0$.
(Note, however, that this reasoning can be applied only for every particular
constraint $\tilde{\cal L}_{2n}Z = 0$, $n \geq -1$,
not to the entire generating functional,
where different terms are summed, multiplied by different powers of
$\epsilon$.)

The second discreapancy is a litlle more serious: it is the occurence of an
extra term $\displaystyle{\frac{(-)^{n+1}}{16\epsilon^{2n}}}$ for all $n\geq
0$ (this difference is present for $n=0$ as well, because ${\cal L}_0$
contains the item $\frac{1}{16}$, which is lacking in (\ref{vircotil}).)
This extra term can not be eliminated by just taking continuum limit:
moreover, it blows up instead of vanishing when $\epsilon \rightarrow 0$.
Remarkably enough, this term disappears when we turn to consideration of
actual Virasoro constraints, not just a formal choice of time variables. It
cancels completely with the other potential source of problem for the
derivation of continuous Ward identites. We proceed now to this most
sophisticated matter in this whole subsection.

The thing is that, as we already mentioned before, the reduction of discrete
Virasoro constraint $L_{2n}Z_N = 0$ contains some non-vanishing terms with the
odd-time  derivatives:
\be
\new
\begin{array}{c}
\left(\sum_{k>0} 2kt_{2k} \frac{\partial}{\partial t_{2k+2n}} + 2
\sum_{k=0}^{n} \frac{\partial^2}{\partial t_{2k}\partial t_{2n-2k}}\right)
Z_N^{\rm red} =                                    \\ =
\left(\sum_{k=0}^{n} \frac{\partial^2}{\partial t_{2k}\partial t_{2n-2k}}
- \sum_{k=0}^{n-1} \frac{\partial^2}
{\partial t_{2k+1}\partial t_{2n-2k-1}}\right)  Z_N^{\rm red}.
\end{array}
\label{colivirder0}
\ee
We added an extra term with the second even-time derivatives to both sides of
the identity, in order to get at the r.h.s. a combination, which has a chance
to vanish in continuum limit. (This formula still needs to be corrected, see
eq.(\ref{colivirder3}) below.)

In order to find rigorous reason for elimination of the terms at the r.h.s we
need to address to explicit formulas from the last subsection of section 4
(no simpler way is known so far). The crucial formula  which we need is:
\be
\frac{\partial^2\phi_n}
{\partial t_k\partial t_l} = \frac{\partial}{\partial t_k}
\frac{\langle n \mid h^l \mid n \rangle}{\langle n \mid n \rangle}
= \left( \sum_{m>n} - \sum_{m<n}\right)
\frac{\langle n \mid h^k \mid m \rangle \langle m \mid h^l \mid n \rangle}
{\langle m \mid m \rangle\langle n \mid n \rangle},
\label{colivirder1}
\ee
and the most important feature of it is its $R$-matrix structure (the
fact that a {\it difference} occurs at the r.h.s.). This structure implies
almost complete cancellation of terms, when we sum over $n$ in order to get
$\log Z_N = \sum_{0}^{N-1} \phi_n$, leaving only a finite sum of the length
{\it independent} of N:
\be
\frac{\partial^2\log Z_N}
{\partial t_k\partial t_l} =
\sum_{0 < j < {\rm min}(k,l)} \left( \sum_{n = N-j}^{N-1}
\frac{\langle n \mid h^k \mid n+j\rangle \langle n+j \mid h^l \mid n \rangle}
{\langle n \mid n \rangle\langle n+j \mid n+j \rangle}\right).
\ee
The finite sum at the r.h.s. can be expressed in terms of $R_n =
e^{\phi_n-\phi_{n-1}}$, which are exactly the quantites to satisfy equations
of Volterra hierarchy and tending to {\it constant} (denoted by  $R_0$ in the
previous section) in continuum limit. Locality property - the finiteness of
the sum at the r.h.s in (\ref{colivirder1}) - implies that this r.h.s. tend to
a constant value as $N \rightarrow \infty$. This constant does not completely
cancels in the difference
\be
\left(\sum_{k=0}^{n} \frac{\partial^2}{\partial t_{2k}\partial t_{2n-2k}}
- \sum_{k=0}^{n-1} \frac{\partial^2}
{\partial t_{2k+1}\partial t_{2n-2k-1}}\right)  \log Z_N^{\rm red},
\label{colivirder2}
\ee
and the remaining contributions appears to be exactly what necessary to cancel
the dangerous term
$\displaystyle{\frac{(-)^{n+1}}{16\epsilon^{2n}}}$
which appeared in the difference between $\tilde{\cal L}_n$ and ${\cal L}_n$.
We refer to ref.\cite{MMMM} for more details about these cancellations, and
the only thing to discuss at the rest of this subsection is the difference
between the r.h.s. of (\ref{colivirder0}) and (\ref{colivirder2}).
In the second expression the second derivatives are taken of $\log Z$, while
they are of $z$ itself in the first one. Of course,
\be
\frac{\partial^2 \log Z_N^{\rm red}}
{\partial t_{2k+1}\partial t_{2n-2k-1}}
=   \frac{1}{Z_N^{\rm red}} \frac{\partial^2  Z_N^{\rm red}}
{\partial t_{2k+1}\partial t_{2n-2k-1}}
\nn
\ee
but this is $not$ true for even derivatives. So, identity (\ref{colivirder0})
yet needs to be transformed a little more in order to contain exactly
(\ref{colivirder1}) at its r.h.s. If this is achieved, the l.h.s. acquires
additional contribution and turns into
\be
\new
\begin{array}{c}
\sum_{k>0} 2kt_{2k} \frac{\partial Z_N^{\rm red}}{\partial t_{2k+2n}} +
\sum_{k=0}^{n} \left( 2\frac{\partial^2 Z_N^{\rm red}}
{\partial t_{2k}\partial t_{2n-2k}}
- \frac{1}{Z_N^{\rm red}}
    \frac{\partial Z_N^{\rm red}}
{\partial t_{2k}}\frac{\partial Z_N^{\rm red}}{\partial t_{2n-2k}}\right)
=  \\
= 4\sqrt{Z_N^{\rm red}} L_{2n}^{\rm red} \sqrt{Z_N^{\rm red}}.
\end{array}
\label{colivirder3}
\ee

As a result of all this reasoning we conclude that the double scaling
continuum limit of reduced 1-matrix model can be described by the following
relation:
\be
\lim_{{\rm d.s.}\ \epsilon \rightarrow 0,\ N \rightarrow\infty}
\sqrt{Z_N^{\rm red}\{t_{\rm even}\}} \ = \ U^{-1}Z_{V = \frac{X^3}{3}}\{T\},
\ee
where factor $U$ is defined in (\ref{Umatrix}),  relation between $t$ and
$T$-variables is given by (\ref{Kazchava}) and
$Z_{V = \frac{X^3}{3}}\{T\}$ is $X^3$-Kontsevich model. The motivation for
this conclusion is that both sides of the equation satisfy the same continuous
Virasoro constraints (1.3).

This whole derivation can be straightforwardly generalized to the case of
multiscaling limit in conformal matrix models and analogous relation contains
roots of the $p$-th degree, see \cite{comamo} for detailed discussion.

\subsection{From Gaussian to $X^3$ Kontsevich model}

We shall now abandon these complicated matters and give a simple illustration
of how the things can work, if expressed in the adequate terms. Namely, as
alternative to the sophisticated procedure, involving explicit
switch to Kazakov
variables and the study of limits of Ward-identites (loop equations), we shall
just use the equivalence of the discrete 1-matrix model and Gaussian
Kontsevich model, proved in the section 3.8 above in order to take the
continuum limit just of this simplest Kontsevich model. This procedure,
suggested in ref.\cite{Toda}  appears
to be just a kind of a standard eveluation of the integral in the large
$N$-limit by the steepest descent method. It is important here that GKM is not
sensitive to the size of the matrix $n$ in Kontsevich integral, therefore this
limit, when expressed in terms of GKM, has nothing to do with the infinitely
large matrices.

Relation to be proved below is
\be
\lim_{d.s.\hbox{ }N\rightarrow \infty }{\cal F}_{\{\hat V\}} =
{\cal F}^2_{\{V\}},
\label{coligako}
\ee
where $\hat V(X) = \frac{1}{2}X^2 - N\log X$ and $V(X) = \frac{1}{3}X^3$.

Very naively, what happens as $N \rightarrow \infty$ is that in the Kontsevich
integral,
\be
\int dX \exp {\rm tr}\left(-\frac{1}{2}X^2 + N\log X + \Lambda X\right)
\ee
a stationary point arises at $X = X_0$, such that
\be
X_0 = \frac{N}{X_0} + \Lambda.
\ee
Expansion of this action in powers of $\tilde X = \gamma^{-1} (X - X_0)$ comes
entirely from the logarithmic piece:
\be
S - S_0 &= \frac{\gamma^2}{2}\tilde X^2 - N\left( \log\left(1 + \frac{\gamma
\tilde X}{X_0}\right) - \frac{\tilde X}{X_0}\right) = \nn \\
&= \frac{\gamma^2}{2}\left( 1 + \frac{N}{X_0^2}\right) \tilde X^2
+ \sum_{k\geq 3} \frac{N}{k}\left(-\gamma\frac{\tilde X}{X_0}\right)^k.
\label{actexpan}
\ee
In the continuum limit $\gamma$ should be adjusted in such a way, that
quadratic term is finite, i.e.
${\gamma \sim \left( 1 + \frac{N}{X_0^2}\right)^{-1/2}}$.
Now, if $\Lambda$ remains finite as $N \rightarrow \infty$, $X_0 \sim
\sqrt{N}$,
$\gamma \sim 1$ and all the terms with $k\geq 3$ in the sum are damped as
$\gamma^kNX_0^{-k} \sim N^{1-\frac{k}{2}}$. This is the naive continuum limit.
However, it is clear, that one can usually ask $\Lambda$ to behave more
adequately - blow up together with growth of $N$ -  and fine tune the way in
which it tends to infinity so that at last the first term with $k=3$ also
survives. For this purpose $\Lambda$ and thus $X_0$ should scale in such a
way, that both quantites
$\displaystyle{\gamma^2\left( 1 + \frac{N}{X_0^2}\right)}$ and
$\displaystyle{\frac{N\gamma^3}{X_0^3}}$ remain finite. This requirement in the
case of the latter expression means that
$\gamma \sim X_0N^{-1/3}$ and then
\be
\gamma^2\left( 1 + \frac{N}{X_0^2}\right) \sim \frac{N + X_0^2}{N^{2/3}}.
\nn
\ee
This is never finite, unless $N + X_0^2 \rightarrow 0$ as $N \rightarrow
\infty$. This in turn implies that $X_0 \sim i\sqrt{N}$ and
$\Lambda \rightarrow 2X_0 \sim  2i\sqrt{N}$ should be pure imaginary. One can
also check that the terms with $k>3$ in the sum (\ref{actexpan}) all tend to
zero in this specific limit. Thus we are left with a model which has only
cubic and quadratic terms in the action. By simple shift of variables
quadratic term can be changed for a linear one and we get a description of the
theory in the vicinity of the stationary point in terms of an $X^3$-Kontsevich
model.

In practice things are a little more complicated, because also reduction to
even-times should be taken into account. However, this does not really add too
many new problems. We need that only even times
${t_{2k} = {1\over 2k} {\rm tr}{1\over \Lambda ^{2k}}}$
remain non-vanishing, while all the odd times  ${t_{2k+1} = {1\over
2k+1} {\rm tr}{1\over \Lambda ^{2k+1}} = 0}$.
This obviously implies that the matrix $\Lambda$ should be of block form:
\be
\Lambda = \left(
\begin{array}{cc}
{\cal M} & 0\\0 & -{\cal M}
\end{array}\right)
\ee
and, therefore, the matrix integration variable is also naturally decomposed
into block form:
\be
X = \left(
\begin{array}{cc}
{\cal X} & {\cal Z}\\{\cal Z} & {\cal Y}
\end{array}
\right) .
\ee
Then
\be
\new
\begin{array}{c}
{\cal F}_{\{\hat V=X^2/2-N\log X\}}
= \int d{\cal X}d{\cal Y}d^2{\cal Z}\ \\
{\rm det} ({\cal X}{\cal Y}-\bar {\cal Z}
{1\over {\cal Y}}{\cal Z}{\cal Y})^N
e^{-{\rm tr}\{|{\cal Z}|^2+{\cal X}^2/2+{\cal Y}^2/2-{\cal M}{\cal X}+
{\cal M}{\cal Y}\}}.
\end{array}
\ee
To take the limit $N \rightarrow \infty $, one should assume certain scaling
behaviour of ${\cal X}$, ${\cal Y}$ and ${\cal Z}$. Moreover, our previous
naive consideration gave us some feeling of the
{\it fine tuned} scaling behaviour can look like. So we take
\be
\new
\begin{array}{c}
{\cal X} = \gamma (i\beta I + x), \\
{\cal Y} = \gamma (-i\beta I + y),      \\
{\cal Z} = \gamma \zeta , \\
{\cal M} = \gamma ^{-1}(i\alpha I + m)
\end{array}
\ee
with some large real $\alpha $, $\beta $ and $\gamma $. If expressed through
these variables, the action becomes:
\be
\new
\begin{array}{c}
{\rm tr}\left(|{\cal Z}|^2 + {\cal X}^2/2 + {\cal Y}^2/2 - {\cal M}{\cal X} +
{\cal M}{\cal Y} - N\log ({\cal X}{\cal Y} - \bar {\cal Z}{1\over
{\cal Y}}{\cal Z}{\cal Y})\right) = \\
= \gamma ^2{\rm tr}\left(\frac{1}{2}(i\beta I + x)^2 +
{1\over 2}{\rm tr}(i\beta I -y)^2 + |z|^2\right) - \\
- {\rm tr}(i\alpha I + m)(2i\beta I + x - y) - \\
- N{\rm tr} \log \ \beta ^2\gamma ^2\left(1 - i {x-y\over \beta } +
{xy\over \beta ^2} - {|\zeta |^2\over \beta ^2}(1 + {\cal O}(1/\beta ))\right)
=
\end{array}
\ee
$$
= [2\alpha \beta  - \beta ^2\gamma ^2 - 2N\ \log \ \beta \gamma ] {\rm tr}\ I -
2i\beta \ {\rm tr}\ m +
\eqno{(A)}
$$
$$
+ i\left(\beta \gamma ^2 - \alpha  + {N\over \beta }\right)
({\rm tr}\ x - {\rm tr}\ y) +
{1\over 2}\left(\gamma ^2 - {N\over \beta ^2}\right)
({\rm tr}\ x^2 + {\rm tr}\ y^2) +
\eqno{(B)}
$$
$$
+ \left(\gamma ^2 + {N\over \beta ^2}\right) {\rm tr} |\zeta |^2 -
\eqno{(C)}
$$
$$
- {\rm tr}\ mx + {\rm tr}\ my + {iN\over 3\beta ^3}{\rm tr}(x^3 - y^3) +
\eqno{(D)}
$$
$$
+ {\cal O}(N/\beta ^4) + {\cal O}\left(|\zeta |^2 {N\over \beta ^3}\right).
\eqno{(E)}
$$
We want to adjust the scaling behaviour of $\alpha $, $\beta $ and $\gamma $
in
such a way that only the terms in the line $(D)$ survive. This goal is
achieved
in several steps.

The line ({\it A}) describes normalization of functional integral, it does not
contain $x$ and $y$. Thus, it is not of interest for us at the moment.

Two terms in the line $(B)$ are eliminated by adjustment of $\alpha $ and
$\gamma $:
\be
\gamma ^2 = {N\over \beta ^2}\hbox{ , }  \alpha  = {2N\over \beta }\hbox{ .}
\ee
As we shall see soon,  $\gamma ^2 = N/\beta ^2$ is large in the limit of
$N\rightarrow \infty $ . Thus, the term $(C)$ implies that the fluctuations of
$\zeta $-field are severely suppressed, and this is what makes the terms of
the
second type in the line $(E)$ negligible. More general, this is the reason for
the integral  $Z_{\{\hat V\}}$
to split into a product of two independent integrals leading to the square
of partition function in the limit $N\rightarrow \infty $ (this splitting is
evident as, if ${\cal Z}$ can be neglected, the only mixing term
$\displaystyle{\log \det
\left(
\begin{array}{cc}
{\cal X} & {\cal Z}\\{\cal Z} & {\cal Y}
\end{array}
\right)} $  turns into  $\log {\cal X}{\cal Y} = \log {\cal X} +
\log {\cal Y}$).

Thus, we remain with a single free parameter $\beta $ which can be adjusted so
that
\be
{\beta ^3\over N}\rightarrow {\rm const}\ \ \
\hbox{   as }\ \ \ N\rightarrow \infty \nn \\
(i.e\hbox{. }  \beta  \sim  N^{1/3},\ \gamma^2 \ \sim N^{1/3},\ \alpha \sim
N^{2/3}  ),
\ee
making the terms in the last line $(E)$ vanishing and the third term in the
line $(D)$ finite.

This proves the statement (\ref{coligako}) in a rather straightforward way.
Unfortunately no generalization of this procedure for other discrete models is
found so far, the main problem beeing identification of GKM-type realizations
of other (for example, conformal) discrete matrix models.

\

\section{Conclusion}

We came to the end of our brief review of the facts, that are already known
about the relation between matrix models and integrable hierarchies. There are
still several topics, which are already discussed in the literature, but not
presented in these notes.

First of all, we did not discuss the relation between matrix models and
theories of topological (Landau-Ginzburg) gravity (LGG). This field is fastly
developing during the last months and will be soon ready to inclusion in this
kind of reviews. The list of things which are already clarified enough,
includes realization of the Ward identities in the form of "recursion
relations" for topological gravity \cite{WitTG}. Also the relation between
quasiclassical hierarchies, arising in the spherical approximation to
topological theories \cite{KriDu}, to integrable structure of Generalized
Kontsevich model is more or less understood \cite{Krich}. Of special
importance is the chapter of this theory, which provides matrix-model
description of module spaces, associated with Riemann surfaces
\cite{Penner},\cite{Ko}.
What still deserves better understanding is axionatic construction of
topological gravity, similar to remarkably simple construction of topological
LG models (before they are coupled to $2d$ gravity) in terms of the
Grothendieck residues and chiral rings \cite{chiri}: see \cite{Rabi} for a
very nice presentation  of the latter case and
\cite{LP} for the first big steps towards similar construction in the former
case. Also relation to the theory of non-conformal LG models \cite{CeVa}
deserves being clarified. A piece which is essentially lacking so far is the
clear description of minimal $(p,q)$-models, coupled to $2d$ gravity in the
case of $p\neq 1$. In this situation Generalized  Kontsevich model is known to
describe nothing more but duality transformation between $(p,q)$ and $(q,p)$
models \cite{KhMa}, rather than the models themselves. This subject is also
connected with the theory of Kac-Schwarz operator \cite{KaSch}. The work in
this direction is extremely important for the understanding of unification of
various string models and of essential symmetries of the future string field
theory (in particular generic BRST and Batalin-Vilkovisky symmetries are
very close analogues of the complete sets of the Ward identites, as described
in the general framework in the beginning of section 2). All these things
would constitute a natural next section of these notes, but we choosed to wait
a little longer untill further clarification is achieved in this fragment of
the theory.

Second, we did not touch at all physical interpretations of matrix models,
which include quantum gravity, Yang-Mills theory and many other possible
applications. This should be a subject of very different reviews, for which
the whole content of these notes is just a piece of techniques involved in the
study of physical phenomena.

Third, the biggest $terra\ incognita$ in this branch of science,
which remained beyond the scope of
these notes, is the theory of non-eigenvalue matrix models, which are related
to physical theories in space-time dimensions $d\geq 2$. It is indeed a $terra\
incognita$, at least from the point of view of the semi-rigorous analysis,
which we are reviewing. The recent breakthrough in this field is due to
appearence of the Kazakov-Migdal model \cite{KazMi} (see also the lates review
\cite{SeWe} and refernces therein), which for the first time
opened the possibility  to treat a wide class of non-eigenvalue models by
exact methods of localization theory (other names for this field, which in
fact is growing up into generic theory of integability  are
Duistermaat-Heckman theorem or Fourier analysis on group manifolds). The work
in this direction is, however, only at the early stages and this is why we
decided not to present the first non-systematized results in these notes.
A part of it which is very close to beeing satisfactrily understood is the
"boudary model" of $c=1$ string ("$d=2$ dilaton gravity") - a very important
one from the point of view of general string theory. For the present state of
knowledge about this model see \cite{VeVe}, and
its relation to integrability theory is partly revealed in refs.
\cite{DiMo} and \cite{KhMa}.

In the domain, which $was$ actually reviewed the weakest points are the theory
of continuum limits and that of the multicomponent hierarchies.
These theories, when developed, can
also help to move in the most important direction, which was many time
mentioned in the text above: towards creation of more general theory of
integrability. The next natural step, when approached from this side should be
generalization of conventional integrable hierarchies, which would lift the
restriction to level $k=1$ simply-laced Kac-Moody algebras and unitary
representations. The emerging theory will of course have much to do with both
localization theory and non-eigenvalue matrix models, and when it is created
we shall find ourselves at a new level of understanding, which will be one
step closer to the goal of construction of the entire building of string
theory (mathematical physics) and will probably provide us with unexpected
new means for investigation of the features of the real physical world
around us.

\section{Acknowledgements}

These notes reflect the content of the lectures, given at the University of
Amsterdam and NIKHEF in February-March 1993. I am deeply indebted to
colleagues at Amsterdam for their patience and attention and especially to
Sander Bais for his hospitality and encouragement.

It is a pleasure to thank my collaborators and friends, mentioned by names in
the end of the Introduction for the lessons that they taught me during our
work on the subject of matrix models.

I also aknowledge the support of FOM during the time when these notes were
being written.

\end{document}